\documentclass{elsarticle}

\usepackage{hyperref}

\usepackage{listings}
\usepackage{mathtools}
\usepackage{tikz}
\usetikzlibrary{arrows.meta, shapes, arrows, automata, fit, matrix,
  positioning, decorations.pathreplacing, calligraphy}
\usepackage{amssymb}
\usepackage{amsmath}
\usepackage{leftidx}
\usepackage{hyperref}
\usepackage{paralist}

\journal{Theoretical Computer Science}

\newtheorem{definition}{Definition}
\newtheorem{lemma}{Lemma}
\newtheorem{theorem}{Theorem}
\newtheorem{proposition}{Proposition}

\newtheorem{fact}{Fact}
\newtheorem{assumption}{Assumption}
\newtheorem{example}{Example}
\newcommand{\proof}[1]{\emph{Proof}: {#1}}

\newcommand{\nat}{\mathbb{N}}
\newcommand{\universe}{\mathbb{U}}
\newcommand{\vars}{\mathbb{X}^1}
\newcommand{\Vars}{\mathbb{X}^2}

\newcommand{\globports}{\mathbb{P}}
\newcommand{\globstates}{\mathcal{Q}_\beh}
\newcommand{\statemachines}{\mathbb{B}}
\newcommand{\preds}{\mathbb{A}}
\newcommand{\isdef}{\stackrel{\scriptscriptstyle{\mathsf{def}}}{=}}
\newcommand{\interv}[2]{[{#1},{#2}]}
\newcommand{\tuple}[1]{\langle {#1} \rangle}

\newcommand{\set}[1]{\{ {#1} \}}

\newcommand{\pow}[1]{\mathrm{pow}({#1})}
\newcommand{\dom}[1]{\mathrm{dom}({#1})}

\newcommand{\card}[1]{|\!| {#1} |\!|}

\newcommand{\exptime}[1]{${#1}\mathsf{EXPTIME}$}

\newcommand{\signature}{\Sigma}
\newcommand{\comps}{\mathfrak{C}}
\newcommand{\ncomps}{N}
\newcommand{\acomp}{\mathsf{C}}
\newcommand{\interacs}{\mathfrak{I}}
\newcommand{\ainterac}{\mathsf{I}}
\newcommand{\ninteracs}{M}

\newcommand{\compstates}[1]{\mathit{states}({#1})}

\newcommand{\interports}{\mathfrak{P}}
\newcommand{\interportsof}[1]{\interports({#1})}
\newcommand{\ports}{\mathfrak{P}}
\newcommand{\portsof}[1]{\ports({#1})}
\newcommand{\compof}[1]{\mathit{comp}({#1})}

\newcommand{\outport}[1]{\mathit{out}\_#1}
\newcommand{\inport}[1]{\mathit{in}\_#1}
\newcommand{\iact}[1]{\mathit{#1}}
\newcommand{\stmt}[1]{\mathsf{#1}}

\newcommand{\beh}{\beta}
\newcommand{\behof}[1]{\beh({#1})}
\newcommand{\states}{Q}
\newcommand{\statesof}[1]{\mathit{states}_\beh({#1})}
\newcommand{\events}{P}

\newcommand{\arrow}[2]{\xrightarrow{{\scriptstyle #1}}_{{\scriptstyle #2}}}

\newcommand{\pre}[1]{\text{$\leftidx{^\bullet}{\text{${#1}$}}$}}
\newcommand{\post}[1]{{#1}^\bullet}
\newcommand{\transof}[1]{T_\beh{}(#1)}

\newcommand{\anet}{\mathsf{N}}
\newcommand{\anetof}[1]{\anet({#1})}
\newcommand{\amarkednet}{\mathcal{N}}
\newcommand{\places}{S}

\newcommand{\trans}{T}

\newcommand{\edges}{E}

\newcommand{\amark}{\mathrm{m}}
\newcommand{\reach}[1]{\mathit{Reach}({#1})}
\newcommand{\step}[1]{\mathit{Step}({#1})}

\newcommand{\dead}[1]{\mathit{Dead}({#1})}

\newcommand{\store}{\mathfrak{s}}
\newcommand{\struc}{\alpha}
\newcommand{\asys}{\mathfrak{S}}
\newcommand{\symm}[1]{\sim_{#1}}

\newcommand{\nodes}[1]{\mathrm{nodes}({#1})}
\newcommand{\children}[2]{\mathrm{childs}_{\scriptscriptstyle{#2}}({#1})}
\newcommand{\leaves}[1]{\mathrm{leaves}({#1})}

\newcommand{\comp}{\bullet}

\newcommand{\adl}{\textsf{CL}}
\newcommand{\mso}{\textsf{MSO}}

\newcommand{\predname}[1]{\mathcal{#1}}
\newcommand{\apred}{\predname{A}}
\newcommand{\bpred}{\predname{B}}
\newcommand{\lpred}{\predname{L}}
\newcommand{\lsent}{\phi_{\lpred}} 
\newcommand{\npred}[1]{\#_{\mathsf{pred}}({#1})}
\newcommand{\pred}[2]{\mathsf{pred}_{#1}({#2})}
\newcommand{\rtrees}[1]{\mathbb{T}({#1})}
\newcommand{\charform}[1]{\mathfrak{F}({#1})}

\newcommand{\emp}{\mathsf{emp}}

\let\Asterisk\undefined
\newcommand{\Asterisk}{\mathop{\scalebox{1.9}{\raisebox{-0.2ex}{$\ast$}}}\hspace*{1pt}}%
\renewcommand{\vec}[1]{\mathbf #1}
\newcommand{\fv}[1]{\mathrm{fv}({#1})}

\newcommand{\nary}{\kappa}
\newcommand{\wsks}{$\mathsf{WS}\nary{S}$}

\newcommand{\wsksmodels}[1]{\models^{#1}_{\scriptscriptstyle{\mathsf{wsks}}}}
\newcommand{\arule}{\rho}
\newcommand{\nrules}{P}
\newcommand{\rewrtree}{\mathit{RewrTree}_\asid}
\newcommand{\rtreeof}[2]{\mathcal{T}^{#2}_{\!{#1}}}
\newcommand{\ismark}{\amark}
\newcommand{\markof}[2]{\sigma^{\scriptscriptstyle{#2}}_{\!{#1}}}
\newcommand{\instof}[1]{\mathit{inst}({#1})}
\newcommand{\pathof}[2]{\pi({#1},{#2})}
\newcommand{\autoformof}[1]{\Delta_{#1}}
\newcommand{\nstates}{L}
\newcommand{\interof}[1]{\mathit{inter}({#1})}
\newcommand{\flow}{\Phi}
\newcommand{\iflow}{\Psi}
\newcommand{\trapinv}{\Theta}
\newcommand{\mutexinv}{\Omega}
\newcommand{\trap}{\theta}
\newcommand{\mutex}{\omega}
\newcommand{\initials}{\mathit{Init}}
\newcommand{\intersects}[2]{\card{{#1} \cap {#2}} \geq 1}
\newcommand{\notintersects}[2]{\card{{#1} \cap {#2}}=0}
\newcommand{\singleintersects}[2]{\card{{#1} \cap {#2}}=1}
\newcommand{\deadlock}{\mathit{DeadLock}}
\newcommand{\errset}[1]{\mathit{ErrSet}_{#1}}
\newcommand{\included}[2]{{#1} \subseteq {#2}}

\newcommand{\unfoldrule}{\leftarrow}
\newcommand{\asid}{\Delta}
\newcommand{\linasid}{\Lambda}
\newcommand{\aprof}{\lambda}
\newcommand{\profof}[1]{\aprof({#1})}

\newcommand{\chain}[2]{\mathit{Chain}_{{#1},{#2}}}

\newcommand{\rbr}{{\bf ]\!]}}
\newcommand{\lbr}{{\bf [\![}}
\newcommand{\sem}[2]{{\lbr #1 \rbr}_{\scriptscriptstyle{#2}}}

\newcommand{\lang}[1]{\mathcal{L}({#1})}

\newcommand{\proj}[2]{{#1}\!\!\downarrow_{\scriptscriptstyle{#2}}}

\newcommand{\tokcomp}{\mathsf{S}}
\newcommand{\tokinterac}{\mathsf{T}}
\newcommand{\toktoken}{\mathfrak{t}}
\newcommand{\toknotok}{\mathfrak{n}}

\newcommand{\tokin}{\textit{in}}
\newcommand{\tokout}{\textit{out}}

\newcommand{\ntype}{\mathsf{N}}
\newcommand{\ltype}{\mathsf{L}}
\newcommand{\rrinterac}{\mathsf{R}}
\newcommand{\iointerac}{\mathsf{I}}
\newcommand{\req}{\mathit{req}}
\newcommand{\reply}{\mathit{reply}}
\newcommand{\inp}{\mathit{in}}
\newcommand{\outp}{\mathit{out}}

\newcommand{\treenode}{\mathit{Node}}
\newcommand{\treeleaf}{\mathit{Leaf}}
\newcommand{\treeroot}{\mathit{Root}}

\newcommand{\codesize}{\footnotesize}

\lstdefinelanguage{JavaScript}{
  keywords={typeof, true, false, catch, function, return, null, catch, switch, var, if, in, while, do, od, else, case, break, when, with, assume},
  ndkeywords={class, export, boolean, throw, implements, import, this},
  sensitive=false,
  comment=[l]{//},
  morecomment=[s]{/*}{*/},
  morecomment=[s]{$}{$},
  morestring=[b]',
  morestring=[b]"
}
\lstnewenvironment{code}[1][]{%
    \lstset{%
	    basicstyle=\codesize\ttfamily, 
	    breaklines=false,
	    commentstyle=\color[HTML]{073642}\ttfamily\itshape,
	    identifierstyle=\color{black},
	    inputencoding=latin1,
	    keywordstyle=\color[HTML]{228B22}\bfseries,
	    language=JavaScript,
	    ndkeywordstyle=\color[HTML]{228B22}\bfseries,
	    numbers=left,
	    numbersep=5pt,
	    xleftmargin=30pt,
	    xrightmargin=10pt,
	    frame=Tb,
	    framexleftmargin=15pt,
	    numberstyle=\scriptsize, 
	    prebreak = \raisebox{0ex}[0ex][0ex]{\ensuremath{\hookleftarrow}},
	    showstringspaces=false,
	    stringstyle=\color[rgb]{0.639,0.082,0.082}\ttfamily,
	    upquote=true,
	    mathescape=true,
    	escapebegin=\color[HTML]{073642},
	    tabsize=2,
	    #1
    }
}{}

\lstnewenvironment{examplecode}[1][]{%
    \lstset{%
      basicstyle=\codesize\ttfamily, 
      breaklines=false,
      commentstyle=\color[HTML]{073642}\ttfamily\itshape,
      identifierstyle=\color{black},
      inputencoding=latin1,
      keywordstyle=\color[HTML]{228B22}\bfseries,
      language=JavaScript,
      ndkeywordstyle=\color[HTML]{228B22}\bfseries,
      numbers=left,
      numbersep=5pt,
      xleftmargin=75pt,
      xrightmargin=60pt,
      frame=Tb,
      framexleftmargin=15pt,
      numberstyle=\scriptsize, 
      prebreak = \raisebox{0ex}[0ex][0ex]{\ensuremath{\hookleftarrow}},
      showstringspaces=false,
      stringstyle=\color[rgb]{0.639,0.082,0.082}\ttfamily,
      upquote=true,
      mathescape=true,
      escapebegin=\color[HTML]{073642},
      tabsize=2,
      #1
    }
}{}

\begin{document}

\begin{frontmatter}

  \title{Verification of Component-based Systems with Recursive Architectures}

  \author{Marius Bozga\corref{mycorrespondingauthor}}
  \ead{marius.bozga@univ-grenoble-alpes.fr}
  \author{Radu Iosif}
  \ead{radu.iosif@univ-grenoble-alpes.fr}
  \author{Joseph Sifakis}
  \ead{joseph.sifakis@univ-grenoble-alpes.fr}
  \address{Univ. Grenoble Alpes, CNRS, Grenoble INP\fnref{grenobleinpfootnote}, VERIMAG, 38000 Grenoble, France}
  \fntext[grenobleinpfootnote]{Institute of Engineering Univ. Grenoble Alpes}
  \cortext[mycorrespondingauthor]{Corresponding author}
  
  \begin{abstract}
We study a sound verification method for parametric component-based
systems.  The method uses a resource logic, a new formal specification
language for distributed systems consisting of a finite yet unbounded
number of components. The logic allows the description of architecture
configurations coordinating instances of a finite number of types of
components, by means of inductive definitions similar to the ones used to
describe algebraic data types or recursive data structures.  For parametric
systems specified in this logic, we show that decision problems such as
reaching deadlock or violating critical section are undecidable, in
general. Despite this negative result, we provide for these decision
problems practical semi-algorithms relying on the automatic synthesis of
structural invariants allowing the proof of general safety properties.  The
invariants are defined using the \wsks\ fragment of the monadic second
order logic, known to be decidable by a classical automata-logic
connection, thus reducing a verification problem to checking satisfiability
of a \wsks\ formula.
  \end{abstract}

  \begin{keyword}
    Resource Logic \sep Component-based Distributed Systems \sep Parameterized Verification
  \end{keyword}
  
\end{frontmatter}

\section{Introduction}

Mastering the complexity of a distributed system requires a deep
understanding of its coordination mechanisms. We distinguish between
\emph{endogenous} coordination, that explicitly uses synchronization
primitives in the code describing the behavior of the components
(e.g.\ semaphores, monitors, compare-and-swap, etc.) and \emph{exogenous}
coordination, that defines global rules describing how the components
interact. These two orthogonal paradigms play different roles in the design
of a system: exogenous coordination is used during high-level model
building, whereas endogenous coordination is considered at a later stage of
development, to implement the model using low-level synchronization.

In this paper we focus on (high-level) \emph{exogenous coordination of
  distributed systems}, consisting of a finite yet unbounded number of interconnected
components.  Communication is assumed to be correct, i.e.\ we abstract from
packet losses and corruptions.  Components behave according to a small set
of finite-state abstractions of sequential programs, whose transitions are
labeled with events. They communicate via interactions (handshaking)
modeled as sets of events that occur simultaneously in multiple
components. For instance, Fig.~\ref{fig:ring}(a) shows a token-ring
system, whose components are depicted by yellow boxes containing state
machines that represent their behavior and whose architecture is the set of
connectors between components depicted by solid lines.

The separation between \emph{behavior} and \emph{coordination} is a
fundamental principle in the design of large-scale distributed systems
\cite{KramerMagee98}. This modular view of a distributed system, in which
the internal computation and the state changes of each component are
encapsulated in a well defined interface, is key to scalable design methods
that exploit a conceptual hierarchy. For instance, a ring is a chain whose
final output port is connected to the initial input port, whereas a chain
consists of a (head) component linked to a separate (tail) chain as shown
in Fig.~\ref{fig:ring}(b).  Moreover, system designers are accustomed to
using architectural patterns, such as rings, pipelines, stars, trees, etc.,
that define interactions between (unboundedly large) sets of components.
Such high-level models of real-life distributed systems are suitable for
reasoning about correctness in the early stages of system design, when
details related to network reliability or the implementation of
coordination by means of low-level synchronization mechanisms
(e.g.\ semaphores, monitors, compare-and-swap, etc.) are abstracted away.

\subsection{Running Example}
\label{sec:running-example}

\begin{figure}[t!]
  \centerline{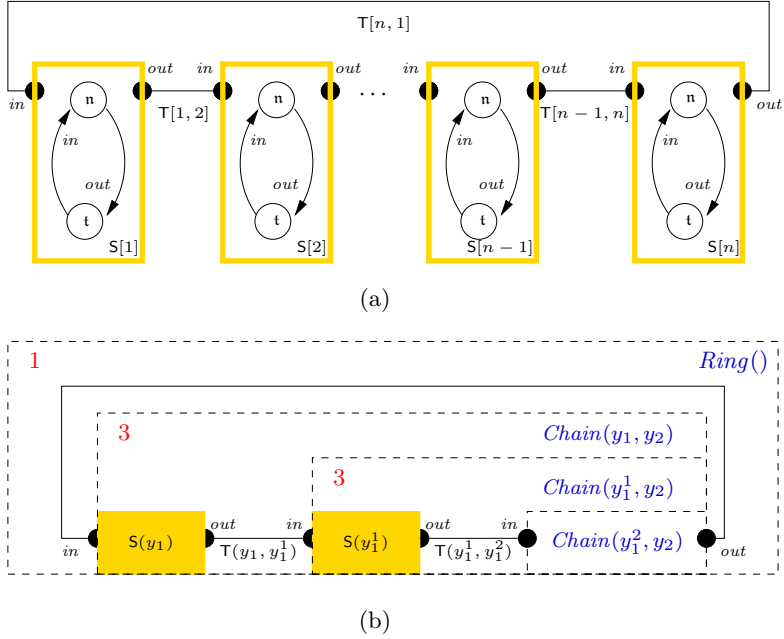}
  \caption{Recursive Specification of a Token-Ring System}
  \vspace*{-\baselineskip}
  \label{fig:ring}
\end{figure}

For starters, we consider the following recursive definition of a
token-ring architecture, composed of a finite but unbounded number of
stations that are instances of the same component type $\tokcomp$,
connected via transfer connectors, of interaction type
$\tokinterac$. The behavior of each station is a machine with two
states, indicating whether the station has a token ($\toktoken$) or
not ($\toknotok$). The number of tokens is constant and equal to the
number of stations that are initially in state $\toktoken$. Each
station not having a token may receive a token from the left, via the
$\tokin$ port (that triggers the transition $\toknotok
\arrow{\tokin}{} \toktoken$) and send its token to the right, via the
$\tokout$ port (that triggers the transition $\toktoken
\arrow{\tokout}{} \toknotok$). We refer to Fig.~\ref{fig:ring}(a) for a
depiction of a token-ring system, defined by the rules:
\begin{eqnarray}
\mathit{Ring}() & \leftarrow & 
\exists y_1 \exists y_2 ~.~ \mathit{Chain}(y_1, y_2) * \tokinterac(y_2,y_1) 
\label{rule:ring} \\
\mathit{Chain}(x_1, x_2) & \leftarrow & \mathit{Comp}(x_1) * \tokinterac(x_1,x_2) *  \mathit{Comp}(x_2)
\label{rule:chain1} \\
\mathit{Chain}(x_1, x_2) & \leftarrow & 
\exists y_1 ~.~ \mathit{Comp}(x_1) * \tokinterac(x_1,y_1) * \mathit{Chain}(y_1, x_2)
\label{rule:chain2} \\
\mathit{Comp}(x_1) & \leftarrow & \tokcomp^\toknotok(x_1)
\label{rule:chain3} \\
\mathit{Comp}(x_1) & \leftarrow & \tokcomp^\toktoken(x_1)
\label{rule:chain4}
\end{eqnarray}
Intuitively, rule \ref{rule:ring} states that a token-ring consists of
a chain of components and an interaction between the $\tokout$ port of
the last and the $\tokin$ port of the first component in the chain. A
chain consists of either two components and an interaction between
their $\tokout$ and $\tokin$ ports, respectively (rule
\ref{rule:chain1}), or a component and an interaction between its
$\tokout$ port and the $\tokin$ port of the first component of a
distinct chain (\ref{rule:chain2}). The rules (\ref{rule:chain3}) and
(\ref{rule:chain4}) intuitively say that a component has type
$\tokcomp$ and can be in state $\toknotok$ or in state $\toktoken$,
respectively. The star symbol $*$ used in the rules
\ref{rule:ring}-\ref{rule:chain4} is a commutative and associative
logical connective that composes sub-systems with disjoint sets of
components and interactions; for instance, in rule \ref{rule:chain1},
the two components declared as $\mathit{Comp}(x_1)$ and
$\mathit{Comp}(x_2)$ are necessarily different, meaning that $x_1$ and
$x_2$ cannot be mapped to the same value. The rules \ref{rule:chain1}
and \ref{rule:chain2} correspond to the base and the inductive case of
a recursive definition of finite chains of length at least two. We
refer to Fig.~\ref{fig:ring}(b) for a depiction of the recursive
unfolding of the above rules.

Note that no transfer of tokens is possible if the number of tokens in
the system is either zero (there is no token to be transfered) or
equals the number of components (there is no room to place a
token). In this case, we say that a system is in a
\emph{deadlock}. The decision problem ``can a token-ring started in a
non-deadlock actually reach a deadlock?'' is challenging, because it
requires a proof that holds for systems of \emph{any} size,
i.e.\ number of components and interactions. We show that, even if
such problems are undecidable, in general, a large number of instances
of these problems can be handled by the methods developed in this
paper.

\subsection{Contributions of this Work}

The parameterized verification method presented in this paper relies on an
idea reported in \cite{DBLP:conf/facs2/BozgaI21}. In addition to a complete
development of the technical results, previously omitted for space reasons,
here we apply the verification method to a specification language based on
a resource logic, that resembles Separation Logic
\cite{DBLP:conf/lics/Reynolds02}, instead of the recursive term algebra
introduced in \cite{DBLP:conf/facs2/BozgaI21}. An extension of this
resource logic has been recently developed for Hoare-style reasoning about
the safety properties of programmed reconfiguration in distributed systems
\cite{DBLP:journals/corr/abs-2107-05253}, to which the work presented here
provides a push-button verification back-end. In combination with the proof
system reported in \cite{DBLP:journals/corr/abs-2107-05253}, the
verification method presented in this paper can automatically prove the
correctness of a distributed system \emph{after the reconfiguration of its
  coordinating architecture}. Since various dialects of (Concurrent)
Separation Logic are being commonly used to specify and reason about
concurrent systems \cite{DBLP:journals/tcs/OHearn07,Sergey16,FarkaN0DF21},
we expect this new logic to be easily accepted by the research and
development community. The contribution of this paper is
three-fold: \begin{enumerate}
\item We introduce a logic-based language, called Configuration Logic
  (\adl), to describe the sets of configurations
  (i.e.\ architectures and local states of components) of distributed
  systems parameterized by (i) the number of components of each type
  that are active in the system, e.g.\ a system with $n$ readers and
  $m$ writers, in which $n$ and $m$ are not known a priori and (ii)
  the pattern in which the interactions occur (e.g.\ a pipeline, ring,
  star or more general hypergraph structures). The language uses
  predicate symbols to hierarchically break the architecture into
  specific patterns. The interpretation of these predicate symbols is
  defined inductively by rewriting rules consisting of formul{\ae}
  that contain predicate atoms, in a way that recalls the usual
  definitions of algebraic datatypes
  \cite{DBLP:journals/jsat/BarrettST07} or heaps
  \cite{DBLP:conf/lics/Reynolds02}.
\item We tackle a parametric safety problem concerning systems
  described in this language, which is essentially checking that the
  reachable states of every instance stays clear of a set of global
  error configurations, such as deadlocks or critical section
  violations. In particular, we show that both the parametric deadlock
  freedom and critical section violation problems are undecidable,
  even for architectures as simple as a chain of components.
\item We develop a verification method that synthesizes parametric
  invariants from the syntactic description of the architecture (in
  \adl) and from the behavior of its components (finite-state
  machines). The invariants and the set of error configurations are
  both described using a decidable fragment of Monadic Second Order
  Logic (\mso), that enables the use of off-the-shelf solvers for
  checking the resulting verification conditions.
\end{enumerate}

\begin{figure}[t!]
  \centerline{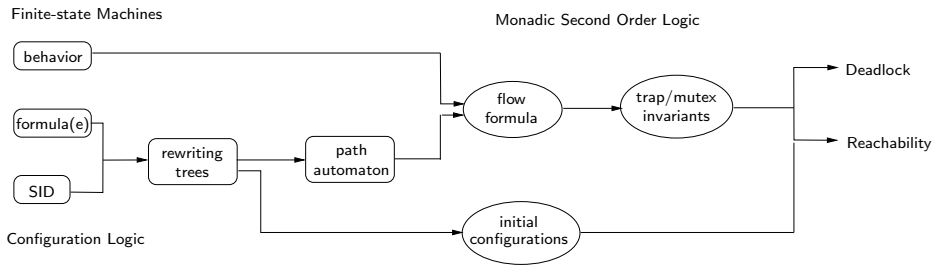}
  \caption{The Synthesis of Verification Conditions}
  \label{fig:verification}
\end{figure}

The stages of the synthesis of verification conditions are depicted in
Fig.~\ref{fig:verification}. The starting point is a set of inductive
definitions and a \adl\ formula describing the initial configurations of
the system, together with a description of component behavior by
finite-state machines.  The \adl\ formula uses inductively defined
predicate symbols defined by sets of rewriting trees, describing the
unfoldings of their inductive definitions.  In particular, these rewriting
trees describe architectures, up to a permutation of the indices of
components of each type.

We use rewriting trees as the backbones of architecture encoding, from
which we derive a \mso\ \emph{flow formula}, that essentially
describes the operational semantics induced by the interactions of the
architecture, relying on (i) a static description of the profile of
the interactions (how they glue component types), and (ii) the
transition systems describing the behavior of component types in the
architecture. The main technical problem in building flow formul{\ae}
is tracking the identities and values of the variables that occur
within predicate symbols, which, in turn, are recursively replaced by
their definitions, in a rewriting tree. For instance, the $y_2$
variable introduced by rule (\ref{rule:ring}) using existential
quantification substitutes the $x_2$ variable from the definition of a
$\mathit{Chain}$ predicate, in rules (\ref{rule:chain1}) and
(\ref{rule:chain2}) several times, before the same variable ($y_2$)
substitutes the $x_1$ variable in rule (\ref{rule:chain3}) or
(\ref{rule:chain4}) (\S\ref{sec:running-example}). This is achieved by
defining a \emph{path automaton} that traverses the rewriting tree
downwards tracking the introduction of a variable by existential
quantification in a rule to the rule where it is instantiated in a
component atom.

The flow formula is subsequently used to define invariants,
i.e.\ over-approx\-i\-ma\-tions of the sets of states reachable from
some initial configuration, in a system defined by some unfolding
(rewriting) of the predicate symbols in \adl\ specifications. In
contrast with the classical approach of invariant inference using,
e.g.\ abstract interpretation \cite{CousotCousot79}, our technique
generates invariants as \mso\ formul{\ae} obtained directly from the
flow formul{\ae}, by a syntactic translation. Analogously, the set of
error configurations (deadlocks and critical section violations) is
obtained directly from the flow formula of the system. The
verification condition asks that the conjunction of the
\mso\ formul{\ae} describing the invariant and error configurations,
respectively, is unsatisfiable; a satisfiable verification condition
might indicate the presence of a spurious error caused by the
over-approximation introduced by the invariant synthesis. Since we
represent the configurations of a system by rewriting trees, we use a
decidable fragment of \mso, interpreted over trees to answer the
verification condition automatically.

The paper is structured as follows. Section \ref{sec:preliminaries}
introduces the definitions of component-based systems and their
operational semantics, Section \ref{sec:logic} describes the resource
logic used to define sets of configurations, Section
\ref{sec:undecidability} describes the parametric decision problems
and gives the general undecidability results, Section
\ref{sec:rewriting} defines rewriting trees formally, together with a
notion of a canonical model, induced by a rewriting tree of a logical
formula, Section \ref{sec:verification} describes the invariant
synthesis and the encoding of the verification conditions in \mso,
Section \ref{sec:conclusion} gives concluding remarks and sketches
directions for future work.

\section{Component-based Systems}
\label{sec:preliminaries}

\subsection{Definitions}

This section introduces the definitions needed to formally describe
our model of component-based systems. We denote by $\nat$ the set of
natural numbers. Given integers $i$ and $j$, we write $\interv{i}{j}$
for the set $\set{i,i+1,\ldots,j}$, assumed to be empty if $i>j$. For
a set $A$ and an integer $i\geq1$, we denote by $A^i$ the $i$-times
Cartesian product of $A$ with itself. By $\pow{A}$ we denote the
powerset of $A$. For a finite set $A$, we denote by $\card{A}$ its
cardinality.

Let $\globports$ be a countable set of \emph{ports}. We consider
classes of component-based systems that share the same
\emph{signature} $\signature = (\comps, \interacs, \interports)$,
where: \begin{compactitem}
\item $\comps = \set{\acomp_1, \ldots, \acomp_\ncomps}$ is a finite
  set of relation symbols of arity one, called \emph{component types},
\item $\interacs = \set{\ainterac_1, \ldots, \ainterac_\ninteracs}$ is
  a finite set of relation symbols of arity $\#(\ainterac_j)\geq1$,
  called \emph{interaction types},
\item $\ports : \comps \cup \interacs \rightarrow \pow{\globports}
  \cup \bigcup_{i\geq1} \globports^i$ is an \emph{interface}
  associating each component type $\acomp\in\comps$ a finite set of
  ports $\portsof{\acomp} \in \pow{\globports}$ and each interaction
  type $\ainterac\in\interacs$ a finite tuple of ports
  $\portsof{\ainterac} \in \globports^{\#(\ainterac)}$.
 \end{compactitem}
W.l.o.g., we assume that $\bigcup_{i=1}^\ncomps \portsof{\acomp_i} =
\globports$ and $\portsof{\acomp_i} \cap \portsof{\acomp_j} = \emptyset$,
for all $1 \leq i < j \leq \ncomps$, i.e.\ each port $p$ belongs to the
interface of exactly one component type, denoted by $\compof{p}$. We denote
by $\interacs^{(k)}$ the subset of interaction types of arity $k$, formally
$\interacs^{(k)} \isdef \{\ainterac \in \interacs ~|~ \#(\ainterac) = k\}$,
for any $k \geq 1$.

\begin{example}[contd. from \S\ref{sec:running-example}]
  \label{ex:token-ring-signature}
  The signature for the token-ring example from Fig.~\ref{fig:ring}a
  is $\signature = \tuple{\set{\tokcomp}, \set{\tokinterac}, \ports}$,
  where $\portsof{\tokcomp} = \set{\tokin, \tokout}$ and
  $\portsof{\tokinterac} = \tuple{\tokout, \tokin}$, i.e.\ the
  interaction type $\tokinterac$ connects an $\tokout$ port to an $\tokin$
  port. \hfill$\blacksquare$
\end{example}

The component and interaction types are interpreted as sets and
relations over a countably infinite \emph{universe} $\universe$ of
indices. The particular nature of indices is not important at this
point; we assume that two indices can only be compared for equality,
with no other associated relation or function. An \emph{architecture}
$\struc$ over the signature $\signature = (\comps, \interacs, \ports)$
associates each component type $\acomp_i$ a set $\struc(\acomp_i)
\subseteq \universe$ and each interaction type $\ainterac_j$ a
relation $\struc(\ainterac_j) \subseteq \universe^{\#(\ainterac_j)}$,
defining: \begin{compactitem}
\item \emph{components} $\acomp_i[u]$, for some $u \in
  \struc(\acomp_i)$, and
\item \emph{interactions} $\ainterac_j[u_1, \ldots,
  u_{\#(\ainterac_j)}]$, for some $\tuple{u_1, \ldots,
  u_{\#(\ainterac_j)}} \in \struc(\ainterac_j)$.
\end{compactitem}
Intuitively, an architecture is a description of the static structure
of a system. Note that an index $u \in \universe$ can refer two (or
more) different components $\acomp_1[u]$ and $\acomp_2[u]$ belonging
to different component types and similarly, a tuple of indices
$\vec{u}$ can refer two (or more) different interactions
$\ainterac_1[\vec{u}]$ and $\ainterac_2[\vec{u}]$ of different
interaction types, with the same arity; interactions may occur
disconnected from components,
e.g.\ $\ainterac[u_1,\ldots,u_{\#(\ainterac)}]$ does not necessarily
mean that each component
$\compof{\tuple{\portsof{\ainterac}}_i}[u_i]$, for all $i \in
\interv{1}{\#(\ainterac)}$ is present in the structure.

\begin{example}[contd. from Example \ref{ex:token-ring-signature}]
  \label{ex:token-ring-structure}
  Letting indices be natural numbers, the structure of
  Fig.~\ref{fig:ring}a is described by the architecture $\struc$ over
  $\signature$, such that $\struc(\tokcomp) = \interv{1}{n}$ and
  $\struc(\tokinterac) = \{\tuple{1,2}, \tuple{2,3}, \ldots,
  \tuple{n,1}\} \rangle$, for some $n \geq 2$. \hfill$\blacksquare$
\end{example}

A \emph{behavior} is described by a finite state machine $M =
(\states, \events, \arrow{}{})$, where $\states$ is a finite set of
states, $\events\subseteq\globports$ is a finite set of ports and
$\arrow{}{} \subseteq \states \times \events \times \states$ is a
transition relation; we write $q \arrow{p}{} q'$ instead of $(q,p,q')
\in \arrow{}{}$. Let $\statemachines$ be the set of finite-state
machines with ports from $\globports$. The \emph{behavior map} $\beh : \comps
\rightarrow \statemachines$ associates each component type
$\acomp\in\comps$ with a state machine $\behof{\acomp} =
(Q_\acomp,\portsof{\acomp},\arrow{}{\acomp})$, whose set of states is
denoted by $\statesof{\acomp} \isdef Q_\acomp$. In the following, we
consider w.l.o.g. that $\statesof{\acomp_i} \cap \statesof{\acomp_j} =
\emptyset$, for all $1 \leq i < j \leq \ncomps$ and define
$\globstates \isdef \bigcup_{i=1}^\ncomps \statesof{\acomp_i}$.

\begin{example}[contd. from Example \ref{ex:token-ring-signature}]
  \label{ex:token-ring-behavior}
  The behavior of the component type $\tokcomp$ in
  Fig.~\ref{fig:ring}a is described by the state machine
  $\behof{\tokcomp} = (\set{\toktoken,\toknotok},
  \set{\tokin,\tokout}, \arrow{}{})$, with transitions $\toknotok
  \arrow{\tokin}{} \toktoken$ and $\toktoken \arrow{\tokout}{}
  \toknotok$. \hfill$\blacksquare$ 
\end{example}

For a port $p$ we denote by $\transof{p}$ the set of transitions labeled by
$p$ in the finite-state machine associated by $\beh$ to $\compof{p}$.  We
extend this notation to tuples of ports by taking
$\transof{\tuple{p_1,\ldots, p_n}} \isdef \transof{p_1} \times \cdots
\times \transof{p_n}$.  That is, the set $\transof{\tuple{p_1,\ldots,
    p_n}}$ contains $n$-tuples of transitions $\tuple{t_1,\ldots, t_n}$
labeled respectively with ports $\tuple{p_1,\ldots, p_n}$.  For an
interaction $\ainterac$ we use $\transof{\ainterac}$ as a shortcut for
$\transof{\portsof{\ainterac}}$. In particular, $\transof{\ainterac}$
contains the set of tuples of component's transitions that are \emph{synchronizing}
by interaction $\ainterac$.

A \emph{system} $\asys$ is a tuple $(\signature, \struc, \beh)$, where
$\struc$ is an architecture and $\beh$ is a behavior associated with
the signature $\signature$. When $\signature$ is clear from the
context, we omit it and denote a system as $(\struc,\beh)$.

\subsection{Operational Semantics}

We represent the operational semantics of a system as a Petri net,
recalled below for self-containment reasons. A \emph{Petri net} is a
tuple $\anet = (\places,\trans,\edges)$, where $\places$ is a set of
\emph{places}, $\trans$ is a set of \emph{transitions}, $\places \cap
\trans = \emptyset$, and $\edges \subseteq (\places \times \trans)
\cup (\trans \times \places)$ is a set of \emph{edges}. Given $x,y \in
\places \cup \trans$, we write $E(x,y)\isdef1$ if $(x,y) \in E$ and
$E(x,y)\isdef0$, otherwise. Let $\pre{x} \isdef \set{y \in \places
  \cup \trans \mid E(y,x)=1}$, $\post{x} \isdef \set{y \in \places
  \cup \trans \mid E(x,y)=1}$ and lift these definitions to sets of
nodes. A \emph{marking} of $\anet$ is a function $\amark : \places
\rightarrow \nat$. A transition $t$ is \emph{enabled} in $\amark$ if
and only if $\amark(s) > 0$ for each place $s \in \pre{t}$.  We write
$\amark \arrow{t}{} \amark'$ whenever $t$ is enabled in $\amark$ and
$\amark'(s) = \amark(s) - E(s,t) + E(t,s)$, for all $s \in \places$
and $t \in \trans$. A sequence of transitions $\sigma = t_1,
\ldots,t_n$ is a \emph{firing sequence}, written $\amark
\arrow{\sigma}{} \amark'$ if and only if either (i) $n=0$ and
$\amark=\amark'$, or (ii) $n\geq1$ and there exist markings $\amark_1,
\ldots, \amark_{n-1}$ such that $\amark \arrow{t_1}{} \amark_1 \ldots
\amark_{n-1} \arrow{t_{n}}{} \amark'$. A marking $\amark$ is a
\emph{deadlock} of a Petri net $\anet = (\places, \trans, \edges)$ if and
only if no transition $t \in \trans$ is enabled in $\amark$ and let
$\dead{\anet}$ denote the set of deadlocks of $\anet$.

A \emph{marked Petri net} is a pair $\amarkednet=(\anet,\amark_0)$,
where $\amark_0$ is the \emph{initial marking} of $\anet$. A firing
sequence is \emph{initial} if it starts in $\amark_0$. A marking
$\amark$ is \emph{reachable} in $\amarkednet$ if there exists an
initial firing sequence ending in $\amark$. Let $\reach{\amarkednet}$
(resp. $\step{\amarkednet}$) be the set of markings reachable (in one
step) in $\amarkednet$. For simplicity, we write
$\reach{\anet,\amark_0}$ and $\step{\anet,\amark_0}$ instead of
$\reach{(\anet,\amark_0)}$ and $\step{(\anet,\amark_0)}$,
respectively. A marked Petri net $\amarkednet$ is
\emph{boolean}\footnote{Boolean Petri nets are sometimes called $1$-safe or
$1$-bounded in the literature.} if $\amark(s) \leq 1$, for each $s \in
\places$ and $\amark \in \reach{\amarkednet}$. All marked Petri nets
considered in the following will be boolean and we blur the
distinction between a marking $\amark : \places \rightarrow \set{0,1}$
and the set $\set{s \in \places \mid \amark(s) = 1}$, by writing
$s\in\amark$ (resp. $s\not\in\amark$) instead of $\amark(s)=1$
(resp. $\amark(s)=0$).

\begin{definition}\label{def:execution-semantics}
  The \emph{operational semantics} of a system $\asys =
  (\signature,\struc,\beh)$ with signature $\signature = (\comps,
  \interacs, \ports)$ is represented by the Petri net
  $\anetof{\asys} = (\places, \trans, \edges)$, where:
\[\begin{array}{rcl}
\places & \isdef & \bigcup_{\acomp\in\comps}
\set{ ~q[u] ~\mid~ q \in \statesof{\acomp},~ u \in \struc(\acomp)~}
\\[3mm]
\trans & \isdef & \bigcup_{\ainterac\in\interacs}
\{ ~ (\ainterac[u_1,\ldots,u_{\#(\ainterac)}], \tuple{t_1,\ldots, t_{\#(\ainterac)}})
  ~\mid~ \tuple{p_1, \ldots, p_{\#(\ainterac)}} = \interportsof{\ainterac}, \\[1mm]
  & & \hspace{1.5cm} \tuple{u_1,\ldots,u_{\#(\ainterac)}} \in \struc(\ainterac),~
 \tuple{t_1,\ldots, t_{\#(\ainterac)}} \in \transof{\ainterac}, \\[1mm]
  & & \hspace{1.5cm} \forall i,j \in \interv{1}{\#(\ainterac)}.~ i\not=j \Rightarrow
  u_i\not=u_j \mbox{ or } \compof{p_i} \not= \compof{p_j} ~ \}
  \\[3mm]
  \edges & \isdef & \bigcup_{\ainterac\in\interacs}
  \{ ~ (q[u_i], (\ainterac[u_1,\ldots,u_{\#(\ainterac)}], \tuple{t_1,\ldots, t_{\#(\ainterac)}}) ), \\[1mm]
  & & \hspace{1cm} ((\ainterac[u_1,\ldots,u_{\#(\ainterac)}], \tuple{t_1,\ldots, t_{\#(\ainterac)}}), q'[u_i]) ~\mid ~\\[1mm]
  & & \hspace{1.5cm} t_i = (q \arrow{p_i}{} q'),~i \in \interv{1}{\#(\ainterac)} ~ \}
\end{array}\]
\end{definition}

The places, transitions and edges of $\anetof{\asys}$ are defined
jointly by the architecture $\struc$ and the behavior $\beh$.  For
every component $\acomp[u]$ in $\alpha$, the Petri net contains the
places $q[u]$, for each state $q$ in $\statesof{\acomp}$.  For every
interaction $\ainterac[u_1,\ldots,u_{\#(\ainterac)}]$ in $\alpha$, the
Petri net contains one transition for every tuple
$\tuple{t_1,\ldots,t_{\#(\ainterac)}}$ of transitions of component
types behavior, which are \emph{synchronizing} according to
$\ainterac$, that is, where their labeling ports
$\tuple{p_1,\ldots,p_{\#(\ainterac)}}$ form the tuple
$\ports(\ainterac)$. Moreover, a transition corresponding to an
interaction $\ainterac[u_1,\ldots,u_{\#(\ainterac)}]$ involves
pairwise distinct components; the last condition in the definition of
$\trans$ above requires that $u_i \neq u_j$ or $\compof{p_i} \neq
\compof{p_j}$. Finally, edges are defined according to the tuple of
synchronizing transitions $\tuple{t_1,\ldots,t_{\#(\ainterac)}}$ and
connect to pre- and post- places, respectively $q[u_i]$ and $q'[u_i]$,
for each involved component $\acomp[u_i]$. For the sake of clarity we
omit writing the tuple $\tuple{t_1,\ldots, t_{\#(\ainterac)}}$ when it
is determined by $\ainterac$, namely for those behaviors where a port
labels exactly one transition, in each state machine, as it is the
case in the example below:

\begin{example}[contd. from Example \ref{ex:token-ring-behavior}]
  \label{ex:token-ring-pn}
  \begin{figure}[htb]
  \centerline{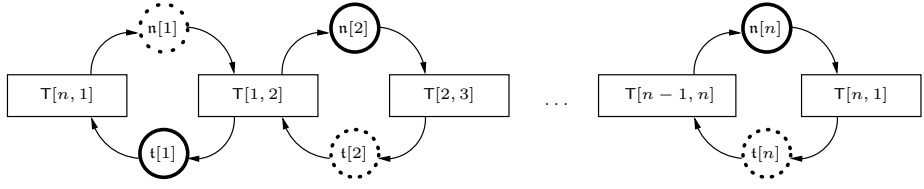}
  \caption{The Execution Semantics of a Token-Ring System}
  \label{fig:ring-pn}
  \end{figure}
  The Petri net describing the execution semantics of the token-ring
  system from Fig.~\ref{fig:ring}(a) is given in
  Fig.~\ref{fig:ring-pn}. Consider the marking of this Petri net
  depicted in dashed lines. From this marking, the sequence of
  interactions $\tokinterac[1,2], \tokinterac[2,3], \ldots,
  \tokinterac[n-1,n], \tokinterac[n,1]$ can be fired any number of
  times, in this order. These interactions correspond to the joint
  execution of the transitions labeled by the ports $\mathit{out}$ and
  $\mathit{in}$, from any two adjacent components $i$ and $(i\mod
  n)+1$ of the ring, respectively. \hfill$\blacksquare$
\end{example}

The executions of a system $\asys = (\signature, \struc, \beh)$ can be
represented by the firing sequences of a marked Petri net involving
only those boolean markings of $\anet(\asys)$, that contain exactly
one state $q \in \statesof{\acomp}$ per component $\acomp[u]$, such
that $u \in \struc(\acomp)$. We formalize and prove this fact below:

\begin{definition}\label{def:precise-marking}
  Given a system $\asys = (\signature, \struc, \beh)$ with signature
  $\signature = (\comps, \interacs, \ports)$, a marking $\amark$ of
  the Petri net $\anet(\asys) = (\places, \trans, \edges)$ is
  \emph{precise} if and only if, for each $\acomp \in \comps$ and each
  $u \in \struc(\acomp)$, we have \(\card{\amark \cap \set{q[u] \mid q
      \in \statesof{\acomp}}} = 1\).
\end{definition}

\begin{proposition}\label{prop:precise-marking-reachability}
  Given a system $\asys$ and a precise initial marking $\amark_0$ of
  $\anetof{\asys}$, every marking $\amark \in \reach{\anetof{\asys},
    \amark_0}$ is precise.
\end{proposition}
\proof{ Let $\asys = (\signature, \struc)$ and $\amark' \in
  \reach{\anetof{\asys}, \amark_0}$ be a marking. The proof goes by
  the length $n\geq0$ of the firing sequence taking $\amark_0$ into
  $\amark'$. For the base case $n=0$, we have that $\amark'=\amark_0$
  is precise, by hypothesis. For the inductive step $n \geq 1$, let
  $\amark$ be the predecessor of $\amark'$ in the sequence, hence
  $\amark$ is precise for $\asys$, by the inductive hypothesis. Then,
  by Def. \ref{def:execution-semantics}, there exists a transition
  $(\ainterac[u_1, \ldots, u_{\#(\ainterac)}], \tuple{t_1,\ldots,t_{\#(\ainterac)}})$ in $T$, such that
  \begin{align}\label{eq:amarkprime}
    \amark' = (\amark \setminus \set{q_1[u_1], \ldots,
      q_{\#(\ainterac)}[u_{\#(\ainterac)}]}) \cup \set{q_1'[u_1], \ldots,
      q_{\#(\ainterac)}'[u_{\#(\ainterac)}]} \end{align} where
  $\portsof{\ainterac} = \tuple{p_1, \ldots, p_{\#(\ainterac)}}$ and $t_i =
  (q_i \arrow{p_i}{} q'_i) \in \arrow{}{\compof{p_i}}$ for all $i \in
  \interv{1}{\#(\ainterac)}$ and moreover $u_i\not=u_j$ or $\compof{p_i}
  \not= \compof{p_j}$ for all distinct $i,j\in \interv{1}{\#(\ainterac)}$.
  
  Let $\acomp \in \comps$ be a component type and let $u \in
  \struc(\acomp)$ be an index. Since $\amark$ is precise, we have
  $\amark \cap \set{q[u] \mid q \in \statesof{\acomp}} =
  \set{q^*[u]}$, for some state $q^*$. We distinguish the following
  cases: \begin{compactitem}
  \item $u=u_i$ for some $i\in \interv{1}{\#(\ainterac)}$ and $q^* = q_i$
    In this case, we have: \[\amark'
    \cap \set{q[u] \mid q \in \statesof{\acomp}} = \set{q_i'[u]}
    \text{, by definition of $\amark'$ in (\ref{eq:amarkprime})} \]
  \item $u=u_i$ for some $i\in \interv{1}{\#(\ainterac)}$ and $q^* \not=
    q_i$.  In this case, as the transition $(\ainterac[u_1, \ldots,
      u_{\#(\ainterac)}], \tuple{t_1,\ldots,t_{{\#(\ainterac)}}})$ fires in
    $\amark$ it follows that $q_i[u] \in \amark$.  But then $q_i[u] \in
    \amark \cap \set{q[u] \mid q \in \statesof{\acomp}} = \{q^*[u]\}$
    implies $q_i = q^*$, which is a contradiction.
  \item $u \not\in \set{u_1, \ldots, u_{\#(\ainterac)}}$. In this
    case, we have:
    \[\amark \cap
    \set{q[u] \mid q \in \statesof{\acomp}} = \amark' \cap \set{q[u]
      \mid q \in \statesof{\acomp}}\]
  \end{compactitem}
  In all non-contradictory cases, we obtain that $\card{\amark \cap
    \set{q[u] \mid q \in \statesof{\acomp}}} = 1$. Because
  $\acomp\in\comps$ and $u\in\struc(\acomp)$ are arbitrary, $\amark$
  is precise for $\asys$. \qed}

\section{The Configuration Logic}
\label{sec:logic}

By \emph{configuration} of a system, we understand the architecture
describing the components and interactions from the system, together
with a snapshot of its current state. Configurations are used to
reason about \emph{parametric} systems, that share a common
architectural pattern (style) and differ in the number of instances of
a certain component type. For instance, a token-ring
(Fig.~\ref{fig:ring}) applies the same architectural pattern (the
output of each component is connected to its right neighbor in a
round-robin fashion) to any number $n\geq2$ of components.

\begin{definition}\label{def:configuration}
  A \emph{configuration} of a system $\asys =
  (\signature,\struc,\beh)$ is a pair $(\struc,\amark)$, where
  $\struc$ is an architecture over the signature $\signature$ and
  $\amark$ is a precise marking of $\anetof{\asys}$.
\end{definition}

We aim at describing sets of configurations recursively
(i.e.\ configurations with more complex structure being obtained by
composing simpler ones), using the following definition of
composition:

\begin{definition}\label{def:composition}
  Two architectures $\struc_1$ and $\struc_2$, over the signature
  $\signature = (\comps, \interacs, \ports)$, are \emph{disjoint} if
  and only if $\struc_1(\acomp) \cap \struc_2(\acomp) = \emptyset$,
  for all $\acomp\in\comps$ and $\struc_1(\ainterac) \cap
  \struc_2(\ainterac) = \emptyset$, for all
  $\ainterac\in\interacs$. If $\struc_1$ and $\struc_2$ are disjoint,
  the \emph{composition} of configurations $(\struc_1,\amark_1)$ and
  $(\struc_2,\amark_2)$ is $(\struc_1,\amark_1) \comp
  (\struc_2,\amark_2) \isdef (\struc_1 \cup \struc_2, \amark_1 \cup
  \amark_2)$, where $\struc_1 \cup \struc_2$ denotes the pointwise
  union of the architectures $\struc_1$ and $\struc_2$:
  \begin{compactitem}
  \item $(\struc_1 \cup \struc_2)(\acomp) = \struc_1(\acomp) \cup
    \struc_2(\acomp)$, for each $\acomp\in\comps$,
  \item $(\struc_1 \cup \struc_2)(\ainterac) = \struc_1(\ainterac)
    \cup \struc_2(\ainterac)$, for each $\ainterac\in\interacs$.
  \end{compactitem}
  The composition $(\struc_1,\amark_1) \comp (\struc_2,\amark_2)$ is
  undefined if $\struc_1$ and $\struc_2$ are not disjoint.
\end{definition}
It is easy to check that the composition of configurations
$(\struc_1,\amark_1)$ and $(\struc_2,\amark_2)$ with disjoint
architectures is again a configuration, in particular $\amark_1 \cap
\amark_2 = \emptyset$ and $\amark_1 \cup \amark_2$ is a precise
marking of the Petri net $\anetof{\signature,\struc_1 \cup
  \struc_2,\beh}$.

We introduce the \emph{configuration logic} (\adl) to describe (possibly
infinite) sets of configurations, via inductive definitions. Let
$\signature = (\comps,\interacs,\ports)$ be a signature and $\beh :
\comps \rightarrow \statemachines$ be a behavior map, fixed for
the rest of this section. Let $\vars$ be a set of first-order
variables and $\preds$ be a countably infinite set of predicate
symbols, where $\#(\apred) \geq 0$ denotes the arity of a predicate
symbol $\apred \in \preds$. The formul{\ae} of \adl\ are inductively
defined by the following syntax:
\[\begin{array}{rcl}
\phi & ::= & \emp \mid \acomp(x) \mid \acomp^q(x) \mid \ainterac(x_1,
\ldots, x_{\#(\ainterac)}) \mid \apred(x_1, \ldots, x_{\#(\apred)})
\mid \phi * \phi \mid \exists x ~.~ \phi
\end{array}\]
where $\acomp\in\comps$ are component types, $q \in \statesof{\acomp}$
are states, $\ainterac\in\interacs$ are interaction types,
$\apred\in\preds$ are predicate symbols and $x_1, x_2, \ldots\in\vars$
denote first-order variables. Atomic formul{\ae} of the form
$\acomp(x)$ or $\acomp^q(x)$, $\ainterac(x_1, \ldots,
x_{\#(\ainterac)})$ and $\apred(x_1, \ldots, x_{\#(\apred)})$ are
called \emph{component}, \emph{interaction} and \emph{predicate
  atoms}, respectively. The logical connective $*$ is an associative
and commutative \emph{separating conjunction} operator.

By $\fv{\phi}$ we denote the set of free variables that do not occur
within the scope of a quantifier; $\phi$ is called
\emph{quantifier-free} (q.f.) if and only if it has no quantifiers and
a \emph{sentence} if and only if $\fv{\phi}=\emptyset$,
respectively. A \emph{substitution} is a partial mapping $\theta :
\vars \rightarrow \vars$ and $\phi\theta$ is the formula obtained by
replacing each variable $x \in \fv{\phi} \cap \dom{\theta}$ by
$\theta(x)$ in $\phi$, where $\dom{\theta} \isdef \set{x \in \vars
  \mid \theta \text{ is defined at } x}$. We denote by $[x_1/y_1,
  \ldots, x_k/y_k]$ the substitution that replaces $x_i$ with $y_i$,
for all $i \in \interv{1}{k}$ and is undefined everywhere else. This
notation is extended to tuples of variables of equal length as
$[\vec{x}/\vec{y}]$, where $\vec{x} = \tuple{x_1, \ldots, x_k}$ and
$\vec{y} = \tuple{y_1, \ldots, y_k}$.

Intuitively, a formula $\emp$ describes configurations with empty
architecture, $\acomp(x)$ (resp. $\acomp^q(x)$) describes
configurations with architectures consisting of a single instance of
the component type $\acomp$, indexed by the value of $x$ (resp. in
state $q$), and $\ainterac(x_1,\ldots,x_k)$ describes a single
interaction of type $\ainterac$, between components indexed by $x_1,
\ldots, x_k$, respectively. The formula $\acomp^{q_1}(x_1) * \ldots *
\acomp^{q_k}(x_k) * \ainterac(x_1,\ldots,x_k)$ describes an
architecture consisting of $k$ \emph{pairwise distinct} instances of
the component type $\acomp$, in states $q_1, \ldots, q_k$,
respectively, joined by an interaction of type $\ainterac$. The
formula $\ainterac(x_1,\ldots,x_k) * \ainterac(x'_1, \ldots, x'_k)$
states the existence of two interactions of type $\ainterac$, with
\emph{distinct} tuples of indices, given by the values of $\tuple{x_1,
  \ldots, x_k}$ and $\tuple{x'_1, \ldots, x'_k}$, respectively,
i.e.\ the values of $x_i$ and $x'_i$ must differ for at least one $i
\in \interv{1}{k}$.

The semantics of \adl\ formul{\ae} is given by a satisfaction relation
$\models_\asid^\store$ between configurations and formul{\ae}. This
relation is parameterized by: \begin{compactitem}
\item a \emph{store} $\store : \vars \rightarrow \universe$, i.e. a
  function mapping variables to indices, and
\item a \emph{set of inductive definitions} (SID) $\asid$ consisting
  of rules of the form $\apred(x_1, \ldots, x_{\#(\apred)}) \leftarrow
  \phi$, where $\phi$ is a \adl\ formula such that $\fv{\phi} =
  \set{x_1, \ldots, x_{\#(\apred)}}$.
\end{compactitem}
The satisfaction relation is defined inductively on the structure of the
formul{\ae}:
\[\begin{array}{rclcl}
(\struc,\amark) & \models^\store_\asid & \emp & \iff & 
\struc(\acomp) = \emptyset \text{, for all $\acomp\in\comps$, } \struc(\ainterac) = \emptyset, \\
&&&& \text{for all $\ainterac\in\interacs$ and $\amark = \emptyset$}
\\[2mm]
(\struc,\amark) & \models^\store_\asid & \acomp(x) & \iff & \struc(\acomp) = \set{\store(x)},~ 
\struc(\acomp') = \emptyset \text{, for all } \\  
&&&& \acomp' \in \comps \smallsetminus \set{\acomp},
\struc(\ainterac) = \emptyset \text{, for all } \ainterac\in\interacs \text{ and } \\
&&&& \amark = \set{q[\store(x)]} \text{, for some } q \in \statesof{\acomp}
\\[2mm]
(\struc,\amark) & \models^\store_\asid & \acomp^q(x) & \iff & \struc(\acomp) = \set{\store(x)},~ 
\struc(\acomp') = \emptyset \text{, for all } \\ 
&&&& \acomp' \in \comps \smallsetminus \set{\acomp}, \struc(\ainterac) = \emptyset \text{, for all } \ainterac\in\interacs \\
&&&& \text{ and } \amark = \set{q[\store(x)]}
\\[2mm]
(\struc,\amark) & \models^\store_\asid & \ainterac(x_1, \ldots, x_{\#(\ainterac_k)}) & \iff & 
\struc(\acomp) = \emptyset \text{, for all } \acomp\in\comps,~  \struc(\ainterac) = \{\langle\store(x_1), \\ 
&&&& \ldots,\store(x_{\#(\ainterac_k)})\rangle\},~ \struc(\ainterac') = \emptyset \text{, for all } \\
&&&& \ainterac' \in \interacs \smallsetminus \set{\ainterac} \text{ and } \amark = \emptyset
\\[2mm]
(\struc,\amark) & \models^\store_\asid & \apred(x_1,\ldots,x_{\#(\apred)}) & \iff &
(\struc,\amark) \models^\store_\asid \phi[y_1/x_1, \ldots, y_{\#(\apred)}/x_{\#(\apred)}], \\
&&&& \text{for some rule } \apred(y_1,\ldots,y_{\#(\apred)}) \leftarrow \phi \in \asid
\end{array}\]
\[\begin{array}{rclcl}
(\struc,\amark) & \models^\store_\asid & \phi_1 * \phi_2 & \iff & \text{there exist configurations } 
(\struc_1,\amark_1), (\struc_2,\amark_2) \\ 
&&&& \text{such that } (\struc,\amark) = (\struc_1,\amark_1) \comp (\struc_2,\amark_2) \text{ and } \\
&&&& (\struc_i,\amark_i) \models^\store_\asid \phi_i \text{, for } i = 1,2.
\\[2mm]
(\struc,\amark) & \models^\store_\asid & \exists x ~.~ \phi_1 & \iff & (\struc,\amark) \models^{\store'}_\asid \phi_1 \text{, for some store } \store' \text{ that agrees} \\ 
&&&& \text{with } \store \text{ on all variables from } \vars \smallsetminus \set{x}
\end{array}\]
To simplify the notation, we consider that $\asid$ is fixed and omit
the $\asid$ subscript in the following. If the formula $\phi$ is a
sentence, we can omit the store $\store$ from the satisfaction
relation $\models^\store$ and write $(\struc,\amark) \models \phi$. In
this case, $(\struc,\amark)$ is said to be a \emph{model} of $\phi$
and denote by $\sem{\phi}{}$ the set of models of the sentence
$\phi$. For two sentences $\phi$ and $\psi$, we say that $\phi$
\emph{entails} $\psi$ if $\sem{\phi}{} \subseteq \sem{\psi}{}$,
written $\phi \models \psi$.

\begin{example}[contd. from \S\ref{sec:running-example}]\label{ex:chains}
  The SID consisting of the rules \ref{rule:ring}-\ref{rule:chain4}
  (\S\ref{sec:running-example}) defines systems with token-ring
  architectures. On the other hand, the rules below define chains of
  $\tokcomp$ and $\tokinterac$ components, with at least $n, t\in\nat$
  components in state $\toknotok$ and $\toktoken$, respectively:
  \[\begin{array}{rclcrcl}
  \chain{0}{0}(x,x) & \unfoldrule & \tokcomp(x) \\
  \chain{0}{1}(x,x) & \unfoldrule & \tokcomp^{\toktoken}(x) & & \chain{1}{0}(x,x) & \unfoldrule & \tokcomp^{\toknotok}(x)
  \end{array}\]
  
  \vspace*{-.5\baselineskip}
  \[\begin{array}{rcl}
  \chain{n}{t}(x,z) & \unfoldrule & \exists y.~\tokcomp^{\toktoken}(x) \ast \tokinterac(x,y) \ast \chain{n}{t\dot{-}1}(y,z) \\
  \chain{n}{t}(x,z) & \unfoldrule & \exists y.~\tokcomp^{\toknotok}(x) \ast \tokinterac(x,y) \ast \chain{n\dot{-}1}{t}(y,z) 
  \end{array}\]
  where $k\dot{-}1 \isdef \max(k-1,0)$, $k\in\nat$. \hfill$\blacksquare$
\end{example}

Below we show that \adl\ sentences define indeed system
configurations.

\begin{proposition}\label{prop:struc-mark}
  Given a sentence $\phi$ of \adl, if $(\struc,\amark) \in
  \sem{\phi}{}$ then $\amark$ is a precise marking of the Petri net
  $\anetof{\signature,\struc,\beh}$.
\end{proposition}
\proof{ We prove a more general statement: for each formula $\phi$ and
  each store $\store$, if $(\struc,\amark) \models^\store \phi$ then
  $\amark$ is a precise marking of $\anetof{\signature,\struc,\beh}$. The proof is by induction on the
  definition of the $\models^\store$ relation, by distinguishing the
  following cases: \begin{itemize}
  \item $\phi \in \set{\emp, \ainterac(x_1, \ldots,
    x_{\#(\ainterac)})}$: $\amark = \emptyset$ and $\struc(\acomp) =
    \emptyset$, for all $\acomp\in\comps$, thus $\amark$ is trivially
    precise for $\anetof{\signature,\struc,\beh}$.
  \item $\phi = \acomp(x)$: $\amark = \set{q[\store(x)]}$, for some $q
    \in \statesof{\acomp}$, $\struc(\acomp) = \set{\store(x)}$ and
    $\struc(\acomp') = \emptyset$, for all $\acomp' \in
    \comps\smallsetminus\set{\acomp}$, hence $\amark$ is precise for
    $\anetof{\signature,\struc,\beh}$.
  \item $\phi = \acomp^q(x)$: $\amark =
    \set{q[\store(x)]}$, $\struc(\acomp) = \set{\store(x)}$
    and $\struc(\acomp') = \emptyset$, for all $\acomp' \in
    \comps\smallsetminus\set{\acomp}$, hence $\amark$ is precise for
    $\anetof{\signature,\struc,\beh}$.
  \item $\phi = \apred(x_1, \ldots, x_{\#(\apred)})$: $(\struc,\amark)
    \models^\store \psi[y_1/x_1, \ldots,
      y_{\#(\apred)}/x_{\#(\apred)}]$, for some rule
    $\apred(y_1,\ldots,y_{\#(\apred)}) \leftarrow \psi\in\asid$ and
    apply the inductive hypothesis.
  \item $\phi = \phi_1 * \phi_2$: if $(\struc,\amark) \models^\store
    \phi_1 * \phi_2$ then there exists configurations $(\struc_i,
    \amark_i) \models^\store \phi_i$, for $i = 1,2$, such that
    $\struc_1$ and $\struc_2$ are disjoint, $\struc = \struc_1 \cup
    \struc_2$ and $\amark = \amark_1 \cup \amark_2$. By the inductive
    hypothesis, $\amark_i$ is a precise marking of
    $\anetof{\signature,\struc_i,\beh}$, for $i = 1,2$. Let $\acomp
    \in \comps$ be a component type and let $u \in \struc(\acomp)$ be
    an index. Then either $u \in \struc_1(\acomp)$ or $u \in
    \struc_2(\acomp)$, but not both. Let us consider the case $u \in
    \struc_1(\acomp) \smallsetminus \struc_2(\acomp)$, the other case being
    symmetric. Then $\amark \cap \set{q[u] \mid q
      \in \compstates{\acomp}} = \amark_1 \cap \set{q[u]
      \mid q \in \compstates{\acomp}}$, hence $\card{\amark
      \cap \set{q[u] \mid q \in
        \compstates{\acomp}}} = \card{\amark_1 \cap
      \set{q[u] \mid q \in \compstates{\acomp}}} =
    1$. Since the choices of $\acomp$ and $u$ were arbitrary, we
    obtain that $\amark$ is precise.
  \item $\phi = \exists x ~.~ \phi_1$: by an application of the
    inductive hypothesis. \qed
  \end{itemize}}



\section{Decision Problems}
\label{sec:undecidability}

A decision problem is a class of yes/no queries of the same kind, that
differ only in their input. All decision problems considered in this
paper are parameterized by a given signature $\signature = (\comps,
\interacs, \ports)$. The queries take as input a sentence $\phi$, a
SID $\asid$, a behavior map $\beh : \comps \rightarrow
\statemachines$ and a tuple of states $\tuple{q_1,\ldots,q_n} \in
\globstates^n$ for some $n\in \nat$. They are defined as
follows: \begin{itemize}
\item $\mathit{deadlock}(\phi,\asid,\beh)$: is there a configuration
  $(\struc,\amark)$, such that $(\struc,\amark) \models_\asid \phi$
  and $\reach{\anetof{\signature,\struc,\beh},\amark} \cap
  \dead{\anetof{\signature,\struc,\beh}} \neq \emptyset$ ?
\item $\mathit{reach}(\phi,\tuple{q_1,\ldots,q_k},\asid,\beh)$: are
  there configurations $(\struc,\amark)$, $(\struc,\amark')$ and
  indices $u_1,\ldots,u_k \in \universe$, such that $(\struc,\amark)
  \models_\asid \phi$, $\amark' \in
  \reach{\anetof{\signature,\struc,\beh},\amark}$ and
  $\set{q_i[u_i]~|~i\in\interv{1}{k}} \subseteq \amark'$, with
  $q_1[u_1], \ldots, q_k[u_k]$ pairwise distinct?
  %
\end{itemize}

The above queries occur typically as correctness conditions in system
verification. For instance, proving that, in every system described by a
formula $\phi$, no freeze configuration is reachable means showing that
$\mathit{deadlock}(\phi,\asid,\beh)$ does not hold. Similarly, proving that
each system described by $\phi$ stays clear of a set of error
configurations e.g., at most one component is in some critical
state $q_{crit}$ at any time amounts to proving that
$\mathit{reach}(\phi,\tuple{q_{crit},q_{crit}},\asid,\beh)$ does not hold.

\subsection{Undecidability Results for Linear Systems}

We show that the problems defined by the sets of queries above, taken
over all inputs, but parameterized by a fixed given signature, are
undecidable. In fact, we shall prove stronger results, in which the
input formul{\ae} define sets of configurations with \emph{linear}
architectures. 

\begin{definition}\label{def:linear-architecture}
  Given a signature $\signature = (\set{\acomp}, \interacs, \ports)$,
  a sentence $\phi$, interpreted over a SID $\asid$, is \emph{linear}
  if and only if $\phi \models_{\asid\cup\linasid} \lsent$, where
  $\linasid$ consists of the rules:
  \[\begin{array}{rcl}
  \lpred(x,x) & \leftarrow & \hspace{7.3mm} \acomp(x) * \Asterisk_{\ainterac\in\interacs^{(1)}} \ainterac(x) \\
  \lpred(x,y) & \leftarrow & \exists z ~.~ \acomp(x) * \Asterisk_{\ainterac\in\interacs^{(1)}} \ainterac(x) *
  \Asterisk_{\ainterac\in\interacs^{(2)}} \ainterac(x,z) * \lpred(z,y)
  \end{array}\]
  where $\lpred$ is a binary predicate symbol and $\lsent$ is the sentence
  $\exists x \exists y ~.~ \lpred(x,y)$.
\end{definition}
For instance, the sentences $\exists x \exists y ~.~
\mathit{Chain}(x,y)$ (\S\ref{sec:running-example}) and $\exists x
\exists y ~.~ \chain{n}{t}(x,y)$ for $n, t \geq 0$ (Example
\ref{ex:chains}) are linear, when taking $\comps = \set{\tokcomp}$ and
$\interacs = \set{\tokinterac}$.

Below we show that undecidability occurs for classes of queries taking
linear sentences as input, over a fixed signature consisting of only
one component type and a fixed set of interaction types, which is not
part of the input of a query.

\begin{theorem}\label{thm:deadlock-reach-safe-undec}
  The following problems are undecidable:
  \[\begin{array}{rcl}
  \mathit{LinearDeadlock} & \isdef & \set{\mathit{deadlock}(\phi,\asid,\beh) \mid
    \phi \models_{\asid\cup\linasid} \lsent} \\
  \mathit{LinearReachability} & \isdef & \set{\mathit{reach}(\phi,\tuple{q_1},\asid,\beh) \mid
    \phi \models_{\asid\cup\linasid} \lsent} \\
  \end{array}\]
\end{theorem}
The idea of the proof is to build a component-based system with linear
architecture that simulates the execution of a Post-Turing machine.  We
present the construction in \S\ref{sec:tm:simulation} and complete the
proof in \S\ref{sec:undecidability:proof}.

\subsection{Simulation of Post-Turing Machines by Linear Systems}
\label{sec:tm:simulation}

A Post-Turing machine \cite{Post36,Davis78} executes sequential
deterministic programs $M$ of the form $1:~stmt_1;~ 2:~ stmt_2;~
\ldots;~ m:~ stmt_m$, where each statement $stmt_i$ is one of
following: $\stmt{write~0}$, $\stmt{write~1}$, $\stmt{go~right}$,
$\stmt{go~left}$, $\stmt{goto~step~\mathit{j}~if~read~0}$,
$\stmt{goto~step~\mathit{j}~if~read~1}$, $\stmt{stop}$, for some $j\in
\interv{1}{m}$. The machine operates on an infinite tape of zeroes and
ones. Initially, the head is pointing at some position on the tape and
the program control is at the first statement. At any step, the
current statement is executed and the tape content, the head position
and control are updated according to that statement.

\begin{figure}[t!]
  \begin{center}
    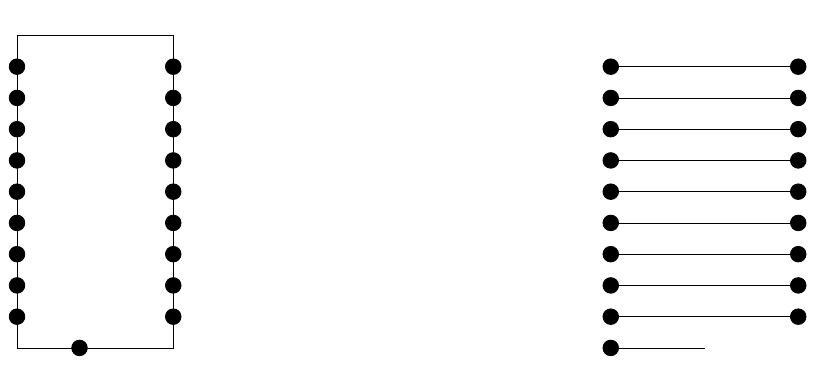 \centerline{\footnotesize(a)}

    \vspace*{\baselineskip}
    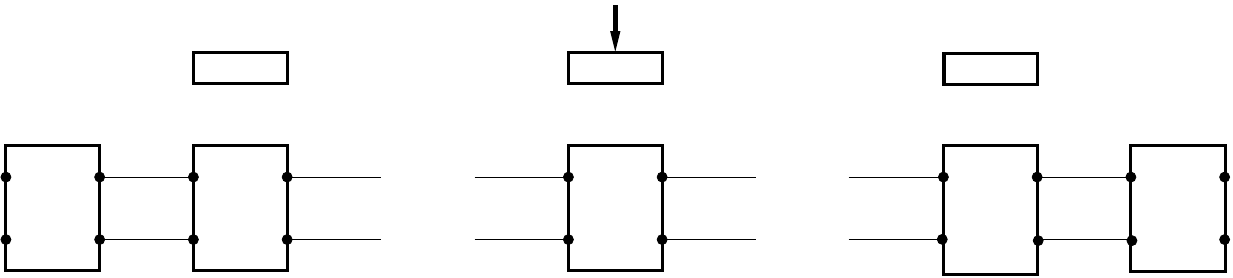 \centerline{\footnotesize(b)}
  \end{center}
  \vspace*{-\baselineskip}
  \caption{Undecidability of Deadlock and Reachability for Linear
    Component-based Systems}\label{fig:simulation}
\end{figure}

We simulate a Post-Turing machine by a component-based system with linear
architecture. Its signature $\signature_{PT} =
(\comps_{PT},\interacs_{PT},\ports_{PT})$ is presented in
Fig.~\ref{fig:simulation}(a) and consists of one component type
$\acomp_{PT}$ and ten interaction types $\interacs_{PT}$, with associated
ports as presented in the figure.  In particular, the signature does not
depend on the program executed by the machine.

The linear system that simulates a Post-Turing machine is depicted in Fig.~\ref{fig:simulation}(b).  First of all, the
machine program $M$ is encoded in the behavior $\beh_M$ of the
component type $\acomp_{PT}$. In particular, this behavior includes
three disjoint state machines, as described in
Table~\ref{tab:tm:behavior}, labeled with the same set of ports.
Second, in the linear system, each component plays a different role
depending on its position, i.e.\ it runs according to one of the
three state machines below:
\begin{compactitem}
\item the leftmost \emph{ctrl} component behaves according to the
  \emph{control state machine} derived from the program $M$: it issues the
  commands to be executed by the tape components depending on the current
  statement, and proceeds further according to the control flow of $M$.  In
  case of \emph{read} commands, it goes to an intermediate state, waiting
  to receive either \emph{read\_0} or \emph{read\_1} answers.  In case of
  \emph{left!} commands, it goes to an error state signaling the overflow
  of the tape on the left side.  The complete definition is given in
  Table~\ref{tab:tm:behavior}~(top).
\item the middle \emph{tape} components behave according to the \emph{tape
  state machine} and mimic the behavior of a single tape cell: the $k$-th
  component state records the $k$-th symbol of the tape $\gamma_k$ and the
  presence ($\top$) or absence ($\bot$) of the machine head at that cell.
  They react to commands (received through
  $in$-ports) by changing their state and/or further issuing commands to
  their neighbors (sent through $out$-ports).  The complete definition is
  given in Table~\ref{tab:tm:behavior}~(middle).
\item the rightmost \emph{sink} component behaves according to the
  \emph{sink state machine} and detects the overflow of the tape on the
  right side and signals it as an error.  This happens when this component
  receives the \emph{right!} command. The complete definition is given in
  Table~\ref{tab:tm:behavior}~(bottom).
\end{compactitem}

\begin{table}[t!]
\small \[\begin{array}{lcl}
\multicolumn{3}{l}{\mbox{control state machine with states } Q_{ctrl}^M \isdef \{ i, (i,\_) ~|~ i \in \interv{1}{m+1}\}
  \mbox{ represents the program }M} \\ \hline
i:~\stmt{write~0} & \hspace{0.5cm} & i \arrow{\outport{write\_0}}{} i\!+\!1 \\
i:~\stmt{write~1} & & i \arrow{\outport{write\_1}}{} i\!+\!1 \\
i:~\stmt{go~right} & & i \arrow{\outport{right}}{} i\!+\!1 \\
i:~\stmt{go~left} & & i \arrow{\outport{left}}{} i\!+\!1,~
  i \arrow{\outport{left!}}{} (i,\_),~ (i,\_)\arrow{\outport{err}}{}(i,\_) \\
i:~\stmt{goto~step~\mathit{j}~if~read~0} & & i \arrow{\outport{read}}{} (i,\_),~
(i,\_)\arrow{\inport{read\_0}}{} j,~ (i,\_) \arrow{\inport{read\_1}}{} i\!+\!1 \\
i:~\stmt{goto~step~\mathit{j}~if~read~1} & & i \arrow{\outport{read}}{} (i,\_),~
(i,\_)\arrow{\inport{read\_1}}{} j,~ (i,\_) \arrow{\inport{read\_0}}{} i\!+\!1 \\
i:~\stmt{stop} & & /* nothing */ \\ \hline
\end{array} \]
\[\begin{array}{lcl}
\multicolumn{3}{l}{\mbox{tape state machine with states } Q_{tape} \isdef \{(\gamma,\pi), (\gamma,\pi,a) ~|~ \gamma\in \{0,1\},~ \pi \in \{\bot,\top\},~ a\in\interacs_{PT}\smallsetminus\{\iact{err}\} \}} \\ \hline
(\gamma,\bot) \arrow{\inport{a}}{} (\gamma,\bot,a)  & &
(\gamma,\bot,a) \arrow{\outport{a}}{} (\gamma,\bot) \hspace{1cm} \forall a\not=\mathit{right!,left!,err} \\
(\gamma,\bot) \arrow{\inport{right!}}{} (\gamma,\top) & & 
(\gamma,\bot,\mathit{left}) \arrow{\outport{left!}}{} (\gamma,\top) \\
(\gamma,\top) \arrow{\inport{write\_\gamma'}}{} (\gamma',\top) \\
(\gamma,\top) \arrow{\inport{read}}{} (\gamma,\top,\mathit{read}) & &
(\gamma,\top,\mathit{read}) \arrow{\outport{read\_\gamma}}{} (\gamma,\top) \\
(\gamma,\top) \arrow{\inport{right}}{} (\gamma,\top,\mathit{right}) & & 
(\gamma,\top,\mathit{right}) \arrow{\outport{right!}}{} (\gamma,\bot) \\ 
(\gamma,\top) \arrow{\inport{left!}}{} (\gamma,\bot) \\  \hline
\end{array}\]
\[\begin{array}{lcl} 
\multicolumn{3}{l}{\mbox{sink state machine with states } Q_{sink} \isdef\{ \mathit{idle}, \mathit{busy}\}} \\ \hline
\mathit{idle} \arrow{\inport{right!}}{} \mathit{busy} & \hspace{2cm} &
\mathit{busy} \arrow{\outport{err}}{} \mathit{busy} \\ \hline
\end{array}\]
\caption{\label{tab:tm:behavior}The behavior
  $\beh_M$ of $\acomp_{PT}$ for a Post-Turing machine executing a program $M$}
\end{table}

\subsection{Proof of Theorem \ref{thm:deadlock-reach-safe-undec}} \label{sec:undecidability:proof}

Let $M$ be the program of a Post-Turing machine and let $w=\gamma_1\gamma_2
\ldots \gamma_n$ be a finite word in $\{0,1\}^n$, assuming moreover
wlog\footnote{$w$ can be augmented with extra zeroes on the right} that
$n\geq 2$.  Consider the signature $\signature_{PT}$ and the behavior
$\beh_M$ as defined in the previous section.  Furthermore, consider the following set
$\asid_w$ of rules, where $\interacs_1 \isdef \interacs_{PT}^{(1)}$, $\interacs_2 \isdef \interacs_{PT}^{(2)}$:
\[\begin{array}{rcl}
\mathit{Zeroes}(x,x) & \leftarrow & \hspace{7.3mm} \acomp_{PT}^{(0,\bot)}(z) * \Asterisk_{\ainterac \in \interacs_1} \ainterac(x) \\
\mathit{Zeroes}(x,y) & \leftarrow & \exists z ~.~  \acomp_{PT}^{(0,\bot)}(x) * \Asterisk_{\ainterac \in \interacs_1} \ainterac(x) *
\Asterisk_{\ainterac \in \interacs_2} \ainterac(x,z) * \mathit{Zeroes}(z,y)
\end{array}\]
\[\begin{array}{rcl}
\mathit{Word\_w}(x,y) & \leftarrow & \exists z_2~ \ldots ~\exists z_{n-1} ~.~
\acomp_{PT}^{(\gamma_1,\top)}(x) * \Asterisk_{\ainterac \in \interacs_1} \ainterac(x) ~* \\ 
& & \hspace{4mm} \Asterisk_{\ainterac \in \interacs_2} \ainterac(x,z_2) * \acomp_{PT}^{(\gamma_2,\bot)}(z_2) *  \Asterisk_{\ainterac \in \interacs_1} \ainterac(z_2) ~*  \\
& & \hspace{4mm} \Asterisk_{\ainterac \in \interacs_2} \ainterac(z_2,z_3) * \acomp_{PT}^{(\gamma_3,\bot)}(z_3) * \Asterisk_{\ainterac \in \interacs_1} \ainterac(z_3) ~*  \\
& & \hspace{8mm} ... \\
& & \hspace{4mm} \Asterisk_{\ainterac \in \interacs_2} \ainterac(z_{n-1},y) * \acomp_{PT}^{(\gamma_n,\bot)}(y) * \Asterisk_{\ainterac \in \interacs_1} \ainterac(y) \\[2mm]
\mathit{Tape\_w}(x,y) & \leftarrow & \exists z_1~ \exists w_1~ \exists z_2~ \exists w_2 ~.~ \mathit{Zeroes}(x,z_1) ~*  \\
& & \hspace{4mm} \Asterisk_{\ainterac \in \interacs_2} \ainterac(z_1, w_1) * \mathit{Word\_w}(w_1,w_2) \\
& & \hspace{4mm} \Asterisk_{\ainterac \in \interacs_2} \ainterac(w_2, z_2) * \mathit{Zeroes}(z_2,y) \\[2mm]
\mathit{Init\_w}(x,y) & \leftarrow & \exists z_1~\exists z_2~.~\acomp_{PT}^{1}(x) *  \Asterisk_{\ainterac \in \interacs_1} \ainterac(x) \\
& & \hspace{4mm} \Asterisk_{\ainterac \in \interacs_2} \ainterac(x,z_1) * \mathit{Tape\_w}(z_1,z_2) ~* \\
& & \hspace{4mm} \Asterisk_{\ainterac \in \interacs_2} \ainterac(z_2,y) * \acomp_{PT}^{idle}(y) * \Asterisk_{\ainterac \in \interacs_1} \ainterac(y)
\end{array} \]
and let $\phi_w \isdef \exists x~\exists y~.~ \mathit{Init}_w(x,y)$.
Intuitively, $\phi_w$ is a linear sentence which defines the valid
(initial) configurations of the linear system encoding the Post-Turing
machine, where moreover $w$ is written on the tape and the machine
head is pointing at the beginning $w$. We prove the following two
assertions equivalent:
\begin{enumerate}
\item[1.] the Post-Turing machine running $M$ terminates on input $w$
\item[2.] the answer to $\mathit{deadlock}(\phi_w,\asid_w,\beh_M)$ is yes
\end{enumerate}
\noindent ``1 $\Rightarrow$ 2'': If the machine terminates then it terminates by visiting
finite portions of the tape to the left and to the right, with respect to
the initial placement of $w$ on the tape.  Henceforth, if the linear component-based
system is started with the amount of tape cells needed on both sides, it
will run without errors (that is, neither left or right tape overflow) until
termination as well.

\noindent ``2 $\Rightarrow$ 1'': First of all, if the linear system
reaches a deadlock from its initial configuration, it means that
neither left or right overflow occur (otherwise, the unary
$\iact{err}$ interactions are enabled and run in an endless loop).
Henceforth, the computation of the linear system involved only the
allocated tape components and followed the execution of the statements
in $M$.  Moreover, by construction of the linear system, the relationship
between the control and the tape state machines ensures that no
blocking can occur between them: tape cells are continuously ready to
receive the commands from the control and to execute them, in order.
Therefore, the system stops in a deadlock only if no more commands are
sent, i.e.\ when the control machine reaches a $\stmt{stop}$ statement
and all tape components are ``idle'', that is, no other commands are
pending for completion.

The termination problem of Post-Turing machines being undecidable
\cite{Davis78}, it implies that the $\mathit{LinearDeadlock}$ problem is
also undecidable.  Moreover, we can use a similar argument to show that the
$\mathit{LinearReachability}$ problem is undecidable as well.  Actually, we
can restrict without loss of generality to programs $M$ containing a unique
$\stmt{stop}$ statement moreover occurring as the last statement $m$ in $M$.
Then, for such programs we can prove the equivalence of the previous
assertions to the following one:
\begin{enumerate}
  \item[3.] the answer to $\mathit{reach}(\phi_w, \tuple{m}, \asid_w, \beh_M)$ is yes
\end{enumerate}
that is, the linear system reaches some configuration where a component is
in state $m$.  In fact, by construction, the only component that could
reach the $m$ state is the leftmost control component, and therefore we can
map back any execution of the linear component-based system reaching $m$ in
the control component, to a terminating execution of the Post-Turing
machine.

\section{Translating \adl\ Specifications into Rewriting Trees}
\label{sec:rewriting}

The inductive interpretation of predicate symbols in the \adl\ logic,
by means of a finite set of definitions, supports the idea of
designing systems hierarchically (top-down). For instance, in
\S\ref{sec:running-example}, we specify a ring system first by a chain
of components, with an interaction between the $\mathit{out}$ port of
the last to the $\mathit{in}$ port of the first component (rule
\ref{rule:ring}). Then a chain consists of one component and an
interaction between the $\mathit{out}$ port of that component and the
$\mathit{in}$ port of the first component of a separate chain (rule
\ref{rule:chain1}), or of two components (rule
\ref{rule:chain2}). Intuitively, one can view these stages of the
definition as rewriting steps, in which a predicate atom is replaced
by one of the rules defining it. In this section, we formalize this
idea by introducing rewriting trees, i.e.\ trees labeled with rules,
that define a partial order in which the rules are applied.

Moreover, the verification method described in \S\ref{sec:verification}
uses rewriting trees as backbones for the encoding of sets of
configurations in \mso\ interpreted over trees
\cite{Thatcher2005GeneralizedFA}. In particular, the component indices from
the set $\universe$ are going to be interpreted as nodes of rewriting trees. As will be
shown below, this interpretation of logical variables loses no generality,
because component indices can only be compared for equality; their
particular nature is of no importance from the point of view of the
verification problems considered. Furthermore, the interpretation of
component indices as tree nodes enables the use of a decidable fragment of
\mso, to encode verification conditions.

The mapping of variables to the nodes of a rewriting tree is uniquely
defined by the occurrences of the component atoms in the tree. More
precisely, because all atomic propositions that occur in the rewriting
tree are joined by separating conjunctions, the variables of all
component atoms $\acomp(x)$ with the same component type $\acomp$ must
be interpreted as different indices, or else the formula corresponding
to the tree would not be satisfiable. Hence, a rewriting tree uniquely
defines an architecture $\struc$ over a given signature $\signature$
and, implicitly, a Petri net $\anetof{\signature,\struc,\beh}$, for a
given behavior map $\beh$ (Def. \ref{def:execution-semantics}).

\subsection{Rewriting Trees}
\label{sec:rewriting-trees}

Trees play an important role in the subsequent developments, hence we
introduce a few formal definitions, for self-containment reasons. Let
$\kappa\geq1$ be an integer constant and let $\interv{1}{\kappa}^*$
denote the set of finite sequences of integers between $1$ and
$\kappa$, called \emph{nodes} in the following. We denote the
concatenation of nodes $w,u\in\interv{1}{\kappa}^*$ as $w \cdot u$ or
simply $wu$, when no confusion arises. A set of nodes $T \subseteq
\interv{1}{\kappa}^*$ is said to be: \begin{compactenum}
\item \emph{prefix-closed} if $wi \in T$, for some $i \in
  \interv{1}{\kappa}$, only if $w \in T$, and
\item \emph{complete} if $wi \in T$, for some $i \in
  \interv{1}{\kappa}$, only if $wj \in T$, for all $j \in
  \interv{1}{i-1}$.
\end{compactenum}
A {\em $\kappa$-ary tree} $\mathcal{T}$ is a function mapping a
complete prefix-closed set of nodes $\nodes{\mathcal{T}}$ into a
finite set of labels. The \emph{root} of $\mathcal{T}$ is the empty
sequence $\epsilon$, the {\em children} of a node $w\in
\nodes{\mathcal{T}}$ are $\children{w}{\mathcal{T}} \isdef \set{wi \in
  \nodes{\mathcal{T}} \mid i \in \interv{1}{\kappa}}$ and the {\em
  parent} of a node $wi \in \nodes{\mathcal{T}}$, where $i \in
\interv{1}{\kappa}$, is $w$. The \emph{leaves} of $\mathcal{T}$ are
$\leaves{\mathcal{T}} \isdef \set{w \in \nodes{\mathcal{T}} \mid w\cdot1
  \not\in \nodes{\mathcal{T}}}$. The \emph{subtree} of $\mathcal{T}$
rooted at $w$ is defined by $\nodes{\proj{\mathcal{T}}{w}} \isdef \{
w' \mid ww' \in \nodes{\mathcal{T}} \}$ and $\proj{\mathcal{T}}{w}(w')
\isdef \mathcal{T}(ww')$, for all $w' \in
\nodes{\proj{\mathcal{T}}{w}}$. A \emph{prefix} of a tree
$\mathcal{T}$ is the restriction of $\mathcal{T}$ to a complete
prefix-closed subset of $\nodes{\mathcal{T}}$. We say that
$\mathcal{T}$ is finite if $\nodes{\mathcal{T}}$ is finite.

Let $\asid$ be a fixed SID in the rest of this section.  For an
arbitrary \adl\ formula $\phi$, we denote by $\npred{\phi}$ the
number of occurrences of predicate atoms and by $\pred{i}{\phi}$
the $i$-th predicate atom from $\phi$, in some predefined total
ordering of the symbols in the syntax tree of $\phi$. A formula
$\phi$ is said to be \emph{predicate-free} if
$\npred{\phi}=0$. Without loss of generality, we assume from now on
that $\npred{\phi} \leq \kappa$, for each \adl\ formula $\phi$
considered in the following. We shall also simplify our technical life
by considering the following assumption:

\begin{assumption}\label{ass:sentence-rule}
  For each sentence $\phi$, there exists a rule $\apred_\phi()
  \leftarrow \phi$ in $\asid$, where $\apred_\phi$ is a predicate
  symbol of zero arity, not occurring in $\phi$ or elsewhere in
  $\asid$.
\end{assumption}
This assumption loses no generality because each query in a decision
problem (\S\ref{sec:undecidability}) considers finitely many
sentences, for which finitely many rules of the above form are added
to $\asid$.

\begin{definition}\label{def:rewriting-tree}
  Given a \adl\ formula $\phi$, a \emph{rewriting tree for $\phi$} is
  a tree $\mathcal{T}$ with labels from $\asid$, such that
  $\mathcal{T}(\epsilon) = \left(\apred(x_1, \ldots, x_{\#(\apred)})
  \leftarrow \phi\right)$, for some predicate symbol $\apred$ and, for
  each $w\in\nodes{\mathcal{T}}$, such that $\mathcal{T}(w) =
  \left(\apred_0(x_1,\ldots,x_{\#(\apred_0)}) \leftarrow
  \psi_0\right)$, the following hold: \begin{compactenum}
  \item\label{it1:rewriting-tree} for all $i \in
    \interv{1}{\npred{\psi_0}}$, if $\pred{i}{\psi_0} =
    \apred_{i}(y^i_1, \ldots, y^i_{\#(\apred_{i})})$, then $wi \in
    \nodes{\mathcal{T}}$ and $\mathcal{T}(wi) = \left(\apred_{i}(x_1,
    \ldots, x_{\#(\apred_{i})}) \leftarrow \psi_{i}\right)$ is a rule
    from $\asid$,
  \item\label{it2:rewriting-tree} $wi \not\in
    \nodes{\mathcal{T}}$, for all $i > \npred{\psi_i}$.
  \end{compactenum}
  We denote by $\rtrees{\phi}{}$ the set of rewriting trees for
  $\phi$.
\end{definition}
By slight abuse of notation, we write $\phi[\psi_1/\varphi_1 \ldots
  \psi_n/\varphi_n]$ for the result of replacing each subformula
$\psi_i$ of $\phi$ by the formula $\varphi_i$, for all $i \in
\interv{1}{n}$. We associate each rewriting tree $\mathcal{T} \in
\rtrees{\phi}$ a \emph{characteristic formula}
$\charform{\mathcal{T}}$, defined inductively on the structure of
$\mathcal{T}$, as follows:
\begin{equation*}
  \charform{\mathcal{T}} \isdef \left\{\begin{array}{l}
  \phi \text{, if } \nodes{\mathcal{T}} = \set{\epsilon} \\[2mm]
  \phi\left[\apred_1(\vec{y}_1) / \charform{\proj{\mathcal{T}}{1}}[\vec{x}_1/\vec{y}_1] ~\ldots~
    \apred_{n}(\vec{y}_{n}) / \charform{\proj{\mathcal{T}}{n}}[\vec{x}_{n}/\vec{y}_{n}]\right], \\
  \text{  if } n = \npred{\phi} \geq 1 \text{ and } \pred{i}{\phi} = \apred_i(\vec{y}_i), \text{ for all } i \in \interv{1}{n}
  \end{array}\right.
\end{equation*}
Intuitively, the characteristic formula of a rewriting tree is the
predicate-free formula obtained by replacing each predicate atom
occurring in a node of the tree by the characteristic formula of its
corresponding subtree, recursively. In the process of building a
characteristic formula, the free variables of the characteristic
formul{\ae} of the subtrees are substituted with the parameters of
their corresponding predicate atoms.

\begin{example}\label{ex:tll}
  Let $\signature = (\set{\ntype,\ltype}, \set{\rrinterac,\iointerac},
  \ports)$ be a signature, where $\portsof{\ntype} =
  \set{\req,\reply}$, $\portsof{\ltype} = \set{\reply,\inp,\outp}$,
  $\portsof{\rrinterac} = \tuple{\req,\reply,\reply}$ and
  $\portsof{\iointerac} = \tuple{\outp,\inp}$. Consider a behavior map
  $\beh$, such that $\statesof{\ntype} = \set{q_0,q_1}$ and
  $\statesof{\ltype} = \set{s_0,s_1,s_2}$. The SID below defines
  tree-shaped architectures, whose leaves are connected in a token
  ring:
  \begin{align}
  \treeroot() & \leftarrow \exists r \exists n_1 \exists \ell_1 \exists r_1 \exists n_2 \exists \ell_2 \exists r_2  ~.~ 
  \ntype^{q_0}(r) * \rrinterac(r,n_1,n_2) * \iointerac(r_1,\ell_2) * \iointerac(r_2,\ell_1) \nonumber\\
  & \hspace*{3.9cm} *~ \treenode(n_1,\ell_1,r_1) * \treenode(n_2,\ell_2,r_2) \label{rule:tll-root} \\[2mm]
  \treenode(n,\ell,r) & \leftarrow \exists n_1 \exists r_1 \exists n_2 \exists \ell_2 ~.~
  \ntype^{q_0}(n) * \rrinterac(n,n_1,n_2) * \iointerac(r_1,\ell_2) \nonumber\\
  & \hspace*{3.9cm} *~ \treenode(n_1,\ell,r_1) * \treenode(n_2,\ell_2,r) \label{rule:tll1} \\[2mm]
  \treenode(n,\ell,r) & \leftarrow \ntype^{q_0}(n) * \rrinterac(n,\ell,r) * \treeleaf(\ell) * \iointerac(\ell,r) * \treeleaf(r) \label{rule:tll2} \\[2mm]
  \treeleaf(x) & \leftarrow \ltype^{s_0}(x) \label{rule:tll-leaf}
  \end{align}
  \begin{figure}[h!]
    \label{fig:tll}
    \centerline{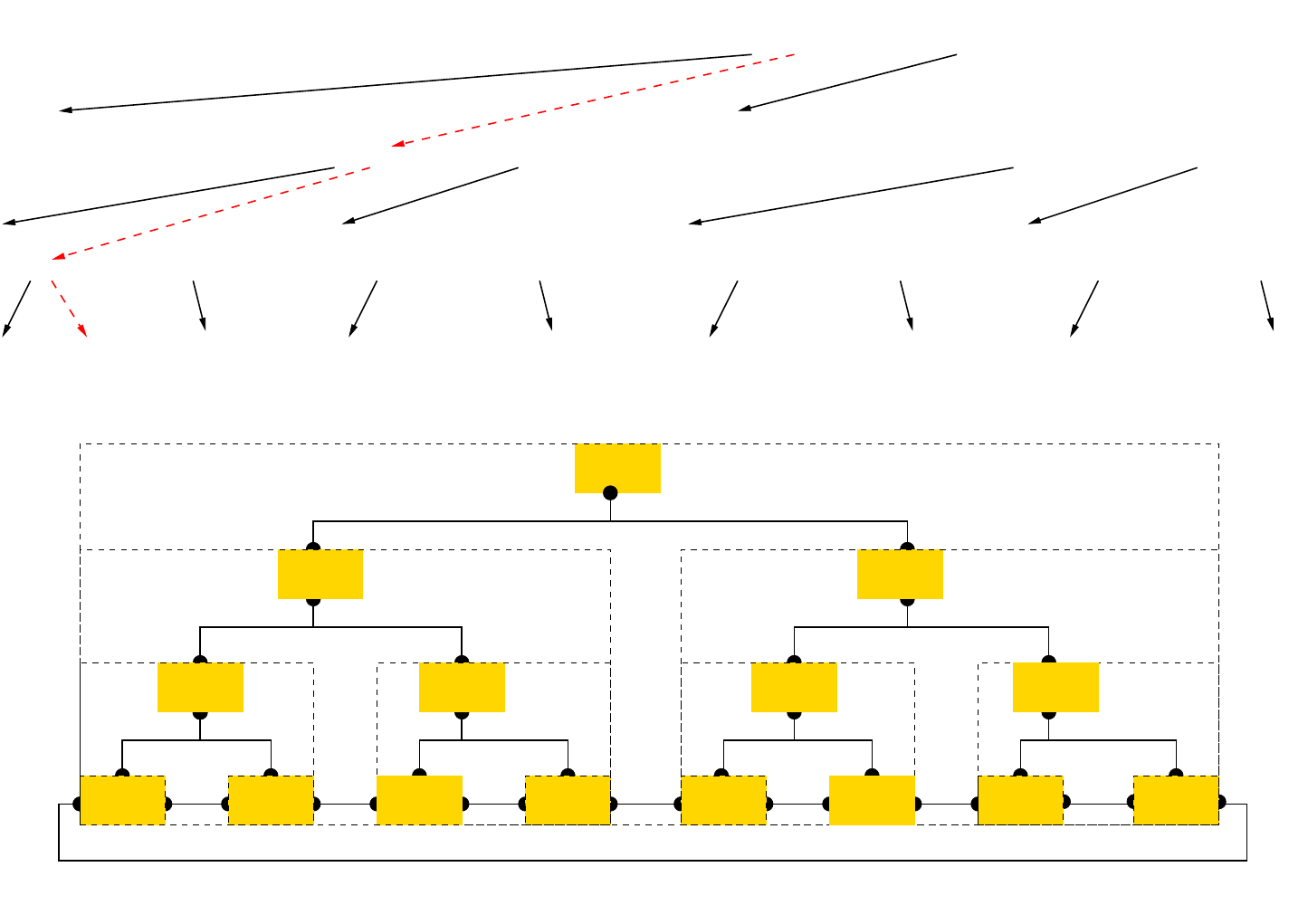}
    \vspace*{-\baselineskip}
    \caption{Trees with Leaves Linked in a Token Ring}
  \end{figure}
  
  Fig.~\ref{fig:tll}(a) shows an unfolding tree for $\phi = \exists r
  \exists n_1 \exists \ell_1 \exists r_1 \exists n_2 \exists \ell_2
  \exists r_2 ~.~ \ntype^{q_0}(r) * \rrinterac(r,n_1,n_2) *
  \iointerac(r_1,\ell_2) * \iointerac(r_2,\ell_1) *
  \treenode(n_1,\ell_1,r_1) * \treenode(n_2,\ell_2,r_2)$, with
  $\apred_\phi = \treeroot$, and Fig.~\ref{fig:tll}(b) shows a model
  of the characteristic formula of this unfolding tree. To avoid name
  clashes, we annotate existentially quantified variables in the
  characteristic formula with the node of the unfolding tree where
  they are introduced, for instance $r^\epsilon, n_1^\epsilon$ and
  $n_2^\epsilon$ are introduced at the root, whereas
  $n_1^1,n_2^1,\ell_2^1,r_1^1$ ($n_1^2,n_2^2,\ell_2^2,r_1^2$) occur
  first at the left (right) child of the root. \hfill$\blacksquare$
\end{example}

The following result shows that each model of a sentence is the model
of the characteristic formula of some rewriting tree for that
sentence, that represents the way in which the model is ``produced''
by unfolding the rules in $\asid$. In general, such rewriting trees
are not unique, as different rewriting trees can have equivalent
characteristic formul{\ae}.

\begin{lemma}\label{lemma:rewriting-trees}
  For each \adl\ sentence $\phi$, we have $\sem{\phi}{} =
  \bigcup_{\mathcal{T} \in \rtrees{\phi}{}}
  \sem{\charform{\mathcal{T}}{}}{}$. 
\end{lemma}
\proof{ We prove a slightly more general statement: for each formula
  $\phi$, each rule $\apred(x_1, \ldots, x_{\#(\apred)}) \leftarrow
  \phi$ from $\asid$, each configuration $(\struc,\amark)$ and each
  store $\store$: \[\begin{array}{rcl} (\struc,\amark) \models^\store
  \phi & \iff & (\struc,\amark) \models^\store \charform{\mathcal{T}}
  \text{, for some } \mathcal{T} \in \rtrees{\phi} \text{, such} \\ &&
  \text{that } \mathcal{T}(\epsilon) = \left(\apred(x_1, \ldots,
  x_{\#(\apred)}) \leftarrow \phi\right)\end{array}\]
  \noindent``$\Rightarrow$'' By induction on the definition of the
  satisfaction relation $\models$, considering the following
  cases: \begin{itemize}
  \item $\phi \in \set{\emp, \acomp^q(x),\ainterac(x_1, \ldots,
    x_{\#(\ainterac)})}$: we define $\mathcal{T}$ by
    $\nodes{\mathcal{T}} \isdef \set{\epsilon}$ and
    $\mathcal{T}(\epsilon) \isdef \left(\apred(x_1, \ldots,
    x_{\#(\apred)}) \leftarrow \phi\right)$. By
    Def. \ref{def:rewriting-tree}, we have $\mathcal{T} \in
    \rtrees{\phi}$ and $\charform{\mathcal{T}} = \phi$, thus
    $(\struc,\amark) \models^\store \charform{\mathcal{T}}$.
  \item $\phi = \exists y_1 \ldots \exists y_n ~.~ \phi_0 *
    \Asterisk_{i=1}^k \apred_i(z^i_1, \ldots, z^i_{\#(\apred_i)})$,
    where $\phi_0$ is quantifier- and predicate-free: since
    $(\struc,\amark) \models^\store \exists y_1 \ldots \exists y_n ~.~
    \phi_0 * \Asterisk_{i=1}^k \apred_i(z^i_1, \ldots,
    z^i_{\#(\apred_i)})$, there exists a store $\store'$ that agrees
    with $\store$ over $\vars \setminus \set{y_1, \ldots, y_n}$ and
    configurations $(\struc_i,\amark_i)$, such that $(\struc,\amark) =
    (\struc_0,\amark_0) \comp \ldots \comp (\struc_k,\amark_k)$,
    $(\struc_0,\amark_0) \models^{\store'} \phi_0$ and
    $(\struc_i,\amark_i) \models^{\store'} \apred_i(z^i_1, \ldots,
    z^i_{\#(\apred_i)})$, for all $i \in \interv{1}{k}$. For each $i
    \in \interv{1}{k}$, since $(\struc_i,\amark_i) \models^{\store'}
    \apred_i(z^i_1, \ldots, z^i_{\#(\apred_i)})$, there exists
    $\apred_i(x_1, \ldots, x_{\#(\apred_i)}) \leftarrow \psi_i$ in
    $\asid$, such that $(\struc_i,\amark_i) \models^{\store'}
    \psi_i[x_1/z^i_1, \ldots, x_{\#(\apred_i)}/z^i_{\#(\apred_i)}]$, by
    the definition of $\models$. Let $\store'_i \isdef \store' \circ
    [x_1/z^i_1, \ldots, x_{\#(\apred_i)}/z^i_{\#(\apred_i)}]$ be a
    store. We have $(\struc_i,\amark_i) \models^{\store'_i} \psi_i$ and,
    by the induction hypothesis, there exists a rewriting tree
    $\mathcal{T}_i \in \rtrees{\psi_i}$, such that
    $\mathcal{T}_i(\epsilon) = \left(\apred_i(x_1, \ldots,
    x_{\#(\apred_i)}) \leftarrow \psi_i\right)$ and $(\struc_i,\amark_i)
    \models^{\store'_i} \charform{\mathcal{T}_i}$, hence
    $(\struc_i,\amark_i) \models^{\store'}
    \charform{\mathcal{T}_i}[x_1/z^i_1, \ldots,
      x_{\#(\apred_i)}/z^i_{\#(\apred_i)}]$. We define the rewriting
    tree $\mathcal{T}$ as: \begin{compactitem}
    \item $\nodes{\mathcal{T}} = \set{\epsilon} \cup \bigcup_{i=1}^k i
      \cdot \nodes{\mathcal{T}_i}$,
    \item $\mathcal{T}(\epsilon) = \apred(x_1, \ldots, x_{\#(\apred)})
      \leftarrow \phi$,
    \item $\mathcal{T}(iw) = \mathcal{T}_i(w)$, for each $i \in
      \interv{1}{k}$ and each $w \in \nodes{\mathcal{T}_i}$.
    \end{compactitem}
    We obtain $\charform{\mathcal{T}} = \exists y_1 \ldots \exists y_n
    ~.~ \phi_0 * \Asterisk_{i=1}^k \charform{\mathcal{T}_i}[x_1/z^i_1,
      \ldots, x_{\#(\apred_i)}/z^i_{\#(\apred_i)}]$, leading to
    $(\struc,\amark) \models^\store \charform{\mathcal{T}}$.
  \end{itemize}

  \noindent''$\Leftarrow$'' By induction on the structure of
  $\mathcal{T} \in \rtrees{\phi}$, such that $\mathcal{T}(\epsilon) =
  \left(\apred(x_1, \ldots, x_{\#(\apred)}) \leftarrow \phi\right)$,
  by distinguishing the following cases: \begin{itemize}
  \item $\nodes{\mathcal{T}} = \set{\epsilon}$: since
    $\mathcal{T}(\epsilon) = \left(\apred(x_1, \ldots, x_{\#(\apred)})
    \leftarrow \phi\right)$, we have $\charform{\mathcal{T}} = \phi$,
    hence $(\struc,\amark) \models^\store \phi$. 
  \item $\children{\epsilon}{\mathcal{T}} = \set{1, \ldots, k}$, for
    some $k \geq 1$: assume w.l.o.g. that $\phi = \exists y_1 \ldots
    \exists y_n ~.~ \phi_0 * \Asterisk_{i=1}^k \apred_i(z^i_1, \ldots,
    z^i_{\#(\apred_i)})$, where $\phi_0$ is a quantifier- and
    predicate-free formula. By the definition of characteristic
    formul{\ae}, we have $\charform{\mathcal{T}} = \exists y_1 \ldots
    \exists y_n ~.~ \phi_0 * \Asterisk_{i=1}^k
    \charform{\proj{\mathcal{T}}{i}}[x_1/z^i_1, \ldots,
      x_{\#(\apred_i)}/z^i_{\#(\apred_i)}]$, where $\mathcal{T}(i) =
    \left(\apred_i(x_1, \ldots, x_{\#(\apred_i)}) \leftarrow
    \psi_i\right)$, hence $\proj{\mathcal{T}}{i} \in
    \rtrees{\psi_i}$. Since $(\struc,\amark) \models^\store
    \charform{\mathcal{T}}$, there exists a store $\store'$ that
    agrees with $\store$ over $\vars \setminus \set{y_1, \ldots, y_n}$
    and configurations $(\struc_0, \amark_0), \ldots,
    (\struc_k,\amark_k)$, such that $(\struc_0,\amark_0)
    \models^{\store'} \phi_0$ and $(\struc_i,\amark_i)
    \models^{\store'} \charform{\proj{\mathcal{T}}{i}}[x_1/z^i_1,
      \ldots, x_{\#(\apred_i)}/z^i_{\#(\apred_i)}]$, for all $i \in
    \interv{1}{m}$. Consider an arbitrary index $i \in \interv{1}{m}$
    and let $\store'_i \isdef \store' \circ [x_1/z^i_1, \ldots,
      x_{\#(\apred_i)}/z^i_{\#(\apred_i)}]$, hence
    $(\struc_i,\amark_i) \models^{\store'_i}
    \charform{\proj{\mathcal{T}}{i}}$. By the inductive hypothesis, we
    have $(\struc_i,\amark_i) \models^{\store'_i} \psi_i$, hence
    $(\struc_i,\amark_i) \models^{\store'_i} \apred_i(x_1, \ldots,
    x_{\#(\apred_i)})$ and $(\struc_i,\amark_i) \models^{\store'}
    \apred_i(z^i_1, \ldots, z^i_{\#(\apred_i)})$ follows. Because the
    above holds for all $i \in \interv{1}{k}$, we conclude that
    $(\struc, \amark) \models^\store \phi$. \qed
  \end{itemize}}

\subsection{Canonical Models}
\label{sec:canonical-model}

In the following, we shall equate the universe of indices $\universe$
with the set $\interv{1}{\kappa}^*$ of tree nodes, i.e.\ finite
sequences of integers between $1$ and $\kappa$. This is without loss
of generality, because both sets are infinitely countable and the
nature of indices plays no role in the interpretation of
\adl\ formul{\ae}. In this setting, we define the
\emph{canonical model} $(\struc_{\mathcal{T}}, \amark_{\mathcal{T}})$
of the characteristic formula $\charform{\mathcal{T}}$ of a rewriting
tree $\mathcal{T}$. In order to simplify its definition, we proceed
under the following assumption:

\begin{assumption}\label{ass:distinct-component-types}
  Each rule of $\asid$ is of one of the forms:
  \begin{align}
  \apred(x_1) & \leftarrow \phi \text{, such that } \phi \neq \emp \tag{I}\label{rule:inst} \\
  \apred(x_1, \ldots, x_{\#(\apred)}) & \leftarrow \exists y_1 \ldots \exists y_n ~.~
  \phi ~*~ \psi ~* \Asterisk_{i=1}^p \apred_i(z_1^i, \ldots, z_{\#(\apred_i)}^i)
  \tag{II}\label{rule:ind}
  \end{align}
  for some $n,m\geq 0$ and $p\geq1$, where: \begin{compactitem}
  \item $\phi \in \set{\acomp^q(x_1) \mid \acomp\in\comps,~ q \in \statesof{\acomp}} \cup \set{\emp}$,  
  \item $\psi$ is a separating conjunction of interaction atoms,
      %
      %
      %
      %
    \item $\bigcup_{i=1}^p\set{z_1^i, \ldots, z_{\#(\apred_i)}^i} =
      \left(\set{x_1, \ldots, x_{\#(\apred)}} \setminus
      \fv{\phi}\right) \cup \set{y_1, \ldots, y_n}$, and
    \item $\set{z_1^i, \ldots, z_{\#(\apred_i)}^i} \cap \set{z_1^j,
      \ldots, z_{\#(\apred_j)}^j} = \emptyset$, for all $1 \leq i < j
      \leq p$.
    \end{compactitem}
\end{assumption}
The direct consequence of this assumption is that, each variable that
occurs in a characteristic formula of a rewriting tree, occurs in
exactly one component atom, in a rule that labels exactly one node
from the rewriting tree. This fact is proved below:

\begin{lemma}\label{lemma:unique-node}
  Let $\phi$ be a \adl\ formula and $\mathcal{T} \in \rtrees{\phi}$
  be a rewriting tree for $\phi$. For each variable $x$, that occurs
  free or existentially quantified in $\charform{\mathcal{T}}$, there
  exists exactly one node $w \in \nodes{\mathcal{T}}$ and one
  component atom $\acomp^q(x)$ in $\mathcal{T}(w)$.
\end{lemma}
\proof{ It is sufficient to prove the statement for the case where $x$
  occurs in the label of the root of $\mathcal{T}$. If it is
  introduced at some node $w \in \nodes{\mathcal{T}} \setminus
  \set{\epsilon}$ by an existential quantifier, the same proof can be
  done taking $\proj{\mathcal{T}}{w}$ instead of $\mathcal{T}$. The
  proof goes by induction on the structure of $\mathcal{T}$. For the
  base case $\nodes{\mathcal{T}} = \set{\epsilon}$, the only
  possibility is that $\mathcal{T}(\epsilon)$ is a rule of the form
  (\ref{rule:inst}) and the proof is immediate. For the inductive
  step, let $\mathcal{T}(\epsilon)$ be the following
  rule: \[\apred(x_1, \ldots, x_{\#(\apred)}) \leftarrow \exists y_1
  \ldots \exists y_n ~.~ \phi ~*~ \psi ~* \Asterisk_{i=1}^p
  \apred_i(v_1^i, \ldots, v_{\#(\apred_i)}^i)\] If $\phi$ is of the
  form $\acomp^q(x_1)$ and $x = x_1$, then $\epsilon$ is the unique
  node such that $x$ occurs in a component atom. Else, $x \in
  \set{x_1, \ldots, x_{\#(\apred)}} ~\cup~ \set{y_1, \ldots, y_n}
  \setminus \fv{\phi}$, thus $x \in \set{v_1^i, \ldots,
    v_{\#(\apred_i)}^i}$, for exactly one $i \in \interv{1}{p}$, by
  Assumption \ref{ass:distinct-component-types}. In this case the
  inductive hypothesis applies, thus $x$ occurs in exactly one
  component atom $\acomp^q(x)$ in $\mathcal{T}(w)$, for exactly one $w
  \in \nodes{\proj{\mathcal{T}}{i}}$. \qed}

Note, however, that Assumption~\ref{ass:distinct-component-types} is
not without loss of generality.  For instance, architectures of the
form $\exists z_1 \ldots \exists z_k ~.~ \ainterac(x, z_1) *
\acomp^q(z_1) * \ainterac(z_1,z_2) * \ldots * \acomp^q(z_k) *
\ainterac(z_k,y)$, having unconnected (loose) interactions, cannot be
defined by SIDs that meet the requirement of Assumption
\ref{ass:distinct-component-types}. On the positive side, all examples
considered in this paper can be written using only rules of this form,
for instance, the definition of the $\chain{n}{t}$ predicates, for
constant integers $n,t \geq 0$ (Example \ref{ex:chains}) or the
definition of the $\treeroot$, $\treenode$ and $\treeleaf$ predicates
(Example \ref{ex:tll}).

The canonical model of a characteristic formula
$\charform{\mathcal{T}}$ is obtained by instantiating each variable
$x$ that occurs free or existentially quantified in
$\charform{\mathcal{T}}$ by the unique node $w \in
\nodes{\mathcal{T}}$, such that $x$ occurs in a single component atom
$\acomp^q(x)$ in $\mathcal{T}(w)$ (Lemma
\ref{lemma:unique-node}). Formally, let $\phi = \exists y_1 \ldots
\exists y_n ~.~ \psi ~* \Asterisk_{i=1}^p \apred_i(z^i_1, \ldots,
z^i_{\#(\apred_i)})$ be a formula, where $\psi$ is quantifier- and
predicate-free, and define, for each rewriting tree $\mathcal{T} \in
\rtrees{\phi}$, a store $\store^\epsilon_{\mathcal{T}} : \fv{\psi}
\cup \bigcup_{i=1}^p \set{z^i_1, \ldots, z^i_{\#(\apred_i)}}
\rightarrow \nodes{\mathcal{T}}$, as follows:
\[\store^\epsilon_{\mathcal{T}}(x) \isdef \left\{\begin{array}{ll}
\epsilon & \text{if $x$ occurs in a component atom from $\psi$} \\
i \cdot \store^\epsilon_{\proj{\mathcal{T}}{i}}(x_j) & \text{if $x = z^i_j$, for some $j \in \interv{1}{\#(\apred_i)}$, where} \\
& \mathcal{T}(i) = \big(\apred_i(x_1, \ldots, x_{\#(\apred_i)}) \leftarrow \varphi_i\big) \text{, for all $i \in \interv{1}{p}$}
\end{array}\right.\]
Note that, because the rules of the SID meet the conditions of
Assumption \ref{ass:distinct-component-types}, the two cases of the
definition of $\store^\epsilon_{\mathcal{T}}$ above are exclusive and,
moreover, $\store^\epsilon_{\mathcal{T}}(x)$ is defined for all free
variables of the matrix $\psi ~* \Asterisk_{i=1}^p \apred_i(z^i_1,
\ldots, z^i_{\#(\apred_i)})$ of $\phi$. We extend the store
$\store^\epsilon_{\mathcal{T}}$ to a store $\store_{\mathcal{T}} :
\fv{\eta} \rightarrow \nodes{\mathcal{T}}$ mapping all variables (free
or existentially quantified) that occur in the characteristic formula
$\charform{\mathcal{T}} = \exists z_1 \ldots \exists z_k ~.~ \eta$,
where $\eta$ is a quantifier- and predicate-free formula:
\begin{equation}\label{eq:store-tree}
  \store_{\mathcal{T}}(x) \isdef \left\{\begin{array}{ll}
  \store^\epsilon_{\mathcal{T}}(x) & \text{if $x \in \fv{\charform{\mathcal{T}}}$} \\
  w \cdot \store^\epsilon_{\proj{\mathcal{T}}{w}}(x) & \text{if $x$ is existentially quantified at  $w \in \nodes{\mathcal{T}}$}
  \end{array}\right.
\end{equation}
In the above definition, we have assumed that all existentially
quantified variables have pairwise distinct names (this is w.l.o.g. as
quantified variables can be $\alpha$-renamed, if necessary). Note that
$\store_{\mathcal{T}}$ is uniquely defined, for each given rewriting
tree $\mathcal{T}$ and induces a unique \emph{canonical model},
defined below:

\begin{definition}\label{def:canonical-model}
  The \emph{canonical model} of a rewriting tree $\mathcal{T}$ is the
  configuration $(\struc_{\mathcal{T}},\amark_{\mathcal{T}})$, defined
  as follows: \begin{compactitem}
  \item $\struc_{\mathcal{T}}(\acomp) \isdef \set{\store_{\mathcal{T}}(x)
    \mid \acomp^q(x) \text{ occurs in } \eta}$,
  \item $\struc_{\mathcal{T}}(\ainterac) \isdef
    \set{\tuple{\store_{\mathcal{T}}(x_1), \ldots,
        \store_{\mathcal{T}}(x_{\#(\ainterac)})} \mid \ainterac(x_1,
      \ldots, x_{\#(\ainterac)}) \text{ occurs in } \eta}$, and
  \item $\amark_{\mathcal{T}} \isdef \set{q[\store_{\mathcal{T}}(x)]
    \mid \acomp^q(x) \text{ occurs in } \eta}$.
\end{compactitem}  
\end{definition}

\begin{lemma}\label{lemma:canonical-model}
  Given a sentence $\phi$, for each rewriting tree $\mathcal{T} \in
  \rtrees{\phi}$, we have $(\struc_{\mathcal{T}},\amark_{\mathcal{T}})
  \models \charform{\mathcal{T}}$.
\end{lemma}
\proof{ Let $\charform{\mathcal{T}} = \exists x_1 \ldots \exists x_n
  ~.~ \eta$. We prove a more general statement, namely if $\phi$ is a
  formula, not necessarily a sentence, then
  $(\struc_{\mathcal{T}},\amark_{\mathcal{T}})
  \models^{\store_{\mathcal{T}}} \eta$. The proof goes by induction on
  the structure of $\mathcal{T}$, distinguishing the cases
  below: \begin{compactitem}
  \item $\nodes{\mathcal{T}} = \set{\epsilon}$ and
    $\mathcal{T}(\epsilon) = \left(\apred(x) \leftarrow
    \acomp^q(x)\right)$ is a rule of type (\ref{rule:inst}): then
    $\struc_{\mathcal{T}}(\acomp) = \set{\store_{\mathcal{T}}(x)} =
    \set{\epsilon}$, $\amark_{\mathcal{T}} =
    \set{q[\store_{\mathcal{T}}(x)]} = \set{q[\epsilon]}$,
    $\struc_{\mathcal{T}}(\acomp') = \emptyset$, for all
    $\acomp'\in\comps\setminus\set{\acomp}$ and
    $\struc_{\mathcal{T}}(\ainterac) = \emptyset$, for all
    $\ainterac\in\interacs$, thus
    $(\struc_{\mathcal{T}},\amark_{\mathcal{T}})
    \models^{\store_{\mathcal{T}}} \acomp^q(x)$.
  \item $\nodes{\mathcal{T}} \neq \set{\epsilon}$ and
    $\mathcal{T}(\epsilon) = \big(\apred(x_1, \ldots, x_{\#(\apred)})
    \leftarrow \exists y_1 \ldots \exists y_n ~.~ \phi * \psi ~*$
    $\Asterisk_{i=1}^p \apred_i(z_1^i, \ldots,
    z_{\#(\apred_i)}^i)\big)$ is a rule of type (\ref{rule:ind}). Let
    $\eta_i$ be the matrix of the characteristic formula
    $\charform{\proj{\mathcal{T}}{i}}$, for all $i \in
    \interv{1}{p}$. By the definition of $\store_{\mathcal{T}}$, we
    have $\store_{\mathcal{T}}(x) = i \cdot
    \store_{\proj{\mathcal{T}}{i}}(x)$, for all $x \in \eta_i$ and $i
    \in \interv{1}{p}$. We define the structures $(\struc_0,\amark_0),
    (\struc_1,\amark_1), \ldots,$ $(\struc_p,\amark_p)$ as
    follows: \begin{compactitem}
    \item $\struc_0(\acomp) \isdef \set{\store_{\mathcal{T}}(x)} =
      \set{\epsilon}$ and $\amark_0 \isdef
      \set{q[\store_{\mathcal{T}}(x)]} = \set{q[\epsilon]}$ if $\phi =
      \acomp^q(x_1)$, else $\struc_0(\acomp) = \emptyset$ and
      $\amark_0 = \emptyset$ and, moreover, $\struc_0(\ainterac)
      \isdef \{\tuple{\store_{\mathcal{T}}(x_1), \ldots,
        \store_{\mathcal{T}}(x_{\#(\ainterac)})} \mid \ainterac(x_1,
      \ldots, x_{\#(\ainterac)}) \text{ occurs in } \psi\}$, thus we
      have $(\struc_0,\amark_0) \models^{\store_{\mathcal{T}}} \phi *
      \psi$.
    \item $\struc_i(\acomp) \isdef \set{\store_{\mathcal{T}}(x) \mid
      \acomp^q(x) \text{ occurs in } \eta_i}$, $\amark_i \isdef
      \{\tuple{\store_{\mathcal{T}}(x_1), \ldots,
        \store_{\mathcal{T}}(x_{\#(\ainterac)})} \mid$ $\ainterac(x_1,
      \ldots, x_{\#(\ainterac)}) \text{ occurs in } \eta_i\}$, for
      each $i \in \interv{1}{p}$. Since $\store_{\mathcal{T}}(x) = i
      \cdot \store_{\proj{\mathcal{T}}{i}}(x)$, for all $x \in \eta_i$
      and $i \in \interv{1}{p}$, we have $\struc_i(\acomp) = \set{i
        \cdot u \mid u \in \struc_{\proj{\mathcal{T}}{i}}(\acomp)}$,
      for all $\acomp \in \comps$, $\amark_i = \set{q[i \cdot u] \mid
        q[u] \in \amark_{\proj{\mathcal{T}}{i}}}$ and
      $\struc_i(\ainterac) = \{\tuple{i \cdot u_1, \ldots, i \cdot
        u_{\#(\ainterac)}} \mid$ $\tuple{u_1, \ldots,
        u_{\#(\ainterac)}} \in
      \struc_{\proj{\mathcal{T}}{i}}(\ainterac)\}$, for all
      $\ainterac\in\interacs$. By the inductive hypothesis, we have
      $(\struc_{\proj{\mathcal{T}}{i}},\amark_{\proj{\mathcal{T}}{i}})
      \models^{\store_{\proj{\mathcal{T}}{i}}} \eta_i$, thus
      $(\struc_i,\amark_i) \models^{\store_{\mathcal{T}}} \eta_i$, for
      all $i \in \interv{1}{p}$. 
    \end{compactitem}
    Since the sets of indices of $\struc_0, \ldots, \struc_p$ and
    $\amark_0, \ldots, \amark_p$ are pairwise disjoint, their
    composition is defined and, moreover, we have
    $(\struc_{\mathcal{T}}, \amark_{\mathcal{T}}) =
    (\struc_0,\amark_0) \comp \ldots \comp
    (\struc_p,\amark_p)$. Consequently, we obtain
    $(\struc_{\mathcal{T}}, \amark_{\mathcal{T}})
    \models^{\store_{\mathcal{T}}} \phi * \psi * \Asterisk_{i=1}^p
    \apred_i(z_1^i, \ldots, z_{\#(\apred_i)}^i)$, as required. \qed
\end{compactitem}}

\begin{example}\label{ex:tll-structure}
  Since $\mathcal{T}$ is a rewriting tree for a sentence, each
  variable occurs bound in $\charform{\mathcal{T}}$ and is mapped by
  the store $\store_{\mathcal{T}}$ into the unique node of
  $\mathcal{T}$ that contains a unique component atom in which that
  variable occurs:
  \[\begin{array}{llll}
  \store_{\mathcal{T}}(r^\epsilon) = \epsilon & \store_{\mathcal{T}}(n^\epsilon_1) = 1 & \store_{\mathcal{T}}(n^\epsilon_2) = 2 \\
  \store_{\mathcal{T}}(n^1_1) = 11 & \store_{\mathcal{T}}(n^1_2) = 12 & \store_{\mathcal{T}}(n^2_1) = 21 & \store_{\mathcal{T}}(n^2_2) = 22 \\
  \store_{\mathcal{T}}(\ell^\epsilon_1) = 111 &  \store_{\mathcal{T}}(r^1_1) = 112 &  \store_{\mathcal{T}}(\ell^1_2) = 121 & \store_{\mathcal{T}}(r^\epsilon_1) = 122 \\
  \store_{\mathcal{T}}(\ell^\epsilon_2) = 211 &  \store_{\mathcal{T}}(r^2_1) = 212 &  \store_{\mathcal{T}}(\ell^2_2) = 221 & \store_{\mathcal{T}}(r^\epsilon_2) = 222
  \end{array}\]
  The configuration $(\struc_{\mathcal{T}},\amark_{\mathcal{T}})$
  corresponding to the rewriting tree $\mathcal{T}$ from
  Fig.~\ref{fig:tll}(a) is given below:
  \[\hspace*{-2mm}\arraycolsep=1.4pt\begin{array}{rll}
  \struc_{\mathcal{T}}(\ntype) & = & \set{\epsilon,1,2,11,12,21,22} \\ 
  \struc_{\mathcal{T}}(\ltype) & = & \set{111,112,121,122,211,212,221,222} \\
  \struc_{\mathcal{T}}(\rrinterac) & = & \set{\tuple{\epsilon,1,2},\tuple{1,11,12},\tuple{11,111,112},\tuple{12,121,122},\tuple{21,211,212},\tuple{22,221,222}} \\
  \struc_{\mathcal{T}}(\iointerac) & = & \set{\tuple{111,112},\tuple{112,121},\tuple{121,122},\tuple{211,212},\tuple{212,221},\tuple{221,222},\tuple{222,111}} \\
  \amark_{\mathcal{T}} & = & \{q_0[\epsilon],q_0[1],q_0[2],q_0[11],q_0[12],q_0[21],q_0[22],s_0[111],s_0[112],s_0[121],s_0[122], \\
  && s_0[211],s_0[212],s_0[221],s_0[222]\}  \hfill\blacksquare
  \end{array}\]
\end{example}
  
\subsection{Symmetry Reduction}

As shown in Lemma \ref{lemma:rewriting-trees}, each model of a
sentence is a model of the characteristic formula of some rewriting
tree for that sentence. In this section, we prove a symmetry property
of the models of a characteristic formula corresponding to a given
rewriting tree, that makes them indistinguishable from the point of
view of the decision problems considered previously
(\S\ref{sec:undecidability}). Let us fix a signature $\signature$,
with component types $\comps = \set{\acomp_1, \ldots, \acomp_\ncomps}$
and interaction types $\interacs = \set{\ainterac_1, \ldots,
  \ainterac_\ninteracs}$, and a behavior map $\beh$, in the rest of
this section.

Intuitively, two architectures are symmetric if they differ only by a
renaming of indices used to interpret the component and interaction
types. For instance, the architectures $\struc_1$ and $\struc_2$,
where $\struc_1(\acomp_1) = \struc_1(\acomp_2) = \set{1}$,
$\struc_1(\ainterac_1) = \set{\tuple{1,1}}$ and $\struc_2(\acomp_1) =
\set{1}$, $\struc_2(\acomp_2) = \set{2}$, $\struc_2(\ainterac_1) =
\set{\tuple{1,2}}$ have the same shape (assuming $\ncomps=2$ and
$\ninteracs=1$). However, this isomorphism cannot be captured by a
global permutation of indices, as it is usually the
case\footnote{Typically, global permutations suffice when only one
component type is considered.} in the literature
\cite{EmersonSistla96}, because the sets $\struc_1(\acomp_1) \cup
\struc_1(\acomp_2)$ and $\struc_2(\acomp_1) \cup \struc_2(\acomp_2)$
have different cardinalities. For this reason, our definition of
symmetry considers one permutation per component type.

Formally, given an architecture $\struc$ over $\signature$ and a tuple
of bijections $\vec{f} = \tuple{f_1, \ldots, f_\ncomps}$, where each
$f_i : \universe \rightarrow \universe$ renames the indices of the
component type $\acomp_i$, for all $i\in\interv{1}{\ncomps}$, we
define:
\[\begin{array}{rcl}
\hspace*{-2mm}(\vec{f}(\struc))(\acomp_i) & \isdef & f_i(\struc(\acomp_i)) \text{, for all $i \in \interv{1}{\ncomps}$} \\
\hspace*{-2mm}(\vec{f}(\struc))(\ainterac) & \isdef & \{\tuple{f_{i_1}(u_1), \ldots, f_{i_{\#(\ainterac)}}(u_{\#(\ainterac)})} \mid
  \tuple{u_1, \ldots, u_{\#(\ainterac)}} \in \struc(\ainterac),~ \\
&& \hspace*{4cm} \forall k \in \interv{1}{\#(\ainterac)} ~.~ \compof{\tuple{\portsof{\ainterac}}_k} = \acomp_{i_k}\}
\end{array}\]
For a marking $\amark$ of the Petri net
$\anetof{\signature,\struc,\beh}$, we define,
moreover: \[\vec{f}(\amark) \isdef \{q[f_i(u)] \mid q[u] \in \amark,~
q \in \statesof{\acomp_i},~ i \in \interv{1}{\ncomps}\}\] Since the
sets $\statesof{\acomp_i}$, for $i \in \interv{1}{\ncomps}$, are
assumed to be pairwise disjoint, for each state $q$ there is exactly
one component type $\acomp_i$, such that $q \in \statesof{\acomp_i}$.

\begin{definition}\label{def:symmetry}
Two configurations are \emph{symmetric}, denoted $(\struc_1,\amark_1)
\symm{} (\struc_2,\amark_2)$, if and only if there exists a tuple of
bijections $\vec{f} = \tuple{f_1, \ldots, f_\ncomps}$, where $f_i :
\universe \rightarrow \universe$, for all $i \in \interv{1}{\ncomps}$,
such that $\vec{f}(\struc_1) = \struc_2$ and $\vec{f}(\amark_1) =
\amark_2$.
\end{definition}

The main idea of using symmetries is to prove that models of the same
characteristic formula $\charform{\mathcal{T}}$ of some rewriting tree
$\mathcal{T}$ are symmetric and, in particular, symmetric with the
canonical model $(\struc_{\mathcal{T}}, \amark_{\mathcal{T}})$. This
proof is greatly simplified by considering only those architectures in
which all interactions are \emph{tightly connected} to components, in
the following sense:

\begin{definition}\label{def:tight-architecture}
  An architecture $\struc$ is \emph{tight} if and only if, for each
  interaction $\ainterac[u_1, \ldots, u_{\#(\ainterac)}]$ from
  $\struc$ and each $k \in \interv{1}{\#(\ainterac)}$, we have $u_k
  \in \struc(\acomp)$, where $\acomp \in \comps$ is the unique
  component type such that $\tuple{\portsof{\ainterac}}_k \in
  \portsof{\acomp}$.
\end{definition}
Note that interactions from an architecture $\struc$, having
unconnected ports, cannot fire in the Petri net
$\anetof{\signature,\struc,\beh}$
(Def. \ref{def:execution-semantics}), thus having no impact on the
answer of a decision problem (deadlock, reachability). Nevertheless,
the \adl\ logic is redundant in this respect, as it allows to specify
architectures with loose interactions. We get rid of this redundancy
and assume a syntactic restriction on formul{\ae}, which guarantees
the tightness of models:

\begin{definition}\label{def:tight-formulae}
  A \emph{profile} is a function $\aprof : \preds \rightarrow
  \bigcup_{i\geq1} \comps^i$, that associates each predicate symbol
  $\apred(x_1, \ldots, x_{\#(\apred)})$ of non-zero arity with a tuple
  of component types of length $\#(\apred)\geq1$. A formula $\phi$ is
  \emph{tight for a profile $\aprof$} if and only if it contains, for
  each interaction atom $\ainterac(x_1, \ldots, x_{\#(\ainterac)})$
  and each $k \in \interv{1}{\#(\ainterac)}$: \begin{compactitem}
  \item a component atom $\acomp^q(x_k)$, such that
    $\tuple{\portsof{\ainterac}}_k \in \portsof{\acomp}$, or
  \item a predicate atom $\apred(y_1, \ldots, y_{\#(\apred)})$, such
    that $x_k = y_\ell$ and $\tuple{\portsof{\ainterac}}_k \in
    \portsof{\tuple{\profof{\apred}}_\ell}$, for some $\ell \in
    \interv{1}{\#(\apred)}$.
  \end{compactitem}
  A SID $\asid$ is \emph{tight} if and only if there exists a profile
  $\aprof_\asid$, such that, for each rule $\apred(x_1, \ldots,
  x_{\#(\apred)}) \leftarrow \phi$ from $\asid$, the formula $\phi$ is
  tight for $\aprof_\asid$ and contains, for each $k \in
  \interv{1}{\#(\apred)}$: \begin{compactitem}
  \item a component atom $\acomp^q(x_k)$, such that
    $\tuple{\aprof_\asid(\apred)}_k = \acomp$, or
  \item a predicate atom $\bpred(y_1, \ldots, y_{\#(\apred)})$, such
    that $x_k = y_\ell$ and $\tuple{\aprof_\asid(\apred)}_k =
    \tuple{\aprof_\asid(\bpred)}_\ell$, for some $\ell \in
    \interv{1}{\#(\bpred)}$.
  \end{compactitem}
  A formula $\phi$, interpreted over a tight SID $\asid$, is
  \emph{tight} if and only if it is tight for the profile
  $\aprof_\asid$.
\end{definition}
For instance, one can check that the predicate atoms
$\chain{n}{t}(x,y)$ are tight, by taking the profile
$\aprof_\asid(\chain{n}{t}) = \tuple{\tokcomp,\tokcomp}$, for all
constants $n,t \geq 0$ (Example \ref{ex:chains}). Analogously, the
predicate atoms $\treenode(n,\ell,r)$ are tight, by taking the profile
$\aprof_\asid(\treenode) = \tuple{\ntype,\ltype,\ltype}$ and
$\aprof_\asid(\treeleaf) = \tuple{\ltype}$ (Example \ref{ex:tll}). The
above conditions guarantee the tightness of architectures described by
tight formul{\ae}:

\begin{lemma}\label{lemma:tight-models}
  For each model $(\struc,\amark)$ of a tight sentence, the
  architecture $\struc$ is tight.
\end{lemma}
\proof{ Let $\phi$ be a tight sentence, such that $(\struc,\amark)
  \models \phi$. By Lemma \ref{lemma:rewriting-trees}, there exists a
  rewriting tree $\mathcal{T} \in \rtrees{\phi}$, such that
  $(\struc,\amark) \models \charform{\mathcal{T}}$. Let
  $\ainterac(z_1, \ldots, z_{\#(\ainterac)})$ be an interaction atom
  from $\charform{\mathcal{T}}$ and $k \in
  \interv{1}{\#(\ainterac)}$. Also, let $w \in \nodes{\mathcal{T}}$ be
  the node where this interaction atom was introduced, such that
  $\mathcal{T}(w) = \left(\apred(x_1, \ldots, x_{\#(\apred)})
  \leftarrow \varphi\right)$ and $\ainterac(y_1, \ldots,
  y_{\#(\ainterac)})$ occurs in $\varphi$, where $y_1, \ldots,
  y_{\#(\ainterac)}$ are substituted by $z_1, \ldots, z_{\#(\apred)}$,
  respectively, during the construction of
  $\charform{\mathcal{T}}$. Since the SID $\asid$, that interprets the
  formula $\phi$, is tight, the rule body $\varphi$ is tight, hence
  one of the following holds: \begin{compactitem}
  \item there exists a component atom $\acomp^q(y_k)$ in $\varphi$,
    such that $\tuple{\portsof{\ainterac}}_k = \acomp$, or
  \item there exists a predicate atom $\bpred(\xi_1, \ldots,
    \xi_{\#(\bpred)})$ in $\varphi$, such that $y_k = \xi_\ell$ and
    $\tuple{\portsof{\ainterac}}_k =
    \tuple{\aprof_\asid(\bpred)}_\ell$, where $\aprof_\asid$ is the
    profile from Def. \ref{def:tight-formulae}. In this case, we apply
    induction to prove the existence of a component atom
    $\acomp^q(\xi_k)$ in $\charform{\proj{\mathcal{T}}{i}}$, such that
    $\tuple{\portsof{\ainterac}}_k = \acomp$, where $i \in
    \interv{1}{\npred{\varphi}}$ is the child of $w$ corresponding to
    $\bpred(\xi_1, \ldots, \xi_{\#(\bpred)})$ in $\mathcal{T}$.
  \end{compactitem}
  In both cases, $\charform{\mathcal{T}}$ contains a component atom
  $\acomp^q(z_k)$, such that $\tuple{\portsof{\ainterac}}_k = \acomp$,
  hence $(\struc,\amark)$ is tight, by
  Def. \ref{def:tight-architecture}. \qed}

The next result proves that each model of a tight sentence is
necessarily symmetric to a canonical model of that sentence:

\begin{lemma}\label{lemma:models-symm}
  Given a tight sentence $\phi$, for each $\mathcal{T} \in
  \rtrees{\phi}$, if $(\struc,\amark) \models \charform{\mathcal{T}}$,
  then $(\struc,\amark) \symm{}{}
  (\struc_{\mathcal{T}},\amark_{\mathcal{T}})$.
\end{lemma}
\proof{ Since $(\struc_{\mathcal{T}},\amark_{\mathcal{T}}) \models
  \charform{\mathcal{T}}$, by Lemma \ref{lemma:canonical-model}, it is
  sufficient to prove that $(\struc_1,\amark_1) \symm{}{}
  (\struc_2,\amark_2)$, for any two configurations
  $(\struc_i,\amark_i)$, such that $(\struc_i,\amark_i) \models
  \charform{\mathcal{T}}$. Let $\charform{\mathcal{T}} \isdef \exists
  x_1 \ldots \exists x_n ~.~ \eta$, where $\eta$ is a quantifier- and
  predicate-free formula. Then $(\struc_i,\amark_i) \models
  \charform{\mathcal{T}}$ if and only if there exist stores
  $\store_i$, such that $(\struc_i,\amark_i) \models^{\store_i} \eta$,
  for $i = 1,2$. We define the sets $U^i_j \isdef \set{\store_i(x)
    \mid \acomp^q_j(x) \text{ occurs in } \eta}$ and the bijections
  $f_j : U^1_j \rightarrow U^2_j$ as $f_j(\store_1(x)) = \store_2(x)$,
  for all variables $x$, such that $\acomp^q_j(x)$ occurs in
  $\eta$. These bijections are extended to bijections $\overline{f}_j
  : \universe \rightarrow \universe$, by the following fact: 
  \begin{fact}
    Given finite sets $U_1,U_2 \subseteq \universe$, such that $f :
    U_1 \rightarrow U_2$ is a bijection, there exists a bijection
    $\overline{f} : \universe \rightarrow \universe$, such that
    $\overline{f}(u) = f(u)$, for all $u \in U_1$.
  \end{fact}
  \proof{ Since $f$ is a bijection, we have $\card{U_1} = \card{U_2}$
    and, since $\card{U_i} = \card{U_i \setminus U_{3-i}} + \card{U_1
      \cap U_2}$, for $i=1,2$, we obtain $\card{U_1 \setminus U_2} =
    \card{U_2 \setminus U_1}$. Hence there exists a bijection
    $\overline{f} : \universe \rightarrow \universe$, such that
    $\overline{f}(u) = f(u)$, for all $u \in U_1$, $\overline{f}(U_2
    \setminus U_1) = U_1 \setminus U_2$ and $\overline{f}(\universe
    \setminus (U_1 \cup U_2)) = f(\universe \setminus (U_1 \cup
    U_2))$. \qed}

  \noindent For $\vec{f} \isdef \tuple{\overline{f}_1, \ldots,
    \overline{f}_\ncomps}$, we prove the following
  points: \begin{compactitem}
  \item $\vec{f}(\struc_1) = \struc_2$: by the definition of
    $\overline{f}_j$, we have $\overline{f}_j(\struc_1(\acomp_j)) =
    \struc_2(\acomp_j)$, for all $j \in \interv{1}{\ncomps}$.  Let
    $\ainterac\in\interacs$ and we prove $\struc_2(\ainterac) =
    \{\tuple{\overline{f}_{i_1}(u_1), \ldots,
      \overline{f}_{i_{\#(\ainterac)}}(u_{\#(\ainterac)})} \mid
    \tuple{u_1, \ldots, u_{\#(\ainterac)}} \in \struc_1(\ainterac),~
    \forall k \in \interv{1}{\#(\ainterac)} ~.~
    \compof{\tuple{\portsof{\ainterac}}_k} = \acomp_{i_k}\}$.
    ``$\subseteq$'' Let $\tuple{u_1, \ldots, u_{\#(\ainterac)}} \in
    \struc_2(\ainterac)$. Then there exists an interaction atom
    $\ainterac(x_1, \ldots, x_{\#(\ainterac)})$ in $\eta$, such that
    $u_i = \store_2(x_i)$, for $i \in \interv{1}{\#(\ainterac)}$.
    Since $\phi$ is a tight sentence and $(\struc_2,\amark_2) \models
    \phi$, by Lemma \ref{lemma:tight-models}, the architecture
    $\struc_2$ is tight (Def.  \ref{def:tight-architecture}).  Then,
    for each $k \in \interv{1}{\#(\ainterac)}$, we have $u_k \in
    \struc_2(\acomp_{i_k})$, where $\acomp_{i_k} \in \comps$ is the
    unique component type such that $\tuple{\portsof{\ainterac}}_k \in
    \portsof{\acomp_{i_k}}$.  By the previous point, there exists
    $u'_k \in \struc_1(\acomp_{i_k})$, such that
    $\overline{f}_{i_k}(u'_k) = u_k = \store_2(x_k)$. By the
    definition of $\overline{f}_{i_k}$, we have $u'_k =
    \store_1(x_k)$, hence $\tuple{u'_1, \ldots, u'_{\#(\ainterac)}}
    \in \struc_1(\ainterac)$. ``$\supseteq$'' Let $\tuple{u_1, \ldots,
      u_{\#(\ainterac)}} \in \struc_1(\ainterac)$, such that
    $\compof{\tuple{\portsof{\ainterac}}_k} = \acomp_{i_k}$, for all
    $k \in \interv{1}{\#(\ainterac)}$. Then there exists an
    interaction atom $\ainterac(x_1, \ldots, x_{\#(\ainterac)})$ in
    $\eta$, such that $u_k = \store_1(x_k)$, for all $k \in
    \interv{1}{\#(\ainterac)}$. Since $\phi$ is a tight sentence and
    $(\struc_1,\amark_1) \models \phi$, by Lemma
    \ref{lemma:tight-models}, the architecture $\struc_1$ is tight,
    thus $\store_1(x_k) = u_k \in \struc_1(\acomp_{i_k})$. By the
    definition of $\overline{f}_{i_k}$, we have $\store_2(x_k) =
    \overline{f}_{i_k}(u_k)$, for all $k \in
    \interv{1}{\#(\ainterac)}$, hence $\tuple{\overline{f}_{i_1}(u_1),
      \ldots, \overline{f}_{i_{\#(\ainterac)}}(u_{\#(\ainterac)})} \in
    \struc_2(\ainterac)$.
  \item $\vec{f}(\amark_1) = \amark_2$: we prove $\amark_2 =
    \set{q[\overline{f}_i(u)] \mid q[u] \in \amark_1, q \in
      \statesof{\acomp_i}, i \in
      \interv{1}{\ncomps}}$. ``$\subseteq$'' Let $q[u] \in
    \amark_2$. Since $(\struc_2,\amark_2) \models^{\store_2} \eta$,
    there exists a component atom $\acomp_i^q(x)$ in $\eta$, such that
    $u = \store_2(x) \in \struc_2(\acomp_i)$ and $\acomp_i$ is the
    unique component type, such that $q \in \statesof{\acomp_i}$. By
    the definition of $\overline{f}_i$, we have $\store_1(x) \in
    \struc_1(\acomp_i)$ and $\store_2(x) =
    \overline{f}_i(\store_1(x))$. Moreover, since $(\struc_1,\amark_1)
    \models^{\store_1} \eta$, we have $q[\store_1(x)] \in
    \amark_1$. ``$\supseteq$'' Let $q[u] \in \amark_1$ and $q \in
    \statesof{\acomp_i}$, for some $i \in \interv{1}{\ncomps}$. Since
    $(\struc_1, \amark_1) \models^{\store_1} \eta$, there exists a
    component atom $\acomp^q(x)$ in $\eta$, such that $u =
    \store_1(x)$. By the definition of $\overline{f}_i$, we have
    $\overline{f}_i(\store_1(x)) = \store_2(x)$, hence
    $q[\overline{f}_i(u)] = q[\store_2(x)] \in \amark_2$, since
    $(\struc_2,\amark_2) \models^{\store_2} \eta$. \qed
  \end{compactitem}}

Moreover, the symmetry relation is preserved by the Petri net
describing the operational semantics of the system (Def. \ref{def:execution-semantics}):

\begin{lemma}\label{lemma:pn-symm}
   Given a tight architecture $\struc$ and a behavior $\beh$, over a
   signature $\signature = (\set{\acomp_1, \ldots,
     \acomp_\ncomps},\interacs,\ports)$, for any tuple of bijections
   $\vec{f} = \tuple{f_1, \ldots, f_\ncomps}$, where $f_i : \universe
   \rightarrow \universe$, for all $i \in \interv{1}{\ncomps}$, the
   following hold: \begin{compactenum}
   \item\label{it1:pn-symm} $\amark' \in
    \reach{\anetof{\signature,\struc,\beh}, \amark} \iff \vec{f}(\amark')
    \in \reach{\anetof{\signature, \vec{f}(\struc), \beh}, \vec{f}(\amark)}$,
   \item\label{it2:pn-symm} $\amark \in
     \dead{\anetof{\signature,\struc,\beh}} \iff \vec{f}(\amark) \in
     \dead{\anetof{\signature,\vec{f}(\struc),\beh}}$,
   \end{compactenum}
   for any two markings $\amark,\amark'$ of the Petri net
   $\anetof{\signature,\struc,\beh}$.
\end{lemma}
\proof{ By Def. \ref{def:execution-semantics}, let
  $\anetof{\signature,\vec{f}(\struc),\beh}$ be the Petri net
  $(\places^{\vec{f}}, \trans^{\vec{f}}, \edges^{\vec{f}})$, where:
  \[\begin{array}{rcl}
  \places^{\vec{f}} & = & \bigcup_{i=1}^\ncomps \set{q[u] \mid q
    \in \compstates{\acomp_i},~ u \in (\vec{f}(\struc))(\acomp_i)} \\[1mm]
  & = & \bigcup_{i=1}^\ncomps \set{q[f_i(u)] \mid q \in
    \compstates{\acomp_i},~ u \in \struc(\acomp_i)}
  \\[2mm]
  \trans^{\vec{f}} & = & \bigcup_{\ainterac\in\interacs}
  \{ ~ (\ainterac[u_1,\ldots,u_{\#(\ainterac)}], \tuple{t_1,\ldots, t_{\#(\ainterac)}})
  ~\mid~ \tuple{p_1, \ldots, p_{\#(\ainterac)}} = \interportsof{\ainterac}, \\[1mm]
  & & \hspace{1.2cm} \tuple{u_1,\ldots,u_{\#(\ainterac)}} \in (\vec{f}(\struc))(\ainterac),~
  \tuple{t_1,\ldots,t_{\#(\ainterac)}} \in \transof{\ainterac}, \\ [1mm]
  & & \hspace{1.2cm} \forall i,j \in \interv{1}{\#(\ainterac)}.~ i\not=j \Rightarrow
  u_i\not=u_j \mbox{ or } \compof{p_i} \not= \compof{p_j} ~ \} \\[2mm]
  & = & \bigcup_{\ainterac\in\interacs}
  \{ ~ (\ainterac[f_{k_1}(u_1),\ldots,f_{k_{\#(\ainterac)}}(u_{\#(\ainterac)})], \tuple{t_1,\ldots, t_{\#(\ainterac)}})
  ~\mid~ \tuple{p_1, \ldots, p_{\#(\ainterac)}} = \interportsof{\ainterac}, \\[1mm]
  & & \hspace{1.2cm} \tuple{\acomp_{k_1},\ldots \acomp_{k_{\#(\ainterac)}}} = \tuple{\compof{p_1},\dots,\compof{p_{\#(\ainterac)}}}, \\ [1mm]
  & & \hspace{1.2cm} \tuple{u_1,\ldots,u_{\#(\ainterac)}} \in \struc(\ainterac),~
  \tuple{t_1,\ldots,t_{\#(\ainterac)}} \in \transof{\ainterac}, \\ [1mm]
  & & \hspace{1.2cm} 
  \forall i,j \in \interv{1}{\#(\ainterac)}.~ i\not=j \Rightarrow f_{k_i}(u_i)\not=f_{k_j}(u_j) \mbox{ or } \compof{p_i} \not= \compof{p_j}~ \}
  \end{array}\]
  \[\begin{array}{rcl}
  \edges^{\vec{f}} & = & \bigcup_{\ainterac\in\interacs}
  \{ ~ (q[f_{k_i}(u_i)], (\ainterac[f_{k_1}(u_1),\ldots,f_{k_{\#(\ainterac)}}(u_{\#(\ainterac)})], \tuple{t_1,\ldots, t_{\#(\ainterac)}}) ), \\[1mm]
  & & \hspace{1cm} ((\ainterac[f_{k_1}(u_1),\ldots,f_{k_{\#(\ainterac)}}(u_{\#(\ainterac)})], \tuple{t_1,\ldots, t_{\#(\ainterac)}}), q'[f_{k_i}(u_i)]) ~\mid ~\\[1mm]
  & & \hspace{1.2cm} t_i = (q \arrow{p_i}{\compof{p_i}} q'),~i \in \interv{1}{\#(\ainterac)} ~ \}
  \end{array}\]
  We obtain the following equivalence: \[\begin{array}{c}
  \amark \arrow{(\ainterac[u_1, \ldots,
      u_{\#(\ainterac)}], \tuple{t_1, \ldots, t_{\#(\ainterac)}})}{}
  \amark' \text{ in $\anetof{\signature,\struc,\beh}$} \\
  \iff \\
  \vec{f}(\amark) \arrow{(\ainterac[f_{k_1}(u_1), \ldots,
      f_{k_{\#(\ainterac)}}(u_{\#(\ainterac)})], \tuple{t_1, \ldots,
      t_{\#(\ainterac)}})}{} \vec{f}(\amark') \text{ in
    $\anetof{\signature,\vec{f}(\struc),\beh}$}
  \end{array}\]
  where $k_1, \ldots, k_{\#(\ainterac)} \in \interv{1}{\ncomps}$ are
  such that $\acomp_{k_i} = \compof{\tuple{\portsof{\ainterac}}_i}$,
  for all $i \in \interv{1}{\#(\ainterac)}$. Point (\ref{it1:pn-symm})
  uses this fact inductively, on the length of the firing sequence
  leading from $\amark$ to $\amark'$. For point (\ref{it2:pn-symm}) we
  use, moreover, that $q[u] \in \amark \iff q[f_i(u)] \in
  \vec{f}(\amark)$ and $q[u] \in \pre{(\ainterac[u_1, \ldots,
      u_{\#(\ainterac)}],\tuple{t_1, \ldots, t_{\#(\ainterac)}})} \iff
  q[f_i(u)] \in \pre{(\ainterac[f_{k_1}(u_1), \ldots,
      f_{k_{\#(\ainterac)}}(u_{\#(\ainterac)})],\tuple{t_1, \ldots,
      t_{\#(\ainterac)}})}$, where $\acomp_i$ is the unique component
  type such that $q \in \statesof{\acomp_i}$.  \qed}

\begin{theorem}\label{thm:symmetry}
  Given a tight sentence $\phi$, interpreted over a SID $\asid$, a
  behavior map $\beh$ and a tuple of states $\tuple{q_1, \ldots,
    q_k}$, the following equivalences hold:
  \[\begin{array}{rcl}
  \mathit{deadlock}_t(\phi,\asid,\beh) & \iff &
  \mathit{deadlock}(\phi,\asid,\beh) \\
  \mathit{reach}_t(\phi,\tuple{q_1,\ldots,q_k},\asid,\beh)
  & \iff & \mathit{reach}(\phi,\tuple{q_1,\ldots,q_k},\asid,\beh)
  \end{array}\]
  where $\mathit{deadlock}_t(\phi,\asid,\beh)$ and
  $\mathit{reach}_t(\phi,\tuple{q_1,\ldots,q_k},\asid,\beh)$
  are defined below: \begin{itemize}
  \item $\mathit{deadlock}_t(\phi,\asid,\beh)$: does there
    exist rewriting tree $\mathcal{T} \in \rtrees{\phi}$, such that
    $\reach{\anetof{\signature,\struc_{\mathcal{T}},\beh},\amark_{\mathcal{T}}}
    \cap \dead{\anetof{\signature,\struc_{\mathcal{T}},\beh}} \neq
    \emptyset$?
  \item
    $\mathit{reach}_t(\phi,\tuple{q_1,\ldots,q_k},\asid,\beh)$:
    does there exist a rewriting tree $\mathcal{T} \in \rtrees{\phi}$
    and a configuration $(\struc_{\mathcal{T}}, \amark) \in
    \reach{\anetof{\signature,\struc_{\mathcal{T}},\beh},
      \amark_{\mathcal{T}}}$, such that
    $\{q_i[u_i]$ $\mid i\in\interv{1}{k}\} \subseteq \amark$, with
    $q_1[u_1], \ldots, q_k[u_k]$ pairwise distinct?
  \end{itemize}
\end{theorem}
\proof{ We prove the following points: \begin{compactitem}
  \item $\mathit{deadlock}_t(\phi,\asid,\beh) \Rightarrow
    \mathit{deadlock}(\phi,\asid,\beh)$: assume that
    $\mathit{deadlock}_t(\phi,\asid,\beh)$ has a positive
    answer and let $\mathcal{T} \in \rtrees{\phi}$ be a rewriting
    tree, such that
    $\reach{\anetof{\signature,\struc_{\mathcal{T}},\beh},\amark_{\mathcal{T}}}
    \cap \dead{\anetof{\signature,\struc_{\mathcal{T}},\beh}} \neq
    \emptyset$, the existence of which is stated by
    $\mathit{deadlock}_t(\phi,\asid,\beh)$. By Lemma
    \ref{lemma:canonical-model},
    $(\struc_\mathcal{T},\amark_{\mathcal{T}}) \models
    \charform{\mathcal{T}}$ and, by Lemma \ref{lemma:rewriting-trees},
    we obtain $(\struc_{\mathcal{T}},\amark_{\mathcal{T}}) \models
    \phi$, thus $\mathit{deadlock}(\phi,\asid,\beh)$ has a positive
    answer.
  \item $\mathit{deadlock}(\phi,\asid,\beh) \Rightarrow
    \mathit{deadlock}_t(\phi,\asid,\beh)$: assume that
    $\mathit{deadlock}(\phi,\asid,\beh)$ has a positive answer and let
    $(\struc,\amark)$ be a model of $\phi$ and $\amark'$ be a marking
    such that $\amark' \in
    \reach{\anetof{\signature,\struc,\beh},\amark} \cap
    \dead{\anetof{\signature,\struc,\beh}}$. By Lemma
    \ref{lemma:rewriting-trees}, there exists a rewriting tree
    $\mathcal{T} \in \rtrees{\phi}$, such that $(\struc,\amark)
    \models \charform{\mathcal{T}}$. By Lemma \ref{lemma:models-symm},
    since $\phi$ is a tight sentence, we obtain that $(\struc,\amark)
    \symm{}{} (\struc_{\mathcal{T}},\amark_{\mathcal{T}})$ and let
    $\vec{f} = \tuple{f_1,\ldots,f_\ncomps}$ be the bijections from
    Def. \ref{def:symmetry}, such that
    $(\struc_{\mathcal{T}},\amark_{\mathcal{T}}) =
    (\vec{f}(\struc),\vec{f}(\amark))$. By Lemma \ref{lemma:pn-symm},
    we obtain that $\vec{f}(\amark') \in
    \reach{\anetof{\signature,\vec{f}(\struc),\beh},\vec{f}(\amark)}
    \cap \dead{\anetof{\signature,\vec{f}(\struc),\beh}}$, thus
    $\mathit{deadlock}_t(\phi,\asid,\beh)$ has a positive
    answer.
  \item
    $\mathit{reach}_t(\phi,\tuple{q_1,\ldots,q_k},\asid,\beh)
    \Rightarrow\mathit{reach}(\phi,\tuple{q_1,\ldots,q_k},\asid,\beh)$:
    assume that \\
    $\mathit{reach}_t(\phi,\tuple{q_1,\ldots,q_k},\asid,\beh)$
    has a positive answer and let $\mathcal{T} \in \rtrees{\phi}$ be a
    rewriting tree and $\amark$ be a marking of
    $\anetof{\signature,\struc_{\mathcal{T}},\beh}$, such that
    $(\struc_{\mathcal{T}}, \amark) \in
    \reach{\anetof{\signature,\struc_{\mathcal{T}},\beh},
      \amark_{\mathcal{T}}}$ and $\set{q_i[u_i]~|~i\in\interv{1}{k}}
    \subseteq \amark$, where $q_1[u_1], \ldots, q_k[u_k]$ are pairwise
    distinct places. By Lemma \ref{lemma:canonical-model},
    $(\struc_\mathcal{T},\amark_{\mathcal{T}}) \models
    \charform{\mathcal{T}}$ and, by Lemma \ref{lemma:rewriting-trees},
    we obtain $(\struc_{\mathcal{T}},\amark_{\mathcal{T}}) \models
    \phi$, thus
    $\mathit{reach}(\phi,\tuple{q_1,\ldots,q_k},\asid,\beh)$ has a
    positive answer.
  \item $\mathit{reach}(\phi,\tuple{q_1,\ldots,q_k},\asid,\beh)
    \Rightarrow\mathit{reach}_t(\phi,\tuple{q_1,\ldots,q_k},\asid,\beh)$:
    assume that \\
    $\mathit{reach}(\phi,\tuple{q_1,\ldots,q_k},\asid,\beh)$ has a
    positive answer and let $(\struc,\amark)$ be a model of $\phi$ and
    $\amark' \in \reach{\anetof{\signature,\struc,\beh}, \amark}$ be a
    marking, such that $\set{q_i[u_i] \mid i \in \interv{1}{k}}
    \subseteq \amark'$, where $q_1[u_1], \ldots, q_k[u_k]$ are
    pairwise distinct places of $\anetof{\signature,\struc,\beh}$. By
    Lemma \ref{lemma:rewriting-trees}, there exists a rewriting tree
    $\mathcal{T} \in \rtrees{\phi}$, such that $(\struc,\amark)
    \models \charform{\mathcal{T}}$. By Lemma \ref{lemma:models-symm},
    since $\phi$ is a tight sentence, we obtain that $(\struc,\amark)
    \symm{}{} (\struc_{\mathcal{T}},\amark_{\mathcal{T}})$ and let
    $\vec{f} = \tuple{f_1,\ldots,f_\ncomps}$ be the bijections from
    Def. \ref{def:symmetry}, such that
    $(\struc_{\mathcal{T}},\amark_{\mathcal{T}}) =
    (\vec{f}(\struc),\vec{f}(\amark))$. By Lemma \ref{lemma:pn-symm},
    we obtain that $\vec{f}(\amark') \in
    \reach{\anetof{\signature,\vec{f}(\struc),\beh},\vec{f}(\amark)}$. Moreover,
    if $\acomp_{\ell_i}$ is the unique component type, such that $q_i
    \in \statesof{\acomp_{k_i}}$, for all $i \in \interv{1}{k}$, since
    $\set{q_i[u_i] \mid i \in \interv{1}{k}} \subseteq \amark'$, we
    obtain $\set{q_i[f_{\ell_i}(u_i)] \mid i \in \interv{1}{k}}
    \subseteq \vec{f}(\amark')$ and
    $\mathit{reach}_t(\phi,\tuple{q_1,\ldots,q_k},\asid,\beh)$
    has a positive answer. \qed
  \end{compactitem}}

\section{Parametric Verification using Structural Invariants}
\label{sec:verification}

In this section we present a sound (but necessarily incomplete) method
to address practical instances of the (undecidable) decision problems
of deadlock and reachability (Theorem
\ref{thm:deadlock-reach-safe-undec}). The method proceeds by
synthesizing a sufficient verification condition as a formula in a
decidable fragment of \mso, interpreted over
trees of bounded arity $\nary$, with second-order set variables
ranging over finite sets.  This logic is known as the Weak Second-order
calculus of $\nary$ Successors (\wsks). Well-known automata-theoretic
decision procedures exist for \wsks\ \cite{Thatcher2005GeneralizedFA}
and we rely on optimized provers \cite{Mona} to check the verification
conditions.


During the unfolding of the predicate symbols, according to the
structure of a rewriting tree $\mathcal{T}$, an existentially
quantified variable introduced by a rule might be renamed several
times (the renaming occurs when replacing a predicate atom
$\apred(y_1,\ldots,y_{\#(\apred)})$ with one of its definitions
$\varphi[x_1/y_1 \ldots x_{\#(\apred)}/y_{\#(\apred)}]$, for some rule
$\apred(x_1,\ldots,x_{\#(\apred)})\leftarrow\varphi$ from the SID)
before it is instantiated by a component atom $\acomp^q(x)$, that
occurs in some node $w \in \nodes{\mathcal{T}}$. Due to Assumption
\ref{ass:distinct-component-types} on the syntax of the rules in the
SID, this node is unique and the canonical store
$\store_{\mathcal{T}}$ maps the variable to that node,
i.e.\ $\store_{\mathcal{T}}(x) = w$. In order to determine this value,
we must track $x$ along the path in the rewriting tree that leads from
the node where it was introduced (by an existential quantifier, or the
root of the tree if the variable is a top-level parameter) to the node
where the component atom $\acomp^q(x)$ occurs.

\begin{example}\label{ex:track}
  Let us consider the rewriting tree $\mathcal{T}$ depicted in
  Fig.~\ref{fig:tll}(a). The variable $\ell^\epsilon_1$ introduced by
  an existential quantifier at the root of the tree replaces the
  $\ell$ parameter of the rule that labels the left successor of the
  root, i.e.\ the node $1$, that replaces the $\ell$ parameter of the
  label of the node $11$, which finally replaces the $x$ parameter of
  the label of the node $111$, that contains a component atom
  $\treeleaf^{s_0}(x)$. Hence, the value of $\ell^\epsilon_1$ in the
  canonical store is $\store_{\mathcal{T}}(\ell^\epsilon_1) =
  111$. \hfill$\blacksquare$
\end{example}

Tracking variables is done by \emph{path automata} that traverse the
tree downwards, following the substitutions of a particular
variable. A path automaton is then transformed into an equivalent
\wsks\ \emph{path formula}, that defines the set of trees over which
the automaton has an accepting run. The path formula is used to define
a \emph{flow formula} that describes the pre- and post-sets for each
transition of the Petri net
$\anetof{\signature,\struc_{\mathcal{T}},\beh}$ corresponding to the
canonical architecture $\struc_{\mathcal{T}}$, defined by the
rewriting tree $\mathcal{T}$, and the given behavior map
$\beh$. Moreover, path formul{\ae} allow to define \wsks\ formul{\ae}
describing the initial and final configurations of the system.

In particular, the flow formula is used to derive invariants
(i.e.\ over-approxima\-tions of the set of reachable configurations)
directly from the structure of the Petri net
$\anetof{\signature,\struc_{\mathcal{T}},\beh}$. These invariants are
defined by the sets of places of
$\anetof{\signature,\struc_{\mathcal{T}},\beh}$ that do not lose (trap
invariants) or generate extra tokens (mutex invariants), respectively. All
verification conditions for the problems considered in this paper boil
down to checking the satisfiability of a \wsks\ formula.


\subsection{Weak Second Order Calculus of $\kappa$ Successors}

We define the logic \wsks, which is a fragment of \mso\ interpreted over a finite prefix of an infinite $\kappa$-ary
tree. As a remainder, the prefix of a tree $\mathcal{T}$ is the
restriction of $\mathcal{T}$ to a prefix-closed and complete subset of
$\nodes{\mathcal{T}}$ (\S\ref{sec:rewriting-trees}).

Let $\Vars = \set{X,Y,Z,\ldots}$ be a countably infinite set of
second-order variables, ranging over subsets of
$\interv{1}{\kappa}^*$. The formul{\ae} of \wsks\ are defined
inductively by the following syntax:
\[\begin{array}{rclr}
\tau & ::= & \epsilon \mid x \mid \tau.i & \text{ terms} \\
\xi & ::= & \tau = \tau \mid X(\tau) \mid \xi \wedge \xi \mid \neg\xi
\mid \exists x ~.~ \xi \mid \exists X ~.~ \xi & \text{ formul{\ae}}
\end{array}\]
where $x \in \vars$, $X \in \Vars$ and $i \in \interv{1}{\kappa}$. As
usual, we write $x \neq y \isdef \neg x = y$, $\xi_1 \vee \xi_2
\isdef \neg(\neg\xi_1 \wedge \neg\xi_2)$, $\xi_1 \rightarrow \xi_2
\isdef \neg\xi_1 \vee \xi_2$, $\xi_1 \leftrightarrow \xi_2 \isdef
(\xi_1 \rightarrow \xi_2) \wedge (\xi_2 \rightarrow \xi_1)$,
$\forall x ~.~ \xi \isdef \neg\exists x ~.~ \neg\xi$ and $\forall X
~.~ \xi \isdef \neg\exists X ~.~ \neg\xi$. For a constant $n \in
\nat$, we define:
\begin{align}
  X = \set{y_1, \ldots, y_n} & \iff \forall x ~.~ X(x) \leftrightarrow \bigvee_{i=1}^n x = y_i \label{eq:setof} \\
  \card{X} \geq n & \iff \exists y_1 \ldots \exists y_n ~. \bigwedge_{1 \leq i < j \leq n} y_i \neq y_j \wedge \bigwedge_{i=1}^n X(y_i) \label{eq:cardgeq} \\
  \card{X} = n & \iff \card{X} \geq n \wedge \neg \card{X} \geq n+1 \label{eq:cardeq}
\end{align}

The variables of a \wsks\ formula are interpreted by a store $\nu :
\vars \cup \Vars \rightarrow \interv{1}{\kappa}^* \cup
2^{\interv{1}{\kappa}^*}$, such that $\nu(x) \in
\interv{1}{\kappa}^*$, for each $x \in \vars$ and $\nu(X) \subseteq
\interv{1}{\kappa}^*$ is a \emph{finite set}, for each $X \in
\Vars$. The terms of \wsks\ are interpreted inductively, as
$\nu(\epsilon) \isdef \epsilon$ and $\nu(\tau.i) \isdef \nu(\tau)\cdot
i$. We denote by $\wsksmodels{\nu} \xi$ the fact that the
\wsks\ formula $\xi$ is \emph{valid} for the valuation $\nu$. This
relation defined below, by induction on the structure of the
formul{\ae}:
\[\begin{array}{lcl}
\wsksmodels{\nu} \tau_1 = \tau_2 & \iff & \nu(\tau_1) = \nu(\tau_2) \\
\wsksmodels{\nu} X(\tau) & \iff & \nu(\tau) \in \nu(X) \\
\wsksmodels{\nu} \xi_1 \wedge \xi_2 & \iff & \wsksmodels{\nu} \xi_1 \text{ and } \wsksmodels{\nu} \xi_2 \\
\wsksmodels{\nu} \neg\xi & \iff & \text{not } \wsksmodels{\nu}{\xi} \\
\wsksmodels{\nu} \exists x ~.~ \xi & \iff & \wsksmodels{\nu[x\leftarrow w]} \xi \text{, for some node } w \in \interv{1}{\kappa}^* \\
\wsksmodels{\nu} \exists X ~.~ \xi & \iff & \wsksmodels{\nu[X\leftarrow S]} \xi \text{, for some finite set } S \subseteq \interv{1}{\kappa}^*
\end{array}\]
A \wsks\ formula $\xi$ is \emph{satisfiable} if and only if there
exists a store $\nu$, such that $\wsksmodels{\nu} \xi$. Such a store
is said to be a \emph{model} of $\xi$. Note that, because we have
assumed the successor functions $.i$ to be total, for all $i \in
\interv{1}{\kappa}$, formul{\ae} defining infinite sets, such as
e.g.\ $\forall x ~.~ X(x) \rightarrow X(x.1)$, are unsatisfiable,
under the interpretation of second-order variables as finite sets. 

\subsection{Rewriting Trees and Configurations}
\label{sec:trees-configurations}

We begin by building a \wsks\ formula that describes an infinite
$\kappa$-ary tree, whose finite prefix is a rewriting tree
$\mathcal{T} \in \rtrees{\phi}$, for a \adl\ formula $\phi$. Let
$\asid = \set{\arule_1, \ldots, \arule_\nrules}$ be a fixed SID in the
following, such that $\arule_1$ is the rule that labels the root of
$\mathcal{T}$, by convention.  We use a designated tuple of second
order variables $\vec{R} = \tuple{R_1, \ldots, R_\nrules}$, where each
variable $R_i$ is interpreted as the set of tree nodes labeled with
the rule $\arule_i$ in some rewriting tree, for all $i \in
\interv{1}{\nrules}$. With this convention, the $\rewrtree(\vec{R})$
formula (Fig.~\ref{fig:rtree}) defines rewriting
trees: \begin{itemize}
\item line (\ref{eq:rtree1}) states that the sets that interpret the
  second-order variables $\vec{R}$ are pairwise disjoint and that
  $R_1$ is a singleton containing the root of the rewriting tree,
\item line (\ref{eq:rtree2}) states that the union of the sets
  $\vec{R}$ is prefix-closed, i.e.\ the parent of each node from the
  interpretation of a variable $R_i$ belongs to the interpretation of
  some variable $R_j$, for $i,j \in \interv{1}{\nrules}$,
\item lines (\ref{eq:rtree3}) and (\ref{eq:rtree4}) encode the
  conditions \ref{it1:rewriting-tree} and \ref{it2:rewriting-tree} of
  Def. \ref{def:rewriting-tree}, respectively, i.e.\ for that every
  predicate atom $\apred'(y_1, \ldots, y_{\#(\apred')})$ from a rule
  that labels a node $w$ in the rewriting tree there is exactly one
  child of that node, labeled with a rule $\apred'(x_1, \ldots,
  x_{\#(\apred')}) \leftarrow \psi$ and $w$ has no other children.
\end{itemize} 
It is not very hard to show that, for each model $\nu$ of
$\rewrtree(\vec{R})$ there exists a unique rewriting tree, denoted by
$\rtreeof{\nu}{\vec{R}}$, such that $\nodes{\rtreeof{\nu}{\vec{R}}} =
\bigcup_{i=1}^\nrules \nu(R_i)$ and $\rtreeof{\nu}{\vec{R}}(w) =
\arule_i \iff w \in \nu(R_i)$, for all $i \in \interv{1}{\nrules}$ and
$w \in \nodes{\rtreeof{\nu}{\vec{R}}}$.

\begin{figure}[t!]
 \begin{center}
   \begin{align}
     \rewrtree(\vec{R}) \isdef 
     \forall x ~.~ \hspace*{-4mm} \bigwedge_{1 \leq i < j \leq \nrules} \hspace*{-4mm}
     \big(\neg R_i(x) \vee \neg R_j(x)\big) \wedge \big(R_1(x) \leftrightarrow x = \epsilon\big) ~\wedge \label{eq:rtree1} \\
     \forall x ~.~ \bigwedge_{i=1}^\nrules \bigwedge_{\ell=1}^{\kappa} 
     R_i(x.\ell) \rightarrow \bigvee_{j=1}^\nrules R_j(x) ~\wedge \label{eq:rtree2} \\
     \forall x ~.~  \hspace*{-4mm} \bigwedge_{\scriptstyle{i : \arule_i = \left(\apred(x_1, \ldots, x_{\#(\apred)}) \leftarrow \varphi\right)}} 
     \hspace*{-2mm} \bigwedge_{j=1}^{\npred{\varphi}} R_i(x) \rightarrow \hspace*{-7mm} \bigvee_{\begin{array}{l}
         \scriptstyle{j : \pred{j}{\varphi}=\apred'(y_1, \ldots, y_{\#(\apred')})} \\[-.5mm]
         \scriptstyle{\ell : \arule_\ell = \left(\apred'(x_1, \ldots, x_{\#(\apred')}) \leftarrow \psi\right)}    
     \end{array}} \hspace*{-12mm} R_{\ell}(x.j) \label{eq:rtree3} ~\wedge \\
     \forall x ~.~  \hspace*{-4mm} \bigwedge_{\scriptstyle{i : \arule_i = \left(\apred(x_1, \ldots, x_{\#(\apred)}) \leftarrow \varphi\right)}} 
     \bigwedge_{j=\npred{\varphi}+1}^{\kappa}
     \hspace*{-2mm} R_i(x) \rightarrow \bigwedge_{\ell=1}^\nrules \neg R_\ell(x.j) \label{eq:rtree4}
   \end{align}
 \end{center}
 \vspace*{-\baselineskip}
 \caption{Rewriting Trees}
 \label{fig:rtree}
\end{figure}

As discussed in \S\ref{sec:rewriting}, a rewriting tree $\mathcal{T}$
defines a canonical configuration $(\struc_{\mathcal{T}},
\amark_{\mathcal{T}})$, where $\struc_{\mathcal{T}}$ is the
architecture consisting of the components and interactions
corresponding to the atoms of $\charform{\mathcal{T}}$ and
$\amark_{\mathcal{T}}$ is a marking of the Petri net
$\anetof{\signature,\struc_{\mathcal{T}},\beh}$, for a given signature
$\signature=(\set{\acomp_1, \ldots, \acomp_\ncomps}, \set{\ainterac_1,
  \ldots, \ainterac_\ninteracs}, \ports)$ and behavior map $\beh$. In
the rest of this section, we consider $\signature$ and $\beh$ to be
fixed and recall that $\globstates$ denotes the finite set of states
used to describe the behaviors $\behof{\acomp_1}, \ldots,
\behof{\acomp_\ncomps}$.

Our next concern is defining the precise markings of
$\anetof{\signature,\struc_{\mathcal{T}},\beh}$
(Def. \ref{def:precise-marking}) by a \wsks\ formula. We define sets
of places of $\anetof{\signature,\struc_{\mathcal{T}},\beh}$ by means
of a tuple of second-order variables $\vec{X} \isdef \tuple{ X_q \mid
  q \in \globstates}$. Any valuation $\nu$ of the variables $\vec{X}$
corresponds to a set of places $\markof{\nu}{\vec{X}} \isdef \set{q[u]
  \mid u \in \nu(X_q)}$. When needed, we shall use distinct copies of
$\vec{X}$, such as $\vec{X}' \isdef \tuple{X'_q \mid q \in
  \globstates}$, $\vec{Y} \isdef \tuple{Y_q \mid q \in \globstates}$
and $\vec{Z} \isdef \tuple{Z_q \mid q \in \globstates}$. The
\wsks\ formula below defines the precise markings of a Petri net
$\anetof{\signature,\struc_{\mathcal{T}},\beh}$, assuming that
$\mathcal{T}$ is the rewriting tree corresponding to some model of
$\rewrtree(\vec{R})$:

\begin{align}
  \ismark(\vec{X},\vec{R}) \isdef~ & \forall x ~.~ \bigwedge_{i=1}^\ncomps \Big(\bigwedge_{q \neq q' \in \statesof{\acomp_i}}
  \hspace*{-4mm} \big(\neg X_q(x) \vee \neg X_{q'}(x)\big) ~\wedge\label{eq:mark1} \\
  & \hspace*{15.5mm} \big(\bigvee_{q \in \statesof{\acomp_i}} \hspace*{-2mm} X_q(x) \leftrightarrow
  \hspace*{-4mm}\bigvee_{\arule_j \in \instof{\acomp_i}}\hspace*{-2mm} R_j(x)\big)\Big) \label{eq:mark2}
\end{align}
Intuitively, line (\ref{eq:mark1}) above states that a component of
type $\acomp_i$ may not be in two different states from
$\statesof{\acomp_i}$, and line (\ref{eq:mark2}) states that the index
of a component of type $\acomp_i$ must be a node of the rewriting tree
labeled with a rule in which a component atom of the form
$\acomp_i^q(x)$ occurs (see Assumption
\ref{ass:distinct-component-types} on the syntax of the rules labeling
a rewriting tree). We write $\instof{\acomp}$ for the subset of
$\asid$ consisting of those rules that contain a component atom of the
form $\acomp^q(x)$. The correctness of the encoding is stated and
proved below:

\begin{lemma}\label{lemma:tree-marking}
  For any model $\nu$ of $\rewrtree(\vec{R}) \wedge
  \ismark(\vec{X},\vec{R})$, the set $\markof{\nu}{\vec{X}}$ is a
  precise marking of the Petri net $\anetof{\signature,
    \struc_{\rtreeof{\nu}{\vec{R}}}, \beh}$.
\end{lemma}
\proof{ Let $\acomp \in \comps$ be a component type and $u \in
  \struc_{\rtreeof{\nu}{\vec{R}}}(\acomp)$ be an index.  We prove that
  $\card{\markof{\nu}{\vec{X}} \cap \set{q[u] \mid
      q\in\statesof{\acomp}}} = 1$, by proving the following
  points: \begin{compactitem}
  \item $\markof{\nu}{\vec{X}} \cap \set{q[u] \mid
    q\in\statesof{\acomp}}\neq \emptyset$: since $\wsksmodels{\nu}
    \rewrtree(\vec{R})$, we have that $\rtreeof{\nu}{\vec{R}}$ is a
    rewriting tree, such that $\nodes{\rtreeof{\nu}{\vec{R}}} =
    \bigcup_{i=1}^{\nrules} \nu(R_i)$ and $\nu(R_i)$ is the set of
    nodes of $\rtreeof{\nu}{\vec{R}}$ that are labeled with the rule
    $\arule_i$, for all $i \in \interv{1}{\nrules}$. Since $u \in
    \struc_{\rtreeof{\nu}{\vec{R}}}(\acomp) \subseteq
    \nodes{\rtreeof{\nu}{\vec{R}}}$, there exists a rule $\arule_i$,
    such that $\acomp \in \instof{\arule_i}$ and $u \in \nu(R_i)$,
    hence by line (\ref{eq:mark2}), we have $u \in \nu(X_q)$,
    i.e.\ $q[u]\in\markof{\nu}{\vec{X}}$, for some $q \in
    \statesof{\acomp}$. Consequently, $q[u] \in \markof{\nu}{\vec{X}}
    \cap \set{q[u] \mid q\in\statesof{\acomp}}$.
  \item $\card{\markof{\nu}{\vec{X}} \cap \set{q[u] \mid
      q\in\statesof{\acomp}}} < 2$: suppose that $q[u],q'[u]
    \in\markof{\nu}{\vec{X}}$, for $q \neq q' \in
    \statesof{\acomp}$. Hence $u \in \nu(X_q) \cap \nu(X_{q'})$, which
    contradicts line (\ref{eq:mark1}) and the fact that
    $\wsksmodels{\nu}{\ismark(\vec{X},\vec{R})}$. \qed
  \end{compactitem}}

\subsection{Path Automata}
\label{sec:path-automata}

The next step in building \wsks\ verification conditions that are
sufficient for proving deadlock freedom and unreachability of several
control states at once (\S\ref{sec:undecidability}) is tracking the
variables that occur in a rule labeling a node of a rewriting tree
$\mathcal{T}$, down to the component atoms which set their values, in
the canonical architecture $\struc_{\mathcal{T}}$. These values are
used to define the interactions of $\struc_{\mathcal{T}}$ and, in
particular, to encode the pre-set ($\pre{t}$) and post-set ($\post{t}$)
of a transition ($t$) from the Petri net
$\anetof{\signature,\struc_{\mathcal{T}},\beh}$, by a \emph{flow
formula}, described next (\S\ref{sec:flow-formulae}).

As briefly mentioned (see Example \ref{ex:track}) a variable
``traverses'' the rewriting tree of a given formula downwards, while
replacing the formal parameters of the rule labeling the child. This
traversal is described by a special type of tree automaton, that walks
down the tree, following the identity of the tracked variable. These
automata are a restricted version of \emph{tree walking automata}
\cite{Bojanczyk08}.

A \emph{path} in a tree $\mathcal{T}$ is a sequence $n_1, \ldots,
n_\ell \in \nodes{\mathcal{T}}$, such that $n_{i+1}$ is a child of
$n_i$, for all $i \in \interv{1}{\ell-1}$. If $m$ is a descendant of
$n$, we denote by $\pathof{n}{m}$ the unique path starting in $n$ and
ending in $m$. Note that $\pathof{n}{m}$ is determined by $n$ and the
sequence of \emph{directions} $d_1, \ldots, d_p \in
\interv{1}{\kappa}$, such that $m = n \cdot d_1 \cdot \ldots \cdot
d_p$. A \emph{path automaton} is a tuple $A = (S,I,F,\delta)$, where
$S$ is a set of states, $I, F \subseteq S$ are the sets of initial and
final states, respectively, and $\delta \subseteq S \times
\interv{1}{\kappa} \times S$ is a set of transitions $s \arrow{d}{}
s'$, where $s,s' \in S$ are states and $d \in \interv{1}{\kappa}$ is a
direction. A run of $A$ over the path $\pathof{n}{m}$, where $m = n
\cdot d_1 \cdot \ldots \cdot d_p$, is a sequence of states $s_1,
\ldots, s_{p+1} \in S$, such that $s_1 \in I$ and $s_i \arrow{d_i}{}
s_{i+1}$, for all $i \in \interv{1}{p}$. The run is accepting if and
ony if $s_{p+1} \in F$. The \emph{language} $\lang{A}$ of $A$ is the
set of paths in $\mathcal{T}$ over which $A$ has an accepting run.

A path automaton $A = (\set{s_1, \ldots, s_\nstates}, I, F, \delta)$
corresponds, in the sense of Lemma \ref{lemma:automata-wsks} below, to
the formula in Fig.~\ref{fig:path-automaton}, which is, moreover,
built directly from the syntactic description of $A$.
\begin{figure}[t!]
\begin{align}
\autoformof{A}(x,y,\vec{Y}) \isdef & 
\bigwedge_{1 \leq i \neq j \leq L} \forall z . \big(\neg Y_i(z) \vee \neg Y_j(z)\big)
\wedge \bigvee_{s_i \in I} Y_i(x) ~\wedge~ \bigvee_{s_j \in F} Y_j(y) \label{eq:path-auto1} \\
& \wedge \bigwedge_{i=1}^L \big(\forall z ~.~ z \neq y \wedge Y_i(z) \rightarrow \nonumber\\[-4mm]
& \hspace*{5mm} \bigvee_{\scriptstyle{j : s_i \arrow{d}{} s_j}} Y_j(z.d)
\vee \bigvee_{\scriptstyle{j : s_i \arrow{d}{} s_j}} \exists z' ~.~ z'.d = z \wedge Y_j(z')\big) \label{eq:path-auto2} \\
& \wedge \bigwedge_{j=1}^L \big(\forall z ~.~ z \neq x \wedge Y_j(z) \rightarrow \nonumber\\[-4mm]
& \hspace*{5mm} \bigvee_{\scriptstyle{i : s_i \arrow{d}{} s_j}} \exists z' ~.~ z'.d = z \wedge Y_i(z')
\vee \bigvee_{\scriptstyle{i : s_i \arrow{d}{} s_j}} Y_i(z.d)\big) \label{eq:path-auto3}
\end{align}
\caption{Definition of the Path Automaton formula $\autoformof{A}(x,y,\vec{Y})$}
\label{fig:path-automaton}
\end{figure}
Here $\vec{Y} = \tuple{Y_1, \ldots, Y_\nstates}$ are second order
variables interpreted as the sets of tree nodes labeled by the
automaton with the states $s_1, \ldots, s_\nstates$,
respectively. Intuitively, the first three conjuncts of the above
formula (line \ref{eq:path-auto1}) encode the facts that $\vec{Y}$ are
disjoint (no tree node is labeled by more than one state during the
run) and that the run starts in an initial state with node $x$ and
ends in a final state with node $y$. The fourth conjunct (line
\ref{eq:path-auto2}) states that, for every non-final node on the
path, if the automaton visits that node by state $s_i$, then the node
has a $d$-child visited by state $s_j$, where $s_i \arrow{d}{} s_j$ is
a transition of the automaton. The fifth conjunct (line
\ref{eq:path-auto3}) is the reversed flow condition on the path,
needed to ensure that the sets $\vec{Y}$ do not contain useless nodes,
being thus symmetric to the fourth. The following result stems from
the classical automata-logic connection (see
\cite[\S2.10]{KhoussainovNerode} for a textbook presentation):

\begin{lemma}\label{lemma:automata-wsks}
  Given nodes $n,m \in \interv{1}{\kappa}^*$ and directions $d_1,
  \ldots, d_p \in \interv{1}{\kappa}$ such that $m = n \cdot d_1 \cdot
  \ldots \cdot d_p$, for each valuation $\nu$, such that $\nu(x) = n$
  and $\nu(y) = m$, we have \(d_1 \ldots d_p \in \lang{A} \iff
  \wsksmodels{\nu} \exists \vec{Y} ~.~
  \autoformof{A}(x,y,\vec{Y})\enspace.\)
\end{lemma}

Our purpose is to infer, directly from the syntax of the rules in
$\asid$, path automata that recognize the path between the node where
a variable is introduced (either as a formal parameter of a rule or by
an existential quantifier) and the node where the variable is
instantiated, in each given rewriting tree. Formally, for each rule
$\arule = \left(\apred(x_1, \ldots, x_{\#\apred}) \leftarrow \exists
y_1 \ldots \exists y_m ~.~ \psi\right)$, such that $\psi$ is a
quantifie-free formula, and each variable $x \in \fv{\psi}$, we
consider the path automaton $A^x_\arule = (S,I,F,\delta)$,
where: \begin{itemize}
\item $S \isdef \set{s_{\arule'}^z \mid \arule' = \left(\apred'(x_1,
  \ldots, x_{\#\apred'}) \leftarrow \exists y_1 \ldots \exists y_m ~.~
  \phi\right),~ \phi \text{ is q.f.},~ z \in \fv{\phi}}$; the
  intuition is that the automaton visits the state $s_{\arule'}^z$
  while tracking variable $z$ in the body of the rule $\arule'$, that
  labels the node of the rewriting tree which is currently visited by
  the automaton,
\item the initial and final states are $I \isdef \set{s_{\arule}^x}$
  and $F \isdef \{s_{\arule'}^z \mid \arule' = \big(\apred'(x_1,
  \ldots, x_{\#\apred'})$ $\leftarrow \exists y_1 \ldots \exists y_m
  ~.~ \acomp^q(z) * \phi\big),~ \phi \text{ is q.f.},~
  \acomp\in\comps, q \in \statesof{\acomp}\}$; the automaton starts to
  track $x$ in $\arule$ and moves down in the rewriting tree, finally
  tracking a variable $z$ that occurs in a component atom in
  $\arule'$,
\item the transitions are \(s_{\arule}^{y_j} \arrow{d}{}
  s_{\arule'}^{x_j}\), for all rules $\arule = \left(\apred(x_1,
  \ldots, x_{\#(\apred)}) \leftarrow \varphi\right)$ and $\arule' =
  \left(\apred'(x_1, \ldots, x_{\#(\apred')}) \leftarrow
  \varphi'\right)$ from $\asid$, all directions $d \in
  \interv{1}{\npred{\varphi}}$, such that $\pred{d}{\varphi} =
  \apred'(y_1,\ldots,y_{\#(\apred')})$, and all $j \in
  \interv{1}{\#(\apred')}$; if $\arule$ labels the parent of the node
  labeled by $\arule'$ in the rewriting tree, the automaton moves down
  one step, from tracking $y_j$ in $\arule$ to tracking $x_j$ in
  $\arule'$.
\end{itemize}

\begin{example}\label{ex:tll-auto}
  \begin{figure}[t!]
    \centerline{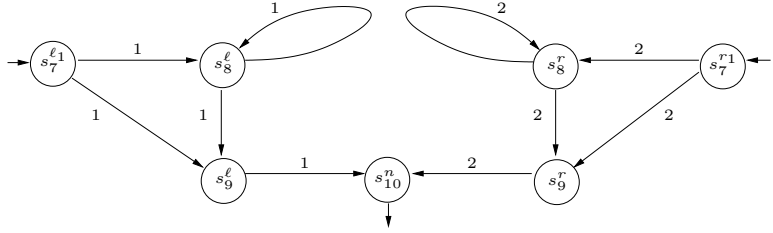}
    \caption{Path Automata Recognizing the Paths to Component Atoms
      from Example \ref{ex:tll}}
    \label{fig:tll-auto}
  \end{figure}

  The path automata recognizing the paths to component atoms, for the
  variables $\ell^\epsilon_1$ and $r^\epsilon_1$ in the rewriting tree
  from Example \ref{ex:tll} are depicted in
  Fig.~\ref{fig:tll-auto}. The initial states are
  $s_{\text{\ref{rule:tll-root}}}^{\ell_1}$ and
  $s_{\text{\ref{rule:tll-root}}}^{r_1}$, respectively, and the final
  state is $s_{\text{\ref{rule:tll-leaf}}}^{n}$ for both automata. The
  rule labels are the ones from Example
  \ref{ex:tll}. \hfill$\blacksquare$
\end{example}

The lemma below shows that these automata recognize the set of paths
corresponding to the recursive substitutions of a given variable down
to the node where it occurs in a component atom, in all rewriting
trees built using the rules from the given SID:

\begin{lemma}\label{lemma:path-automaton}
  Given a rule $\arule = \left(\apred(x_1, \ldots, x_{\#(\apred)})
  \leftarrow \exists y_1 \ldots \exists y_m ~.~ \phi\right)$, such
  that $\phi$ is a quantifier-free formula, and a variable $x \in
  \fv{\phi}$, for each sequence $w \in \interv{1}{\kappa}^*$, the
  following are equivalent: \begin{compactenum}
  \item\label{it1:path-automaton} there exists a rewriting tree
    $\mathcal{T}$ and nodes $n,m \in \nodes{\mathcal{T}}$, such that
    $\mathcal{T}(n) = \arule$, $m = n \cdot w$ and
    $\store_{\mathcal{T}}(x) = m$ (see \ref{eq:store-tree} for the
    definition of $\store_{\mathcal{T}}$).
  \item\label{it2:path-automaton} $w \in \lang{A_{\arule}^x}$.
  \end{compactenum}
\end{lemma}
\proof{ (\ref{it1:path-automaton}) $\Rightarrow$
  (\ref{it2:path-automaton}) By the definition of
  $\store_{\mathcal{T}}$, we have $\store_{\mathcal{T}}(x) =
  \store^\epsilon_{\proj{\mathcal{T}}{n}}(x)$ and
  $\store^\epsilon_{\proj{\mathcal{T}}{n}}(x) = m$ if there
  exist: \begin{compactitem}
  \item a sequence of nodes $n = n_1, n_2, \ldots, n_p = m \in
    \nodes{\mathcal{T}}$, such that $n_{i+1} = n_i \cdot d_i$, for
    some $d_i \in \interv{1}{\kappa}$, for all $i \in
    \interv{1}{p-1}$,
  \item a sequence of variables $x = z_1, \ldots, z_p \in \vars$, such
    that $z_i$ replaces $z_{i+1}$ in the construction of the
    characteristic formula $\charform{\mathcal{T}}$, for all $i \in
    \interv{1}{p-1}$, and occurs in a component atom from
    $\mathcal{T}(m)$.
  \end{compactitem}
  By the definition of rewriting trees, each node $n_i$ is labeled by
  a rule $\arule_i = \left(\apred_i(x_1, \ldots, x_{\#(\apred_i)})
  \leftarrow \exists y_1 \ldots \exists y_m ~.~ \phi_i\right)$, where
  $\phi_i$ is quantifier-free, and there exists $j \in
  \interv{1}{\#(\apred_i)}$ such that $z_i$ occurs on the $j$-th
  position in $\pred{d_i}{\phi_i}$ and $z_{i+1}$ occurs $j$-th within
  the head $\apred_{i+1}(x_1, \ldots, x_{\#(\apred_{i+1})})$ of
  $\arule_{i+1}$. By the definition of $A_{\arule}^x$, there exist
  transitions $s_{\arule_i}^{z_i} \arrow{d_i}{}
  s_{\arule_{i+1}}^{z_{i+1}}$, for all $i \in
  \interv{1}{p-1}$. Moreover, $s_\arule^x = s_{\arule_1}^{z_1}$ and
  $s_{\arule_p}^{z_p}$ are initial state and final states of
  $A_\arule^x$, respectively. Since $m = n \cdot d_1 \ldots d_p = n
  \cdot w$, we obtain that $w = d_1 \ldots d_p \in \lang{A_\arule^x}$.

  \noindent (\ref{it2:path-automaton}) $\Rightarrow$
  (\ref{it1:path-automaton}) Let $w = d_1 \ldots d_p$, where $d_1,
  \ldots, d_p \in \interv{1}{\kappa}$, and $s_{\arule_i}^{z_i}
  \arrow{d_i}{} s_{\arule_{i+1}}^{z_{i+1}}$, for all $i \in
  \interv{1}{p-1}$, be transitions of $A_{\arule}^x$, such that
  $s_{\arule_1}^{z_1} = s_{\arule}^x$ is the initial state and
  $s_{\arule_p}^{z_p}$ is a final state of $A_{\arule}^x$. Then, by
  the definition of $A_{\arule}^x$, the nodes $w_i \isdef n \cdot d_1
  \ldots d_i$ are labeled by rules $\arule_i \isdef
  \left(\apred_i(x_1, \ldots, x_{\#(\apred_i)}) \leftarrow
  \varphi_i\right)$, such that $z_i$ occurs in $\pred{d_i}{\varphi_i}$
  on the same position as $z_{i+1}$, which occurs in $\apred_i(x_1,
  \ldots, x_{\#(\apred_{i+1})})$. Then, there exists a rewriting tree
  $\mathcal{T}$ and nodes $n,m\in\nodes{T}$, such that $m = n \cdot
  d_1 \ldots d_o$ and $A^x_{\arule}$ has an accepting run over $d_1
  \ldots d_p$. By the definition of $\store_{\mathcal{T}}$, we have
  $\store_{\mathcal{T}}(x) =
  \store^\epsilon_{\proj{\mathcal{T}}{n}}(x) = m$. \qed}

The path automata $A_{\arule}^z$ are used to define the
$\mathit{Path}_{\arule}^z(x,y,\vec{R})$ formul{\ae} below:
\[\begin{array}{rcl}
\mathit{Path}_{\arule}^z(x,y,\vec{R}) & \isdef &
\exists \vec{Y} ~.~ \autoformof{A_{\arule}^z}(x,y,\vec{Y})
\wedge \Upsilon(\vec{Y}, \vec{R}) 
\\ 
\Upsilon(\vec{Y},\vec{R}) & \isdef & \bigwedge_{i : \arule_i = \left(\apred'(x_1,
  \ldots, x_{\#(\apred')}) \leftarrow
  \varphi'\right)} \bigwedge_{z \in \fv{\varphi'}} \forall x ~.~
Y_{\arule_i}^z(x) \rightarrow R_i(x)
\end{array}\]
The formula $\Upsilon(\vec{Y},\vec{R})$ above states that all nodes
labeled with a state $q_{\arule_i}^z$ during the run must be also
labeled with $\arule_i$ in the rewriting tree given as input to the
path automaton. Here we denote by $Y_{\arule_i}^z$ the second-order
variable corresponding to the state $q_{\arule_i}^z$ in a path
automaton $A^x_{\arule}$. Intuitively, the formula
$\mathit{Path}_{\arule}^z(x,y,\vec{R})$ states that, in each rewriting
tree defined by a model $\nu$ of the $\rewrtree(\vec{R})$ formula,
there exists a path from $\nu(x)$ to $\nu(y)$, where $\nu(x)$ is a
node labeled with $\arule$, in which $z$ occurs free or existentially
quantified, and $y$ is the node where a variable necessarily equal to
$z$ occurs in a component atom. 

\subsection{Flow Formul{\ae}}
\label{sec:flow-formulae}

At this point, we introduce a \wsks\ formula $\Phi(\vec{X}, \vec{X}',
\vec{R})$ that defines the \emph{structure} of transitions in the
Petri net $\anetof{\signature,\struc_{\rtreeof{\nu}{\vec{R}}},\beh}$,
where $\nu$ is a model of $\rewrtree(\vec{R})$. By ``structure'' here
we mean that $\nu(\vec{X})$ and $\nu(\vec{X}')$ encode the pre- and
post-sets of some transition in
$\anetof{\signature,\struc_{\rtreeof{\nu}{\vec{R}}},\beh}$, as
explained in \S\ref{sec:trees-configurations}.


Fig.~\ref{fig:flow} shows the flow formula of a given SID, consisting
of the rules $\arule_1, \ldots, \arule_\nrules$. We denote by
$\interof{\arule}$ the set of interaction atoms that occur in the
body of a rule $\arule$. Essentially, the formula (\ref{eq:flow}) is
split into a disjunction of formul{\ae}
$\iflow_{\ell,\ainterac(x_1,\ldots,x_{\#(\ainterac)})}$, one for each rule
$\arule_\ell \in \asid$ and each interaction atom $\ainterac(x_1,
\ldots, x_{\#(\ainterac)})$ that occurs in the body of $\arule_\ell$. To understand
the $\iflow_{\ell,\ainterac(x_1,\ldots,x_{\#(\ainterac)})}$ formul{\ae}, recall that
each of the variables $x_i$ is mapped to the (unique) node of the
rewriting tree containing an instance atom $\acomp^q(x)$, such that
$x_i$ replaces $x$ in the characterizing formula of the rewriting
tree $\rtreeof{\nu}{\vec{R}}$, provided that $\nu$ is a model of
$\rewrtree(\vec{R})$. In order to find this node, we track the
variable $x_i$ from the node labeled by the rule $\arule$, to the node
where this instance atom occurs. This is done by the
$\mathit{Path}^{x_i}_{\arule_\ell}(y_0,y_i,\vec{R})$ formula
(\ref{eq:flow-path}), that holds whenever $y_0$ and $y_i$ are respectively
mapped to the source and the destination of a path from a node
$\nu(y_0) \in \nodes{\rtreeof{\nu}{\vec{R}}}$, with label
$\arule_\ell$ to a node $\nu(y_i) \in \nodes{\rtreeof{\nu}{\vec{R}}}$,
such that the component atom that gives the value of $x_i$\footnote{In
  the canonical store $\store_{\rtreeof{\nu}{\vec{R}}}$, see
  \S\ref{sec:canonical-model} for its definition.} occurs in
$\rtreeof{\nu}{\vec{R}}(\nu(y_i))$. Hence, $x_i$ is interpreted as
$\nu(y_i)$ in the canonical model of the characteristic formula of
$\rtreeof{\nu}{\vec{R}}$. 

\begin{figure}[t!]
\begin{center}
\begin{align}
\flow(\vec{X}, \vec{X}', \vec{R}) & \isdef 
\bigvee_{\ell=1}^\nrules \bigvee_{\ainterac(x_1,\ldots,x_{\#(\ainterac)}) \in \interof{\arule_\ell}}
\hspace*{-8mm}\iflow_{\ell,\ainterac(x_1,\ldots,x_{\#(\ainterac)})}(\vec{X}, \vec{X}', \vec{R}) \label{eq:flow}
\\
\iflow_{\ell,\ainterac(x_1,\ldots,x_{\#(\ainterac)})}(\vec{X},\vec{X}',\vec{R}) & \isdef \exists y_0 \ldots \exists y_{\#(\ainterac)} ~.~ R_\ell(y_0) \wedge 
\bigwedge_{i=1}^{\#(\ainterac)}\!\! \mathit{Path}^{x_i}_{\arule_\ell}(y_0,y_i,\vec{R}) \wedge \label{eq:flow-path} \\
& \hspace{-33.5mm} \big( \bigvee_{\tuple{\tau_1,\ldots,\tau_{\#(\ainterac)}} \in \transof{\ainterac}}
\bigwedge_{q \in \globstates} X_q = \set{y_i \mid i\in\interv{1}{\#(\ainterac)}, \tau_i = (q_i\arrow{p_i}{} q_i'), q_i = q} \wedge \label{eq:flow-pre} \\
& \hspace{-1mm} X'_q = \set{y_i \mid i\in\interv{1}{\#(\ainterac)}, \tau_i=(q_i\arrow{p_i}{}q_i'), q_i' = q} \big) \wedge \label{eq:flow-post} \\
& \hspace*{-30mm} \bigwedge_{1 \leq i < j \leq \#(\ainterac)} \big( y_i \neq y_j \vee \compof{\tuple{\portsof{\ainterac}}_i} = \compof{\tuple{\portsof{\ainterac}}_j} \big)
\end{align}
\end{center}
\vspace*{-\baselineskip}
\caption{The Flow Formula}
\label{fig:flow}
\end{figure}

The correctness of the encoding is formalized and proved by the
following:

\begin{lemma}\label{lemma:flow-formula}
  Given a model $\nu$ of $\rewrtree(\vec{R})$, we have
  $\wsksmodels{\nu} \flow(\vec{X},\vec{X}',\vec{R})$ if and only if
  $\markof{\nu}{\vec{X}} = \pre{t}$ and $\markof{\nu}{\vec{X}'} =
  \post{t}$, for some transition $t$ of
  $\anetof{\signature,\struc_{\rtreeof{\nu}{\vec{R}}},\beh}$.
\end{lemma}
\proof{ ``$\Rightarrow$'' If $\wsksmodels{\nu}
  \flow(\vec{X},\vec{X}',\vec{R})$ then there exists a rule
  $\arule_\ell \in \asid = \set{\arule_1, \ldots, \arule_\ncomps}$ and
  $\ainterac(x_1,\ldots,x_{\#(\ainterac)}) \in \interof{\arule_\ell}$,
  such that $\wsksmodels{\nu}
  \iflow_{\ell,\ainterac(x_1,\ldots,x_{\#(\ainterac)})}(\vec{X},
  \vec{X}', \vec{R})$. Then, there exist nodes $w_0, \ldots, w_n \in
  \nodes{\rtreeof{\nu}{\vec{R}}}$, such that $\nu(y_i) = w_i$, for all
  $i \in \interv{0}{n}$ and let $\nu' = \nu[y_0 \leftarrow w_0,
    \ldots, y_n \leftarrow w_n]$. Since, moreover,
  $\wsksmodels{\nu'}{R_\ell(y_0)}$, we obtain that
  $\rtreeof{\nu}{\vec{R}}(w_0) = \arule_\ell$, hence the variables
  $x_1,\ldots,x_{\#(\ainterac)}$ occur in the body of the rule
  $\arule_\ell$ and are assigned to the unique nodes $w'_1, \ldots,
  w'_{\#(\ainterac)} \in \nodes{\rtreeof{\nu}{\vec{R}}}$, for which
  $\pathof{w_0}{w'_i} \in \lang{A^{x_i}_{\arule_\ell}}$, for all $i
  \in \interv{1}{\#(\ainterac)}$, by Lemma
  \ref{lemma:path-automaton}. Since
  $\wsksmodels{\nu'}{\mathit{Path}^{x_i}_{\arule_\ell}(y_0,y_i,\vec{R})}$,
  by Lemma \ref{lemma:automata-wsks}, we obtain $\pathof{w_0}{w_i} \in
  \lang{A^{x_i}_{\arule_\ell}}$, thus $w_i = w'_i$, for all $i \in
  \interv{1}{\#(\ainterac)}$, by the uniqueness of these nodes. Let
  $\tuple{\tau_1,\ldots,\tau_{\#(\ainterac)}}$ be a
  tuple of transitions from $\transof{\ainterac}$ synchronizing by $\ainterac$. Since
  $\wsksmodels{\nu'}{\iflow_{\ell,\ainterac(x_1,\ldots,x_{\#(\ainterac)})}(\vec{X},
    \vec{X}', \vec{R})}$, we have $w_i \neq w_j$, for each $1 \leq i <
  j \leq \#(\ainterac)$, such that
  $\compof{\tuple{\portsof{\ainterac}}_i} =
  \compof{\tuple{\portsof{\ainterac}}_j}$. By
  Def. \ref{def:execution-semantics}, we have that $t =
  (\ainterac[w_1,\ldots,w_{\#(\ainterac)}], \tuple{\tau_1, \ldots,
    \tau_{\#(\ainterac)}})$ is a transition of
  $\anetof{\signature,\struc_{\rtreeof{\nu}{\vec{R}}},\beh}$ and,
  moreover, $\pre{t} = \set{q_1, \ldots, q_{\#(\ainterac)}} =
  \markof{\nu}{\vec{X}}$ and $\post{t} = \set{q'_1, \ldots,
    q'_{\#(\ainterac)}} = \markof{\nu}{\vec{X}'}$, because
  $\wsksmodels{\nu'}{X_q = \set{y_i \mid
      i\in\interv{1}{\#(\ainterac)}, \tau_i = (q_i\arrow{p_i}{} q_i'), q_i = q}}$ and
  $\wsksmodels{\nu'}{X'_q = \set{y_i \mid
      i\in\interv{1}{\#(\ainterac)}, \tau_i = (q_i\arrow{p_i}{} q_i'), q'_i = q}}$, respectively.

  \vspace*{\baselineskip}
  \noindent ``$\Leftarrow$'' Let $t =
  (\ainterac[w_1,\ldots,w_{\#(\ainterac)}], \tuple{\tau_1, \ldots,
    \tau_{\#(\ainterac)}})$ be a transition of
  $\anetof{\signature,\struc_{\rtreeof{\nu}{\vec{R}}},\beh}$, where
  $\tau_i = q_i \arrow{\tuple{\portsof{\ainterac}}_i}{} q'_i$ is a
  transition of $\behof{\compof{\tuple{\portsof{\ainterac}}_i}}$, for
  each $i \in \interv{1}{\#(\ainterac)}$. Then there exists a node
  $w_0 \in \nodes{\rtreeof{\nu}{\vec{R}}}$ and an interaction atom
  $\ainterac(x_1, \ldots, x_{\#(\ainterac)})$ in
  $\rtreeof{\nu}{\vec{R}}(w_0)$, such that
  $\store_{\rtreeof{\nu}{\vec{R}}}(x_i) = w_i$, for all $i \in
  \interv{1}{\#(\ainterac)}$. Assume w.l.o.g. that
  $\rtreeof{\nu}{\vec{R}}(w_0) = \arule_\ell$, for some $\ell \in
  \interv{1}{\nrules}$ and let $\nu' = \nu[y_0 \leftarrow w_0, \ldots,
    y_{\#(\ainterac)} \leftarrow w_{\#(\ainterac)}]$ be a
  valuation. We prove the following points: \begin{compactitem}
  \item $\wsksmodels{\nu'}{R_\ell(y_0)}$: since
    $\wsksmodels{\nu}{\rewrtree(\vec{R})}$,
    $\rtreeof{\nu}{\vec{R}}(w_0) = \arule_\ell$, $\nu'(y_0) = w_0$ and
    $\nu'$ agrees with $\nu$ over $\vec{R}$.
  \item
    $\wsksmodels{\nu'}{\mathit{Path}_{\arule_\ell}^{x_i}(y_0,y_i,\vec{R})}$,
    for all $i \in \interv{1}{\#(\ainterac)}$: since
    $\store_{\rtreeof{\nu}{\vec{R}}}(x_i) = w_i$, we obtain
    $\pathof{w_0,w_i} \in \lang{A^{x_i}_{\arule_\ell}}$ by Lemma
    \ref{lemma:path-automaton} and
    $\wsksmodels{\nu'}{\mathit{Path}_{\arule_\ell}^{x_i}(y_0,y_i,\vec{R})}$
    follows, by Lemma \ref{lemma:automata-wsks}.
  \item $\wsksmodels{\nu'}{y_i \neq y_j}$, for all $1 \leq i < j \leq
    \#(\ainterac)$, such that $\compof{\tuple{\portsof{\ainterac}}_i}
    = \compof{\tuple{\portsof{\ainterac}}_j}$: by
    Def. \ref{def:execution-semantics}, $w_i \neq w_j$ for all
    $i,j\in\interv{1}{\#(\ainterac)}$, such that
    $\compof{\tuple{\portsof{\ainterac}}_i} =
    \compof{\tuple{\portsof{\ainterac}}_j}$.
  \item $\wsksmodels{\nu'}{X_q = \set{y_i \mid
      i\in\interv{1}{\#(\ainterac)}, \tau_i = (q_i\arrow{p_i}{} q_i'), q_i = q}}$, for all $q \in
    \globstates$: because $\markof{\nu}{\vec{X}} = \pre{t} = \set{q_1,
      \ldots, q_{\#(\ainterac)}}$.
  \item $\wsksmodels{\nu'}{X_q = \set{y_i \mid
      i\in\interv{1}{\#(\ainterac)}, \tau_i = (q_i\arrow{p_i}{} q_i'), q'_i = q}}$, for all $q \in
    \globstates$: because $\markof{\nu}{\vec{X}'} = \post{t} =
    \set{q'_1, \ldots, q'_{\#(\ainterac)}}$
  \end{compactitem}
  We obtain that $\wsksmodels{\nu}{\iflow_{\ell,\ainterac(x_1, \ldots,
      x_{\#(\ainterac)})}}$ and, consequently
  $\wsksmodels{\nu}{\flow(\vec{X},\vec{X}',\vec{R})}$. \qed}

\subsection{Structural Invariants}
\label{sec:structural-invariants}

An \emph{invariant} $\mathcal{I}$ of a marked Petri net $\amarkednet =
(\anet,\amark_0)$, where $\anet=(\places,\trans,\edges)$ is a Petri
net, is a set of markings that over-approximates the reachable states
$\reach{\amarkednet}$, that is $\reach{\amarkednet} \subseteq
\mathcal{I} \subseteq \pow{\places}$. The invariant is, moreover, said
to be \emph{inductive} if and only if it is closed under the
transition relation of the Petri net, namely that $\set{\amark' \mid
  \amark \in \mathcal{I},~ t \in \trans,~ \amark \arrow{t}{} \amark'}
\subseteq \mathcal{I}$.

The synthesis of inductive invariants is a notoriously difficult
problem, that has been received much attention in the past. The most
common method of infering inductive invariants is the iteration of an
abstract (over-approximation) transition relation in an abstract
domain, until a fixpoint is reached. Depending on the complexity of
the abstract domain, this approach, known as \emph{abstract
interpretation} \cite{CousotCousot79}, can be quite costly. In
contrast, we consider inductive invariants that can be synthesized
directly from the structure of a Petri net, without iterating (an
abstraction of) its transition relation up to a fixpoint. Such
invariants are called \emph{structural} in the literature
\cite{DBLP:conf/mfcs/Sifakis78}. The next definition introduces two
types of structural invariants:

\begin{definition}\label{def:structural-invariants}
Given a marked Petri net $\amarkednet = (\anet, \amark_0)$, where
$\anet = (\places,\trans,\edges)$, a set of places $\sigma \subseteq
\places$ is said to be a: \begin{itemize}
\item \emph{trap} if $\card{\sigma \cap \amark_0} \geq 1$ and, for
  any $t \in \trans$, if $\card{\sigma \cap \pre{t}} \geq 1$ then
  $\card{\sigma \cap \post{t}} \geq 1$,
\item \emph{mutex} if $\card{\sigma \cap \amark_0} = 1$ and, for any
  $t \in \trans$, we have $\card{\sigma \cap \pre{t}} = \card{\sigma
  \cap \post{t}} \leq 1$.
\end{itemize}
The trap and mutex invariant of $\amarkednet$ are the below sets of
markings, respectively: \begin{itemize}
\item
  $\trapinv(\amarkednet) \isdef \set{\amark \text{ marking of }
  \amarkednet \mid \card{\amark\cap\sigma} \geq 1 \text{, for each
    trap } \sigma \text{ of } \amarkednet}$, 
\item
  $\mutexinv(\amarkednet) \isdef \set{\amark \text{ marking of }
  \amarkednet \mid \card{\amark\cap\sigma} = 1 \text{, for each mutex
  } \sigma \text{ of } \amarkednet}$.
\end{itemize}
\end{definition}

By substituting the reachable set with the intersection of the trap
and mutex invariant, we obtain the following sufficient queries, that
allow to prove the absence of deadlocks and unreachability of a tuple
of states in a system described by a \adl\ formula:

\begin{lemma}\label{lemma:invariant-safety}
  Given a tight sentence $\phi$, interpreted over a SID $\asid$, a
  behavior map $\beh$ and a tuple of states $\tuple{q_1, \ldots,
    q_n}$, the following hold:
  \[\begin{array}{rcl}
  \mathit{deadlock}(\phi,\asid,\beh) & \Rightarrow &
  \mathit{deadlock}^\sharp_t(\phi,\asid,\beh) \\
  \mathit{reach}(\phi,\tuple{q_1,\ldots,q_n},\asid,\beh)
  & \Rightarrow & \mathit{reach}^\sharp_t(\phi,\tuple{q_1,\ldots,q_n},\asid,\beh)
  \end{array}\]
    where $\mathit{deadlock}^\sharp_t(\phi,\asid,\beh)$ and
  $\mathit{reach}^\sharp_t(\phi,\tuple{q_1,\ldots,q_n},\asid,\beh)$
  are defined below: \begin{itemize}
  \item $\mathit{deadlock}^\sharp_t(\phi,\asid,\beh)$:
    does there exist a rewriting tree $\mathcal{T} \in \rtrees{\phi}$,
    such that
    $\trapinv(\anetof{\signature,\struc_{\mathcal{T}},\beh},\amark_{\mathcal{T}})
    \cap\mutexinv(\anetof{\signature,\struc_{\mathcal{T}},\beh},\amark_{\mathcal{T}})
    \cap \dead{\anetof{\signature,\struc_{\mathcal{T}},\beh}} \neq
    \emptyset$?
  \item
    $\mathit{reach}^\sharp_t(\phi,\tuple{q_1,\ldots,q_n},\asid,\beh)$:
    does there exist a rewriting tree $\mathcal{T} \in \rtrees{\phi}$ and a
    configuration $(\struc_{\mathcal{T}}, \amark) \in
    \trapinv(\anetof{\signature,\struc_{\mathcal{T}},\beh},\amark_{\mathcal{T}})
    \cap
    \mutexinv(\anetof{\signature,\struc_{\mathcal{T}},\beh},\amark_{\mathcal{T}})$,
    such that $\set{q_i[u_i]~|~i\in\interv{1}{n}} \subseteq \amark$,
    with $q_1[u_1], \ldots, q_n[u_n]$ pairwise distinct?
  \end{itemize}
\end{lemma}
\proof{ We show first that the trap $\trapinv(\amarkednet)$ and mutex
  $\mutexinv(\amarkednet)$ invariants of a marked Petri net
  $\amarkednet$ are invariants of $\amarkednet$. If $\amark_0 =
  \emptyset$ then $\amarkednet$ has no traps nor mutexes and
  $\reach{\amarkednet} = \trapinv(\amarkednet) =
  \mutexinv(\amarkednet) = \emptyset$, thus $\trapinv(\amarkednet)$
  and $\mutexinv(\amarkednet)$ are trivially inductive invariants. So
  we assume $\amark_0 \neq \emptyset$.

  For the set $\trapinv(\amarkednet)$, let $\sigma$ be an arbitrary
  trap of $\amarkednet$. Since $\amark_0 \neq \emptyset$, we have
  $\card{\sigma \cap \amark_0} \geq 1$, hence $\amark_0 \in
  \trapinv(\amarkednet)$, by the choice of $\sigma$. For each pair of
  markings $\amark$ and $\amark'$, such that $\amark \in
  \trapinv(\amarkednet)$ and $\amark \arrow{t}{} \amark'$, for some
  transition $t \in \trans$, we have $\amark' = (\amark \setminus
  \pre{t}) ~\cup ~\post{t}$. If $\sigma \cap \pre{t} = \emptyset$ then
  $\card{\amark' \cap \sigma} \geq \card{\amark \cap \sigma}$ and
  $\card{\amark \cap \sigma} \geq 1$, because $\amark \in
  \trapinv(\amarkednet)$. Otherwise, $\card{\sigma \cap \pre{t}} \geq
  1$, thus $\card{\sigma \cap \post{t}} \geq 1$ because $\sigma$ is a
  trap of $\amarkednet$, hence $\card{\amark' \cap \sigma} \geq 1$. In
  both cases, we obtain that $\card{\amark' \cap \sigma} \geq 1$,
  hence $\amark' \in \trapinv(\amarkednet)$, because $\sigma$ is an
  arbitrary trap of $\amarkednet$.

  A similar argument shows that $\mutexinv(\amarkednet)$ is an
  inductive invariant of $\amarkednet$. Let $\sigma$ be an arbitrary
  mutex of $\amarkednet$. Then $\card{\sigma \cap \amark_0} = 1$,
  hence $\amark_0 \in \mutexinv(\amarkednet)$, by the choice of
  $\sigma$. For each pair of markings $\amark$ and $\amark'$, such
  that $\amark \in \trapinv(\amarkednet)$ and $\amark \arrow{t}{}
  \amark'$, for some transition $t \in \trans$, we have $\amark' =
  (\amark \setminus \pre{t}) ~\cup ~\post{t}$ and $\card{\sigma \cap
    \pre{t}} = \card{\sigma \cap \post{t}} \leq 1$, hence
  $\card{\amark' \cap \sigma} = 1$ and $\amark' \in
  \mutexinv(\amarkednet)$, because $\sigma$ is an arbitrary mutex of
  $\amarkednet$. Moreover, since inductive invariants are closed under
  intersection, the set $\trapinv(\amarkednet) \cap
  \mutexinv(\amarkednet)$ is also an inductive invariant of
  $\amarkednet$. Finally, because $\reach{\amarkednet}$ is known to be
  the least inductive invariant of $\amarkednet$, we obtain
  $\reach{\amarkednet} \subseteq \trapinv(\amarkednet) \cap
  \mutexinv(\amarkednet)$. The rest of the proof follows from Theorem
  \ref{thm:symmetry} and the above fact. \qed}

\begin{figure}[t!]
\begin{center}
  \begin{align}
    \initials(\vec{Y},\vec{R}) & \isdef \ismark(\vec{Y},\vec{R}) \wedge \forall x ~. \bigwedge_{q \in \globstates} \Big(Y_q(x) \leftrightarrow
    \hspace*{-8mm} \bigvee_{\begin{array}{c}
        \scriptstyle{\ell \in \interv{1}{\nrules} \text{ such that}} \\[-1mm]
        \scriptstyle{\acomp^q(z) \text{ occurs in } \arule_\ell} \end{array}} \hspace*{-8mm} R_\ell(x)\Big) \label{eq:init}
    \\[4mm]
    \trap(\vec{X},\vec{R}) & \isdef \forall \vec{Y} \forall \vec{Z} ~.~
    \flow(\vec{Y},\vec{Z},\vec{R}) \rightarrow (\intersects{\vec{X}}{\vec{Y}} \rightarrow \intersects{\vec{X}}{\vec{Z}}) \label{eq:trap}
    \\ 
    \trapinv(\vec{X},\vec{Y},\vec{R}) & \isdef \ismark(\vec{X},\vec{R}) \wedge
    \forall \vec{Z} ~.~ (\intersects{\vec{Y}}{\vec{Z}} \wedge \trap(\vec{Z},\vec{R})) \rightarrow \intersects{\vec{X}}{\vec{Z}} \label{eq:trapinv}
    \\[4mm]
    \mutex(\vec{X},\vec{R}) & \isdef 
    \forall \vec{Y} \forall \vec{Z} ~.~ \flow(\vec{Y},\vec{Z},\vec{R}) \rightarrow \big(
    (\notintersects{\vec{X}}{\vec{Y}} \leftrightarrow \notintersects{\vec{X}}{\vec{Z}}) ~\wedge\nonumber \\
    & \hspace*{42mm} (\singleintersects{\vec{X}}{\vec{Y}} \leftrightarrow \singleintersects{\vec{X}}{\vec{Z}})\big) \label{eq:mutex}
    \\
    \mutexinv(\vec{X},\vec{Y},\vec{R}) & \isdef \ismark(\vec{X},\vec{R}) \wedge
    \forall \vec{Z} ~.~ (\singleintersects{\vec{Y}}{\vec{Z}} \wedge \mutex(\vec{Y},\vec{R}))
     \rightarrow \singleintersects{\vec{X}}{\vec{Z}} \label{eq:mutexinv}
  \end{align}
\end{center}
\vspace*{-\baselineskip}
\caption{Initial Configurations and Structural Invariants}
\label{fig:inv}
\end{figure}

Given a \adl\ sentence $\phi$ that describes the set of initial
configurations of a system, Fig.~\ref{fig:inv} introduces
\wsks\ formul{\ae} that define the trap and mutex invariants of the
marked Petri nets $(\anetof{\signature,\struc_{\mathcal{T}},\beh},
\amark_0)$, where $\mathcal{T} \in \rtrees{\phi}$ are rewriting trees
and $(\struc_{\mathcal{T}},\amark_0) \models \phi$ are initial
configurations. This construction uses the flow formula
(Fig.~\ref{fig:flow}) derived from the SID that interprets the
predicate symbols from $\phi$ and the behavior map $\beh$ of the
system. Here we denote by $\vec{X}$ (resp. $\vec{Y}$ and $\vec{Z}$)
the tuple of second-order variables $\tuple{X_q \mid q\in\globstates}$
(resp. $\tuple{Y_q \mid q\in\globstates}$ and $\tuple{Z_q \mid
  q\in\globstates}$) and we write $\vec{X} \cap \vec{Y}$
(resp. $\vec{X} \cap \vec{Z}$) for $\bigcup_{q \in \globstates} X_q
\cap Y_q$ (resp. $\bigcup_{q \in \globstates} X_q \cap Z_q$), where
the definitions of set union and intersection are standard in
\mso\ and omitted to avoid clutter.

Let $\nu$ be a model of $\rewrtree(\vec{R})$, meaning that
$\rtreeof{\nu}{\vec{R}}$ is a rewriting tree for $\phi$. We recall
that, by letting $\arule_1 \isdef \left(\apred_\phi() \leftarrow
\phi\right)$ be the first rule in the SID, we capture the fact that
$\rtreeof{\nu}{\vec{R}} \in \rtrees{\phi}$, for each model $\nu$ of
$\rewrtree(\vec{R})$ (see the second conjunct in line \ref{eq:rtree1}
from Fig.~\ref{fig:rtree}). The formula $\initials(\vec{X},\vec{R})$
(\ref{eq:init}) encodes the fact that $\markof{\nu}{\vec{X}}$ is an
initial marking of the Petri net
$\anetof{\signature,\struc_{\rtreeof{\nu}{\vec{R}}},\beh}$ namely,
that each node of the rewriting tree $\rtreeof{\nu}{\vec{R}}$ is the
index of a component of type $\acomp$ in state $q$, if $\acomp^q(z)$
occurs in the rule that labels the node.

We assume further that $\nu$ is a model of $\rewrtree(\vec{R}) \wedge
\initials(\vec{Y},\vec{R})$. The formul{\ae} $\trap(\vec{X},\vec{R})$
(\ref{eq:trap}) and $\trapinv(\vec{X},\vec{Y},\vec{R})$
(\ref{eq:trapinv}) define the traps and the trap invariant of the
marked Petri net
$(\anetof{\signature,\struc_{\rtreeof{\nu}{\vec{R}}},\beh},\markof{\nu}{\vec{Y}})$,
respectively. Similarly, $\mutex(\vec{X},\vec{R})$ (\ref{eq:mutex})
and $\mutexinv(\vec{X},\vec{R})$ (\ref{eq:mutexinv}) define the
mutexes and the mutex invariant of
$(\anetof{\signature,\struc_{\rtreeof{\nu}{\vec{R}}},\beh},\markof{\nu}{\vec{Y}})$,
respectively. We intentionally use the same symbols to denote trap
(mutex) invariants and their defining \wsks\ formul{\ae}, the
distinction between sets of markings and formul{\ae} being clear from
the context. The following lemma shows the correctness of the
definitions from Fig.~\ref{fig:inv}:

\begin{lemma}\label{lemma:structural-invariants}
  If $\wsksmodels{\nu}{\rewrtree(\vec{R}) \wedge
    \initials(\vec{Y},\vec{R})}$ then the following
  hold: \begin{enumerate}
  \item\label{it:trap}
    $\trapinv(\anetof{\signature,\struc_{\rtreeof{\nu}{\vec{R}}},\beh},
    \markof{\nu}{\vec{Y}}) = \set{\markof{\nu'}{\vec{X}} \mid\ 
      \wsksmodels{\nu'}{\trapinv(\vec{X},\vec{Y},\vec{R}),~ 
        \nu'(\vec{Y},\vec{R}) = \nu(\vec{Y},\vec{R})}}$,
  \item\label{it:mutex} $\mutexinv(\anetof{\signature,\struc_{\rtreeof{\nu}{\vec{R}}},\beh},
    \markof{\nu}{\vec{Y}}) = \set{\markof{\nu'}{\vec{X}} \mid\ 
      \wsksmodels{\nu'}{\mutexinv(\vec{X},\vec{Y},\vec{R}),~ 
        \nu'(\vec{Y},\vec{R}) = \nu(\vec{Y},\vec{R})}}$. 
  \end{enumerate}
\end{lemma}
\proof{ We give the proof for point (\ref{it:trap}), the proof for
  (\ref{it:mutex}) uses a similar argument and is left for the
  reader.

  \vspace*{\baselineskip}\noindent``$\subseteq$'' Let $\amark \in
  \trapinv(\anetof{\signature,\struc_{\rtreeof{\nu}{\vec{R}}},\beh},
  \markof{\nu}{\vec{Y}})$ be a marking and $\nu'$ be a valuation such
  that $\markof{\nu'}{\vec{X}} = \amark$ and $\nu'$ agrees with $\nu$
  over $\vec{Y}$ and $\vec{R}$. Clearly, such a valuation exists and,
  in order to show that $\amark \in \set{\markof{\nu'}{\vec{X}}
    \mid\ \wsksmodels{\nu'}{\trapinv(\vec{X},\vec{Y},\vec{R}),~
      \nu'(\vec{Y},\vec{R}) = \nu(\vec{Y},\vec{R})}}$, it suffices to
  prove
  $\wsksmodels{\nu'}{\trapinv(\vec{X},\vec{Y},\vec{R}})$. Because
  $\markof{\nu'}{\vec{X}} = \amark$, we have
  $\wsksmodels{\nu'}{\ismark(\vec{X},\vec{R})}$ and we are left with
  proving $\wsksmodels{\nu'}{\forall \vec{Z} ~.~
    \trap(\vec{Z},\vec{R}) \wedge \intersects{\vec{Y}}{\vec{Z}}
    \rightarrow \intersects{\vec{X}}{\vec{Z}}}$. To this end, let
  $\mu$ be any valuation that agrees with $\nu'$ over $\vec{X}$,
  $\vec{Y}$ and $\vec{R}$, such that
  $\wsksmodels{\mu}{\trap(\vec{Z},\vec{R}) \wedge
    \intersects{\vec{Y}}{\vec{Z}}}$. It suffices to prove that
  $\markof{\mu}{\vec{Z}}$ is a trap of the marked Petri net
  $(\anetof{\signature,\struc_{\rtreeof{\nu}{\vec{R}}},\beh},
  \markof{\nu}{\vec{Y}})$ to obtain
  $\wsksmodels{\mu}{\intersects{\vec{X}}{\vec{Z}}}$, using the fact
  that $\amark \in
  \trapinv(\anetof{\signature,\struc_{\rtreeof{\nu}{\vec{R}}},\beh},
  \markof{\nu}{\vec{Y}})$ and $\markof{\mu}{\vec{X}} = \amark$. Note
  that, since $\mu$ agrees with $\nu'$ over $\vec{X}$, we also have
  $\markof{\mu}{\vec{X}} = \markof{\nu'}{\vec{X}} = \amark$. Since
  $\wsksmodels{\mu}{\intersects{\vec{Y}}{\vec{Z}}}$ and $\mu$ agrees
  with $\nu$ over $\vec{Y}$, we have $\markof{\mu}{\vec{Z}} \cap
  \markof{\nu}{\vec{Y}} \neq \emptyset$. Let $t$ be a transition of
  $\anetof{\signature,\struc_{\rtreeof{\nu}{\vec{R}}},\beh}$, such
  that $\markof{\mu}{\vec{Z}} \cap \pre{t} \neq \emptyset$ and let
  $\mu'$ be a valuation that agrees with $\mu$ over $\vec{Z}$, such
  that $\markof{\mu'}{\vec{Y}'} = \pre{t}$ and $\markof{\mu}{\vec{Z}'}
  = \post{t}$, where $\vec{Y}'$ and $\vec{Z}'$ are distinct copies of
  $\vec{Y}$ and $\vec{Z}$, respectively. By Lemma
  \ref{lemma:flow-formula}, we obtain
  $\wsksmodels{\mu'}{\flow(\vec{Y}',\vec{Z}',\vec{R})}$. Since
  $\wsksmodels{\mu}{\trap(\vec{Z},\vec{R})}$ and $\mu'$ agrees with
  $\mu$ over $\vec{Z}$, we obtain
  $\wsksmodels{\mu'}{\intersects{\vec{Z}}{\vec{Z}'}}$, thus
  $\markof{\mu}{\vec{Z}} \cap \markof{\mu'}{\vec{Z}'} =
  \markof{\mu}{\vec{Z}} \cap \post{t} \neq \emptyset$. We have showed
  that $\markof{\mu}{\vec{Z}}$ is a trap of the marked Petri net
  $(\anetof{\signature,\struc_{\rtreeof{\nu}{\vec{R}}},\beh},
  \markof{\nu}{\vec{Y}})$, which concludes this direction of the
  proof.

  \vspace*{\baselineskip}\noindent``$\supseteq$'' Let $\nu'$ be a
  valuation that agrees with $\nu$ over $\vec{Y}$ and $\vec{R}$, such
  that $\wsksmodels{\nu'}{\trapinv(\vec{X},\vec{Y},\vec{R})}$. To show
  that $\markof{\nu'}{\vec{X}} \in
  \trapinv(\anetof{\signature,\struc_{\rtreeof{\nu}{\vec{R}}},\beh},
  \markof{\nu}{\vec{Y}})$, we must prove that $\markof{\nu'}{\vec{X}}
  \cap \theta \neq \emptyset$, for each trap $\theta$ of the marked
  Petri net
  $(\anetof{\signature,\struc_{\rtreeof{\nu}{\vec{R}}},\beh},
  \markof{\nu}{\vec{Y}})$. Let $\theta$ be such a trap and let $\mu$
  be a valuation that agrees with $\nu'$ over $\vec{X}$, $\vec{Y}$ and
  $\vec{R}$, such that $\markof{\mu}{Z} = \theta$. Since
  $\markof{mu}{Z}$ is a trap of the marked Petri net
  $(\anetof{\signature,\struc_{\rtreeof{\nu}{\vec{R}}},\beh},
  \markof{\nu}{\vec{Y}})$, we obtain: \begin{compactitem}
  \item $\wsksmodels{\mu}{\intersects{\vec{Y}}{\vec{Z}}}$, since
    $\wsksmodels{\nu}{\initials(\vec{Y},\vec{R})}$ and $\mu$ agrees
    with $\nu$ over $\vec{Y}$, and
  \item $\wsksmodels{\mu}{\trap(\vec{X},\vec{R})}$, by Lemma
    \ref{lemma:flow-formula}, since $\mu$ agrees with $\nu$ over
    $\vec{X}$ and $\vec{R}$.
  \end{compactitem}
  Since $\wsksmodels{\mu}{\trapinv(\vec{X},\vec{Y},\vec{R})}$, we
  obtain that $\wsksmodels{\mu}{\intersects{\vec{X}}{\vec{Z}}}$, thus
  $\markof{\mu}{\vec{X}} \cap \markof{\mu}{\vec{Z}} =
  \markof{\nu'}{\vec{X}} \cap \theta \neq \emptyset$, which concludes
  this direction of the proof. \qed}

\subsection{Verification Conditions}
\label{sec:verification-conditions}

The final step in generating sufficient verification conditions for
the deadlock and reachability problems (\S\ref{sec:undecidability}) is
the encoding of the sets of error configurations in \wsks. This is
done separately, for the two kinds of errors considered, by the
formul{\ae} in Fig.~\ref{fig:error}. The convention here is to use
$\vec{X}$ to represent the configurations in the error sets. However,
none of the formul{\ae} from Fig.~\ref{fig:error} constrains $\vec{X}$
to be a marking, i.e.\ by conjunction with $\ismark(\vec{X})$, since
the latter formula is already included in the definition of the
structural invariants, by the formul{\ae}
$\trapinv(\vec{X},\vec{Y},\vec{R})$ and
$\mutexinv(\vec{X},\vec{Y},\vec{R})$ (Fig.~\ref{fig:inv}).

We recall that a marking $\amark$ is a deadlock of a Petri net $\anet
= (\places,\trans,\edges)$ if and only if $\pre{t} \not\subseteq
\amark$, for all transitions $t \in \trans$ (if $\pre{t} \subseteq
\amark$ for some transition $t$, then that transition could be fired
from $\amark$). This condition is captured by the
$\deadlock(\vec{X},\vec{R})$ formula (\ref{eq:deadlock}), where
$\flow(\vec{Y},\vec{Z},\vec{R})$ is the flow formula that defines the
pre- and post-sets, for some transition in the Petri net
$\anetof{\signature,\struc_{\rtreeof{\nu}{\vec{R}}},\beh}$, where
$\nu$ is a model of $\rewrtree(\vec{R})$ (Lemma
\ref{lemma:flow-formula}). Here we write $\vec{Y} \subseteq \vec{X}$
as a shorthand for $\forall x ~.~ \bigwedge_{q \in \globstates} Y_q(x)
\rightarrow X_q(x)$, denoting the fact that the set of places defined
by $\vec{Y}$ is included in the one defined by $\vec{X}$.

The set defined by the $\errset{\tuple{q_1, \ldots, q_n}}(\vec{X})$
formula (\ref{eq:errset}) captures the fact that each configuration
encoded by $\vec{X}$ contains pairwise distinct places $q_1[w_1],
\ldots$, $q_n[w_n]$ which is, essentially, the condition required by a
reachability query. Note that, because the tuple of states
$\tuple{q_1, \ldots, q_n}$ is part of the input of the query, the
number $\card{\set{i \in \interv{1}{k} \mid q_i = q}}$ of occurrences
of each state $q$ in the tuple is constant, hence the cardinality of
each set $X_q$ is compared to a constant (\ref{eq:cardgeq}).

\begin{figure}[t!]
\begin{center}
  \begin{align}
    \deadlock(\vec{X},\vec{R}) & \isdef \forall \vec{Y} \forall \vec{Z} ~.~ \flow(\vec{Y},\vec{Z},\vec{R}) \rightarrow \neg \included{\vec{Y}}{\vec{X}} \label{eq:deadlock} \\
    \errset{\tuple{q_1, \ldots, q_n}}(\vec{X}) & \isdef \bigwedge_{q \in \globstates} \card{X_{q}} \geq \card{\set{i \in \interv{1}{k} \mid q_i = q}} \label{eq:errset}
  \end{align}
\end{center}
\vspace*{-\baselineskip}
\caption{Error Configurations for the Deadlock and Reachability Problems}
\label{fig:error}
\end{figure}

The following theorem states the soundness of the verification
conditions built throughout this section:

\begin{theorem}\label{thm:verification}
  Given a tight sentence $\phi$, interpreted over a SID $\asid$, a
  behavior map $\beh$ and a tuple of states $\tuple{q_1, \ldots,
    q_n}$, the following hold: \begin{enumerate}
  \item\label{it:deadlock:verification} $\deadlock(\vec{X},\vec{R})
    \wedge \trapinv(\vec{X},\vec{Y},\vec{R}) \wedge
    \mutexinv(\vec{X},\vec{Y},\vec{R}) \wedge
    \initials(\vec{Y},\vec{R}) \wedge \rewrtree(\vec{R})$ is
    unsatisfiable only if $\mathit{deadlock}(\phi,\asid,\beh)$ has a
    negative answer.
  \item\label{it:reachability:verification} $\errset{\tuple{q_1,
      \ldots, q_n}}(\vec{X}) \wedge \trapinv(\vec{X},\vec{Y},\vec{R})
    \wedge \mutexinv(\vec{X},\vec{Y},\vec{R}) \wedge
    \initials(\vec{Y},\vec{R}) \wedge \rewrtree(\vec{R})$ is
    unsatisfiable only if
    $\mathit{reach}(\phi,\tuple{q_1,\ldots,q_n},\asid,\beh)$ has a
    negative answer.
  \end{enumerate}
\end{theorem}
\proof{ (\ref{it:deadlock:verification}) To prove the contra-positive
  statement, assume that $\mathit{deadlock}(\phi,\asid,\beh)$ has a
  positive answer. By Lemma \ref{lemma:invariant-safety}, the query
  $\mathit{deadlock}^\sharp_t(\phi,\asid,\beh)$ has a
  positive answer as well, thus there exists a rewriting tree
  $\mathcal{T} \in \rtrees{\phi}$ and a marking $\amark \in
  \trapinv(\anetof{\signature,\struc_{\mathcal{T}},\beh},\amark_{\mathcal{T}})
  \cap\mutexinv(\anetof{\signature,\struc_{\mathcal{T}},\beh},\amark_{\mathcal{T}})
  \cap \dead{\anetof{\signature,\struc_{\mathcal{T}},\beh}}$, where
  $(\struc_{\mathcal{T}},\amark_{\mathcal{T}})$ is the canonical model
  of $\phi$ corresponding to $\mathcal{T}$. Let $\nu$ be a valuation
  such that $\rtreeof{\nu}{\vec{R}} = \mathcal{T}$,
  $\markof{\nu}{\vec{Y}} = \amark_{\mathcal{T}}$ and
  $\markof{\nu}{\vec{X}} = \amark$. Since the tuples of second-order
  variables $\vec{X}$, $\vec{Y}$ and $\vec{T}$ are pairwise disjoint,
  such a valuation exists. We prove the following
  points: \begin{compactitem}
  \item $\wsksmodels{\nu}{\initials(\vec{Y},\vec{R})}$ follows
    directly from the definitions of $\amark_{\mathcal{T}}$
    (Def. \ref{def:canonical-model}) and $\initials(\vec{Y},\vec{R})$
    (\ref{eq:init}).
  \item $\wsksmodels{\nu}{\trapinv(\vec{X},\vec{Y},\vec{R})}$ follows
    from point \ref{it:trap} of Lemma
    \ref{lemma:structural-invariants}, because $\markof{\nu}{\vec{X}}
    = \amark$, $\rtreeof{\nu}{\vec{R}} = \mathcal{T}$ and $\amark \in
    \trapinv(\anetof{\signature,\struc_{\mathcal{T}},\beh},\amark_{\mathcal{T}})$.
  \item $\wsksmodels{\nu}{\mutexinv(\vec{X},\vec{Y},\vec{R})}$ follows
    from point \ref{it:mutex} of Lemma
    \ref{lemma:structural-invariants}, because $\markof{\nu}{\vec{X}}
    = \amark$, $\rtreeof{\nu}{\vec{R}} = \mathcal{T}$ and $\amark \in
    \mutexinv(\anetof{\signature,\struc_{\mathcal{T}},\beh},\amark_{\mathcal{T}})$.
  \item $\wsksmodels{\nu}{\deadlock(\vec{X},\vec{R})}$ because
    $\dead{\anetof{\signature,\struc_{\mathcal{T}},\beh}}$ is the set
    of markings $\markof{\nu'}{\vec{X}}$ of
    $\anetof{\signature,\struc_{\mathcal{T}},\beh}$, such that
    $\wsksmodels{\nu'}{\deadlock(\vec{X},\vec{R}) \wedge
      \rewrtree(\vec{R})}$. This follows directly from the definition
    of a deadlock marking and the definition of
    $\deadlock(\vec{X},\vec{R})$ (\ref{eq:deadlock}), using the
    characterization of the pre- and post-sets of the transitions from
    $\anetof{\signature,\struc_{\mathcal{T}},\beh}$ provided by Lemma
    \ref{lemma:flow-formula}.
  \end{compactitem}

  \vspace*{\baselineskip}
  \noindent (\ref{it:reachability:verification}) The proof of this
  point follows a similar argument as for point
  \ref{it:deadlock:verification}, the only difference being that the
  set of markings containing a set $\set{q_i[u_i]~|~i\in\interv{1}{n}}
  \subseteq \amark$, with $q_1[u_1], \ldots, q_n[u_n]$ pairwise
  distinct coincides with the set $\{\markof{\nu'}{\vec{X}}
  \mid\ \wsksmodels{\nu'}{\errset{\tuple{q_1, \ldots, q_n}}(\vec{X})
    \wedge \ismark(\vec{X},\vec{R}) \wedge
    \rewrtree(\vec{R})}\}$. \qed}

Since the \wsks\ logic is decidable with non-elementary complexity, in
general \cite{Thatcher2005GeneralizedFA}, the problem of checking the
satisfiability of the verification conditions provided by Theorem
\ref{thm:verification} is decidable. By inspection of the formul{\ae}
(\ref{it:deadlock:verification}) and
(\ref{it:reachability:verification}), one can show that checking the
verification conditions is in \exptime{4}, where $4$ is the maximum
quantifier alternation depth of these formul{\ae}. In practice,
however, these checks are quite fast, as shown by the preliminary
experiments performed on a similar encoding, reported in our previous
work \cite{DBLP:conf/facs2/BozgaI21}.

\section{Related Work}

This paper reports on a resource logic for the specification of sets
of configurations of parametric distributed systems with unbounded
numbers of components (processes) and a verification method for safety
properties, such as absence of deadlocks and critical section
violations.

Traditionally, verification of unbounded networks of parallel
processes considers known architectural patterns, typically cliques or
rings \cite{GermanSistla92,ClarkeGrumbergBrowne86}. Because the price
for decidability is drastic restriction on architecture styles \cite{BloemJacobsKhalimovKonnovRubinVeithWidder15}, more
recent works propose practical semi-algorithms, e.g.\ \emph{regular
model checking}
\cite{KestenMalerMarcusPnueliShahar01,AbdullaHendaDelzannoRezine07} or
\emph{automata learning} \cite{ChenHongLinRummer17}. Here the
architectural pattern is implicitly determined by the class of
language recognizers: word automata encode pipelines or rings, whereas
tree automata describe trees.

A first attempt at specifying architectures by logic is the
\emph{interaction logic} of Konnov et al. \cite{KonnovKWVBS16}, which
is a combination of Presburger arithmetic with monadic uninterpreted
function symbols, that can describe cliques, stars and rings. More
structured architectures (pipelines and trees) can be described using
a second-order extension \cite{MavridouBBS17}. As such, these
interaction logics are undecidable and have no support for automated
verification. Recently, interaction logics that support the
verification of safety properties, by structural invariant synthesis
have been developed. These logics use fragments of first order logic
with interpreted function symbols that implicitly determine the class
of architectures, such as cliques \cite{BozgaIosifSifakis19a}, or
pipelines, rings and trees \cite{DBLP:conf/tacas/BozgaEISW20}.

From a theoretical point of view, our resource logic with inductive
definitions is strictly more expressive: for instance, a chain of
components where a certain component type occurs on all even (odd)
positions is provably not expressible in first order logic, but can be
easily defined using our language, whose inductive definitions allow
to describe second order constructs in a controlled manner (rather
than using unrestricted second order quantification, as in
\cite{MavridouBBS17}). Moreover, first order logic with successor
functions can describe at most tree-like architectures, whereas our
language describes structures more general than trees\footnote{For
instance, the tree-shaped architecture with leaves linked in a ring
\S\ref{sec:rewriting}.}, using no interpreted function symbols, other
than ports involved in interactions \cite{BozgaIosifSifakis19a,DBLP:conf/tacas/BozgaEISW20}.

Specifying parameterized component-based systems by inductive
definitions is, however, not new. \emph{Network grammars}
\cite{ShtadlerGrumberg89} use context-free grammar rules to describe
systems with linear (pipeline, token-ring) architectures obtained by
composition of an unbounded number of processes. In contrast, we use
predicate symbols of unrestricted arities to describe architectural
patterns that are, in general, more complex than trees. Such complex
structures can be specified recursively using graph grammar rules with
parameter variables \cite{LeMetayer,Hirsch}. To avoid clashes, these
variables must be renamed to unique names and assigned unique indices
at each unfolding step. Our specifications use quantifiers to avoid
name clashes and separating conjunction to guarantee that the
components and interactions obtained by the unfolding of the rules are
unique.

Verification of network grammars against safety properties requires
the synthesis of \emph{network invariants} \cite{WolperLovinfosse89},
computed by rather costly fixpoint iterations
\cite{LesensHalbwachsRaymond97} or by abstracting (forgetting the
particular values of indices in) the composition of a small bounded
number of instances \cite{KestenPnueliShaharZuck02}. Instead, our
method uses lightweight \emph{structural invariants}, that are
synthesized with little computational effort and prove to be efficient
in many practical examples
\cite{DBLP:conf/tacas/BozgaEISW20,DBLP:conf/facs2/BozgaI21}.

\section{Conclusions and Future Work}
\label{sec:conclusion}

The paper proposes a general framework for the practical
semi-algorithmic verification of parametric systems. The framework
integrates previous work of the authors and extends it through
developments in two complementary directions. The first direction is
modeling parametric system architectures from instances of predefined
component types and interaction types. Configuration logic allows the
specification of configurations characterizing architecture styles
with snapshots of their component states. Parametric system behavior
can be obtained, in the form of Petri nets using operational
semantics, from given configuration logic specifications and behaviors
of component types.  We show that even for very simple linear
architectures, essential safety properties such as deadlock-freedom or
mutual explosion are undecidable.

The second direction proposes a method that from given configuration
logic specifications and the finite-state behavior of its components,
leads to formulas of \wsks\ characterizing two types of structural
invariants used to prove deadlock-freedom and mutual exclusion
properties. The generation process avoids the complexity of
traditional fixpoint computation techniques. It synthesizes
constraints on the state space induced by the interactions of the
parametric architecture. Verification boils down to checking the
satisfiability of \wsks\ formul{\ae}, for which optimized solvers
exist. The whole generation process involves transformations from
configuration specifications into rewriting trees and then into
\wsks\ through path automata. Proving its soundness requires some
tedious technical developments; nonetheless, its implementation does
not suffer from the usual limitations due to state space complexity
and the number of components. Experimental results show that the
proposed verification method is scalable and allows proving safety in
a number of non-trivial cases \cite{DBLP:conf/facs2/BozgaI21}.

As future work, we plan on adding support for broadcast in our
specification language and develop further the invariant synthesis
method to take broadcast into account. We also envisage an extension
of the finite-state model of behavior to more complex automata, such
as pushdown or timed automata.

\bibliography{refs}

\begin{thebibliography}{10}
\expandafter\ifx\csname url\endcsname\relax
  \def\url#1{\texttt{#1}}\fi
\expandafter\ifx\csname urlprefix\endcsname\relax\def\urlprefix{URL }\fi
\expandafter\ifx\csname href\endcsname\relax
  \def\href#1#2{#2} \def\path#1{#1}\fi

\bibitem{KramerMagee98}
J.~{Kramer}, J.~{Magee}, Analysing dynamic change in distributed software
  architectures, IEE Proceedings - Software 145~(5) (1998) 146--154.

\bibitem{DBLP:conf/facs2/BozgaI21}
M.~Bozga, R.~Iosif, Specification and safety verification of parametric
  hierarchical distributed systems, in: 17th International Conference, {FACS}
  2021, Virtual Event, October 28-29, 2021, Proceedings, Vol. 13077 of Lecture
  Notes in Computer Science, Springer, 2021, pp. 95--114.

\bibitem{DBLP:conf/lics/Reynolds02}
J.~C. Reynolds, Separation logic: {A} logic for shared mutable data structures,
  in: 17th {IEEE} Symposium on Logic in Computer Science {(LICS} 2002), {IEEE}
  Computer Society, 2002, pp. 55--74.

\bibitem{DBLP:journals/corr/abs-2107-05253}
E.~Ahrens, M.~Bozga, R.~Iosif, J.~Katoen,
  \href{https://arxiv.org/abs/2107.05253}{Local reasoning about parameterized
  reconfigurable distributed systems}, CoRR abs/2107.05253 (2021).
\newline\urlprefix\url{https://arxiv.org/abs/2107.05253}

\bibitem{DBLP:journals/tcs/OHearn07}
P.~W. O'Hearn, Resources, concurrency, and local reasoning, Theor. Comput. Sci.
  375~(1-3) (2007) 271--307.

\bibitem{Sergey16}
I.~Sergey, A.~Nanevski, A.~Banerjee, G.~A. Delbianco, Hoare-style
  specifications as correctness conditions for non-linearizable concurrent
  objects, SIGPLAN Not. 51~(10) (2016) 92--110.

\bibitem{FarkaN0DF21}
F.~Farka, A.~Nanevski, A.~Banerjee, G.~A. Delbianco, I.~F{\'{a}}bregas, On
  algebraic abstractions for concurrent separation logics, Proc. {ACM} Program.
  Lang. 5~({POPL}) (2021) 1--32.

\bibitem{DBLP:journals/jsat/BarrettST07}
C.~W. Barrett, I.~Shikanian, C.~Tinelli, An abstract decision procedure for a
  theory of inductive data types, J. Satisf. Boolean Model. Comput. 3~(1-2)
  (2007) 21--46.

\bibitem{CousotCousot79}
P.~Cousot, R.~Cousot, Systematic design of program analysis frameworks, in: 6th
  Annual ACM SIGPLAN-SIGACT Symposium on Principles of Programming Languages,
  ACM Press, New York, NY, 1979, pp. 269--282.

\bibitem{Post36}
E.~L. Post, {Finite Combinatory Processes-Formulation 1}, The Journal of
  Symbolic Logic 1~(3) (1936) 103--105.

\bibitem{Davis78}
M.~Davis, {What is a Computation?}, in: Mathematics Today Twelve Informal
  Essays, Conference Board of the Mathematical Sciences, 1978, pp. 241--267.

\bibitem{Thatcher2005GeneralizedFA}
J.~Thatcher, J.~Wright, Generalized finite automata theory with an application
  to a decision problem of second-order logic, Mathematical systems theory 2
  (2005) 57--81.

\bibitem{EmersonSistla96}
E.~Emerson, A.~Sistla, Symmetry and model checking, Formal Methods in System
  Design 9 (1996) 105--131.
\newblock \href {https://doi.org/10.1007/BF00625970}
  {\path{doi:10.1007/BF00625970}}.

\bibitem{Mona}
J.~G. Henriksen, J.~L. Jensen, M.~E. J{\o}rgensen, N.~Klarlund, R.~Paige,
  T.~Rauhe, A.~Sandholm, Mona: Monadic second-order logic in practice, in:
  First International Workshop, {TACAS} '95, Vol. 1019 of LNCS, Springer, 1995,
  pp. 89--110.

\bibitem{Bojanczyk08}
M.~Boja\'{n}czyk, Tree-walking automata, in: Language and Automata Theory and
  Applications: Second International Conference, LATA 2008, Tarragona, Spain,
  March 13-19, 2008. Revised Papers, Springer-Verlag, Berlin, Heidelberg, 2008,
  p. 1–2.

\bibitem{KhoussainovNerode}
B.~Khoussainov, A.~Nerode, Automata Theory and its Applications, Springer,
  2001.

\bibitem{DBLP:conf/mfcs/Sifakis78}
J.~Sifakis, Structural properties of petri nets, in: Mathematical Foundations
  of Computer Science 1978, Proceedings, 7th Symposium, Zakopane, Poland,
  September 4-8, 1978, Vol.~64 of Lecture Notes in Computer Science, Springer,
  1978, pp. 474--483.

\bibitem{GermanSistla92}
S.~M. German, A.~P. Sistla, Reasoning about systems with many processes, J.
  {ACM} 39~(3) (1992) 675--735.

\bibitem{ClarkeGrumbergBrowne86}
M.~Browne, E.~Clarke, O.~Grumberg, Reasoning about networks with many identical
  finite state processes, Information and Computation 81~(1) (1989) 13 -- 31.

\bibitem{BloemJacobsKhalimovKonnovRubinVeithWidder15}
R.~Bloem, S.~Jacobs, A.~Khalimov, I.~Konnov, S.~Rubin, H.~Veith, J.~Widder,
  Decidability of Parameterized Verification, Synthesis Lectures on Distributed
  Computing Theory, Morgan {\&} Claypool Publishers, 2015.

\bibitem{KestenMalerMarcusPnueliShahar01}
Y.~Kesten, O.~Maler, M.~Marcus, A.~Pnueli, E.~Shahar, Symbolic model checking
  with rich assertional languages, Theoretical Computer Science 256~(1) (2001)
  93--112.

\bibitem{AbdullaHendaDelzannoRezine07}
P.~A. Abdulla, G.~Delzanno, N.~B. Henda, A.~Rezine, Regular model checking
  without transducers (on efficient verification of parameterized systems), in:
  13th International Conference, {TACAS} 2007, Vol. 4424 of LNCS, Springer,
  2007, pp. 721--736.

\bibitem{ChenHongLinRummer17}
Y.~Chen, C.~Hong, A.~W. Lin, P.~R{\"{u}}mmer, Learning to prove safety over
  parameterised concurrent systems, in: 2017 Formal Methods in Computer Aided
  Design, {FMCAD} 2017, {IEEE}, 2017, pp. 76--83.

\bibitem{KonnovKWVBS16}
I.~V. Konnov, T.~Kotek, Q.~Wang, H.~Veith, S.~Bliudze, J.~Sifakis,
  Parameterized systems in {BIP:} design and model checking, in: 27th
  International Conference on Concurrency Theory, {CONCUR} 2016, Vol.~59 of
  LIPIcs, Schloss Dagstuhl - Leibniz-Zentrum f{\"{u}}r Informatik, 2016, pp.
  30:1--30:16.

\bibitem{MavridouBBS17}
A.~Mavridou, E.~Baranov, S.~Bliudze, J.~Sifakis, Configuration logics: Modeling
  architecture styles, J. Log. Algebr. Meth. Program. 86~(1) (2017) 2--29.

\bibitem{BozgaIosifSifakis19a}
M.~Bozga, R.~Iosif, J.~Sifakis, Checking deadlock-freedom of parametric
  component-based systems, in: 25th International Conference, {TACAS} 2019,
  Vol. 11428 of LNCS, Springer, 2019, pp. 3--20.

\bibitem{DBLP:conf/tacas/BozgaEISW20}
M.~Bozga, J.~Esparza, R.~Iosif, J.~Sifakis, C.~Welzel, Structural invariants
  for the verification of systems with parameterized architectures, in: 26th
  International Conference, {TACAS} 2020, Vol. 12078 of LNCS, Springer, 2020,
  pp. 228--246.

\bibitem{ShtadlerGrumberg89}
Z.~Shtadler, O.~Grumberg, Network grammars, communication behaviors and
  automatic verification, in: Automatic Verification Methods for Finite State
  Systems, International Workshop, Vol. 407 of LNCS, Springer, 1989, pp.
  151--165.

\bibitem{LeMetayer}
D.~Le~Metayer, Describing software architecture styles using graph grammars,
  IEEE Transactions on Software Engineering 24~(7) (1998) 521--533.

\bibitem{Hirsch}
D.~Hirsch, P.~Inverardi, U.~Montanari, Graph grammars and constraint solving
  for software architecture styles, in: Proceedings of the Third International
  Workshop on Software Architecture, ISAW '98, Association for Computing
  Machinery, New York, NY, USA, 1998, p. 69–72.

\bibitem{WolperLovinfosse89}
P.~Wolper, V.~Lovinfosse, Verifying properties of large sets of processes with
  network invariants, in: Automatic Verification Methods for Finite State
  Systems, International Workshop, Vol. 407 of LNCS, Springer, 1989, pp.
  68--80.

\bibitem{LesensHalbwachsRaymond97}
D.~Lesens, N.~Halbwachs, P.~Raymond, Automatic verification of parameterized
  linear networks of processes, in: The 24th {ACM} {SIGPLAN-SIGACT} Symposium
  on Principles of Programming Languages, {ACM} Press, 1997, pp. 346--357.

\bibitem{KestenPnueliShaharZuck02}
Y.~Kesten, A.~Pnueli, E.~Shahar, L.~D. Zuck, Network invariants in action, in:
  {CONCUR} 2002 - Concurrency Theory, 13th International Conference, Vol. 2421
  of LNCS, Springer, 2002, pp. 101--115.

\end{thebibliography}

\end{document}